\documentclass[aps,prd,twocolumn,nofootinbib]{revtex4-2}

\usepackage{orcidlink}
\usepackage{bm}
\usepackage{amsmath}
\usepackage{graphicx}
\usepackage{subcaption}

\usepackage{changes}
\usepackage{latexsym}
\usepackage{amsfonts}
\usepackage{mathrsfs}
\usepackage{CJKutf8}

\usepackage{float}
\usepackage{color}
\usepackage{comment}
\usepackage{hyperref}
\usepackage{multirow}

\hypersetup{
	colorlinks=false,
	linkcolor=blue,
	citecolor=blue,
}

\newcommand{\physrep}{Physics Reports} 
\begin{document}

\begin{CJK*}{UTF8}{gbsn} 
	
\title{The Environmental Effects on Inspiraling Binary Black Hole Systems in the Centers of the LMC and M31} 
	
\author{Meng Xu$^{1,3,4}$\,\orcidlink{0009-0009-7230-631X}}
\author{Zhijin Li$^{1,3,4}$\,\orcidlink{0009-0003-1638-3026}}
\author{Xiao Guo$^{5,1}$\,\orcidlink{0000-0001-5174-0760}}
\email[Corresponding author:~]{guoxiao17@mails.ucas.ac.cn}
\author{Yun-Long Zhang$^{2,1}$\,\orcidlink{0000-0002-7225-8479}}
\email[Corresponding author:~]{zhangyunlong@nao.cas.cn}

\affiliation{$^{1}$School of Fundamental Physics and Mathematical Sciences, Hangzhou Institute for Advanced Study, UCAS, Hangzhou 310024, China}

\affiliation{$^{2}$National Astronomical Observatories, Chinese Academy of Sciences, Beijing 100101, China}

\affiliation{$^{3}$School of Physical Sciences, University of Chinese Academy of Sciences, Beijing 100049, China}

\affiliation{$^{4}$Institute of Theoretical Physics, Chinese Academy of Sciences, Beijing 100190, China}

\affiliation{$^{5}$Institute for Gravitational Wave Astronomy, Henan Academy of Sciences, Zhengzhou 450046, Henan, China}

\begin{abstract}

Binary black hole (BBH) systems residing in the centers of galaxies evolve within complex astrophysical environments. These environments, comprising dark matter (DM) halos and baryonic accretion disks, can significantly alter the orbital dynamics of the binaries and their resulting gravitational wave (GW) emission. In this study, we investigate the dynamical evolution and GW waveforms of BBH systems embedded in the centers of the Large Magellanic Cloud (LMC) and the Andromeda Galaxy (M31). We construct a comprehensive analytical framework that jointly incorporates GW radiation reaction, DM spike effects (including dynamical friction (DF) and accretion, derived from the NFW profile), and accretion disk perturbations. Using this framework, we track the long-term evolution of the binary's semi-latus rectum $p$ and orbital eccentricity $e$. Our simulations reveal that the coexistence of a DM spike and an accretion disk significantly accelerates the inspiral process compared to pure DM or vacuum scenarios. Crucially, to assess the observability of these environmental effects, we calculate the Signal-to-Noise Ratio (SNR) and waveform Mismatch for future Pulsar Timing Arrays (PTAs). Our analysis demonstrates that these systems can achieve robust detectability thresholds ($\text{SNR} \ge 8$) within specific parameter spaces. Furthermore, the substantial Mismatch (reaching $\sim 0.7$ over a 20-year observation in the LMC scenario) indicates that the phase deviations induced by these environmental effects are highly distinguishable from vacuum templates. These findings predict the prospect of using future GW detections to probe complex galactic environments.

\end{abstract}

\maketitle
	
\end{CJK*}

\tableofcontents

\allowdisplaybreaks

\section{Introduction}

Since the first detection of GWs by LIGO in 2015, gravitational wave astronomy has developed rapidly~\cite{2019PhRvX...9c1040A,2017ApJ...848L..12A,2017PhRvL.119p1101A,2016PhRvL.116f1102A}. Standard parameter estimation for these events typically models the binaries as isolated systems evolving in a vacuum. While this approximation is generally robust for stellar-mass BH mergers where environmental densities are negligible, it may not suffice for all astrophysical scenarios. In reality, BHs do not strictly exist in isolation. This is particularly true for Supermassive Black Holes (SMBHs) or Intermediate-Mass Black Holes (IMBHs) residing in the dense cores of galaxies~\cite{s4wh-x6c4}. Instead, they are embedded in complex environments consisting of DM halos and baryonic accretion disks, interacting dynamically with their surroundings.

As these massive binary systems evolve, they emit GWs in the nanohertz (nHz) to microhertz ($\mu$Hz) frequency bands, which are the prime targets for PTAs~\cite{2010CQGra..27h4016S,2011MNRAS.414.3251L,2013CQGra..30x4009S,2014MNRAS.444.3709Z,2016MNRAS.459.1737S,2019BAAS...51c.336T,2023ApJ...951L...8A,2024A&A...685A..94E,2013CQGra..30v4011L,PhysRevLett.118.151104} and future space-based detectors. Unlike the stellar-mass binaries detected by ground-based interferometers, the dynamical evolution of Intermediate-Mass-Ratio Inspirals (IMRIs) or SMBH binaries in galactic centers is significantly perturbed by environmental effects. The interaction with the surrounding medium, through mechanisms such as DF and accretion, causes the orbital evolution to deviate from vacuum predictions~\cite{s4wh-x6c4}. These deviations imprint unique signatures on the GW waveforms~\cite{Zwick_2023}, offering unprecedented opportunities to probe the astrophysical environment, constrain the properties of DM, and test general relativity (GR) in the strong-field regime~\cite{2020ScPC....3....7B,theligoscientificcollaboration2025blackholespectroscopytests,guo2025theoryagnostichierarchicalbayesianframework}.

A key environmental component is DM. Since Fritz Zwicky first predicted the existence of DM while studying the Coma Cluster in 1933~\cite{1933AcHPh...6..110Z}, accumulating evidence has confirmed its ubiquity in the universe~\cite{1970ApJ...159..379R,1978ApJ...225L.107R,2006ApJ...648L.109C,2016A&A...594A..13P}. In the 1990s, Navarro, Frenk, and White (NFW) derived a universal density profile for DM halos~\cite{1996ApJ...462..563N,1997ApJ...490..493N}. Subsequently, the gNFW model generalized this by introducing additional free parameters to provide greater flexibility~\cite{2008Natur.454..735D}:
\begin{equation}
    \rho_{\mathrm{gNFW}}(r) = \frac{\rho_0}{(r/r_\mathrm{s})^{\gamma}(1 + r/r_\mathrm{s})^{3-\gamma}}.
\end{equation}
Within the gravitational influence of a central SMBH or IMBH, this DM halo can be adiabatically compressed, forming a high-density region known as a ``DM spike''~\cite{1999PhRvL..83.1719G}. The presence of such a spike exerts a drag force on the inspiraling secondary BH, primarily via DF~\cite{1943ApJ....97..255C,1999ApJ...513..252O}, which causes a dephasing of the GW signal. Although the secondary BH can also accrete DM mass, a phenomenon first explored in Ref.~\cite{Macedo_2013}, subsequent research demonstrated that this accretion effect is generally weaker than DF~\cite{PhysRevD.97.064003}. Consequently, the DM spike leaves imprints by altering the orbital inspiral period and phase~\cite{2013PhRvL.110v1101E,2014PhRvD..89j4059B,2015PhRvD..91d4045E}. Furthermore, studies on eccentric orbits have found that orbital eccentricity tends to gradually increase in these environments~\cite{PhysRevD.103.023015,PhysRevD.100.043013}, and other effects such as periastron precession have also been investigated~\cite{PhysRevD.106.064003,PhysRevD.107.084027}.

In addition to DM spikes, a significant environmental impact comes from the presence of a (baryonic) accretion disk~\cite{PhysRevD.77.104027,PhysRevD.89.104059}. Recent attempts have been made to map the effects of accretion disks in IMRIs~\cite{PhysRevX.13.021035,Cole_2023,Derdzinski_2020}. While many studies analyze these factors in isolation, these two components likely coexist in galactic centers. A more comprehensive modeling approach is required. For instance, in~\cite{Becker:2021ivq,Becker:2022wlo}, the environmental impacts of accretion disks and DM spikes were compared, modeling IMRIs on eccentric Keplerian orbits by including GW emission, DF from DM spikes, and gas interactions with the accretion disk.

To assess the detectability of these environmental effects, it is essential to ground the analysis in specific astrophysical targets.~\citet{2025ApJ...978..104G} suggest that, within plausible parameter spaces, IMBHs and SMBHs may exist in the centers of the LMC and the M31~\cite{2025ApJ...978..104G}. Such central BHs could capture lower-mass companions, forming inspiraling systems with orbital periods ranging from months to years, precisely falling within the sensitivity window of PTAs. Affected by the DM environment, the GW signals from such systems may exhibit characteristics distinct from those in DM-free scenarios, providing a potential approach to probe the DM distribution in galactic central regions.

In this paper, we construct a comprehensive framework to study the dynamical evolution of a secondary BH orbiting a central BH located in the centers of the LMC and M31. Unlike previous works, we consider orbital dynamics jointly influenced by the reaction of GWs, a central DM spike, and an accretion disk. Focusing on the DM density distribution in these galaxies, we use the NFW profile ($\gamma=1$) to describe the halo and derive a modified DM spike model accounting for central BH enhancement. We combine two accretion scenarios, $\alpha$ accretion disk with a DM spike and $\beta$ accretion disk with a DM spike, and investigate their effects on the evolution of the orbital semi-latus rectum $p$ and eccentricity $e$ during the inspiral process. By utilizing the specific masses and orbital parameters relevant to LMC and M31, we compare GW waveform deviations induced by these environments against the vacuum scenario. We anticipate that typical BBH systems under these diverse environmental influences will exhibit large eccentricities and significant phase deviations. These findings may enable the measurement of the DM density distribution, thereby providing constraints on DM models and deepening our understanding of the nature of DM.

The structure of this paper is as follows: In Section~\ref{sec:profile}, we introduce the NFW profile of DM distribution in LMC and M31, and derive the spike-like modification formed under the gravitational influence of the central BH.In Section~\ref{sec:equ}, we focuse on the dynamical effects of the DM spike on the evolution of orbital parameters and present the numerical results. In Section~\ref{sec:acc}, we introduce the modeling of (baryonic) accretion disk of the central BH, and examine its impact on the behavior of the secondary BH. In Section~\ref{sec:mismatch}, we calculate the GW waveforms and illustrate the detectability of signal under the influence of various environmental effects. Finally, conclusions are drawn in Section~\ref{sec:con}, followed by a summary of additional discussions in Section~\ref{sec:dis}.

\section{DM density profile }
\label{sec:profile}

\begin{figure}
    \centering
    \includegraphics[width=1\linewidth]{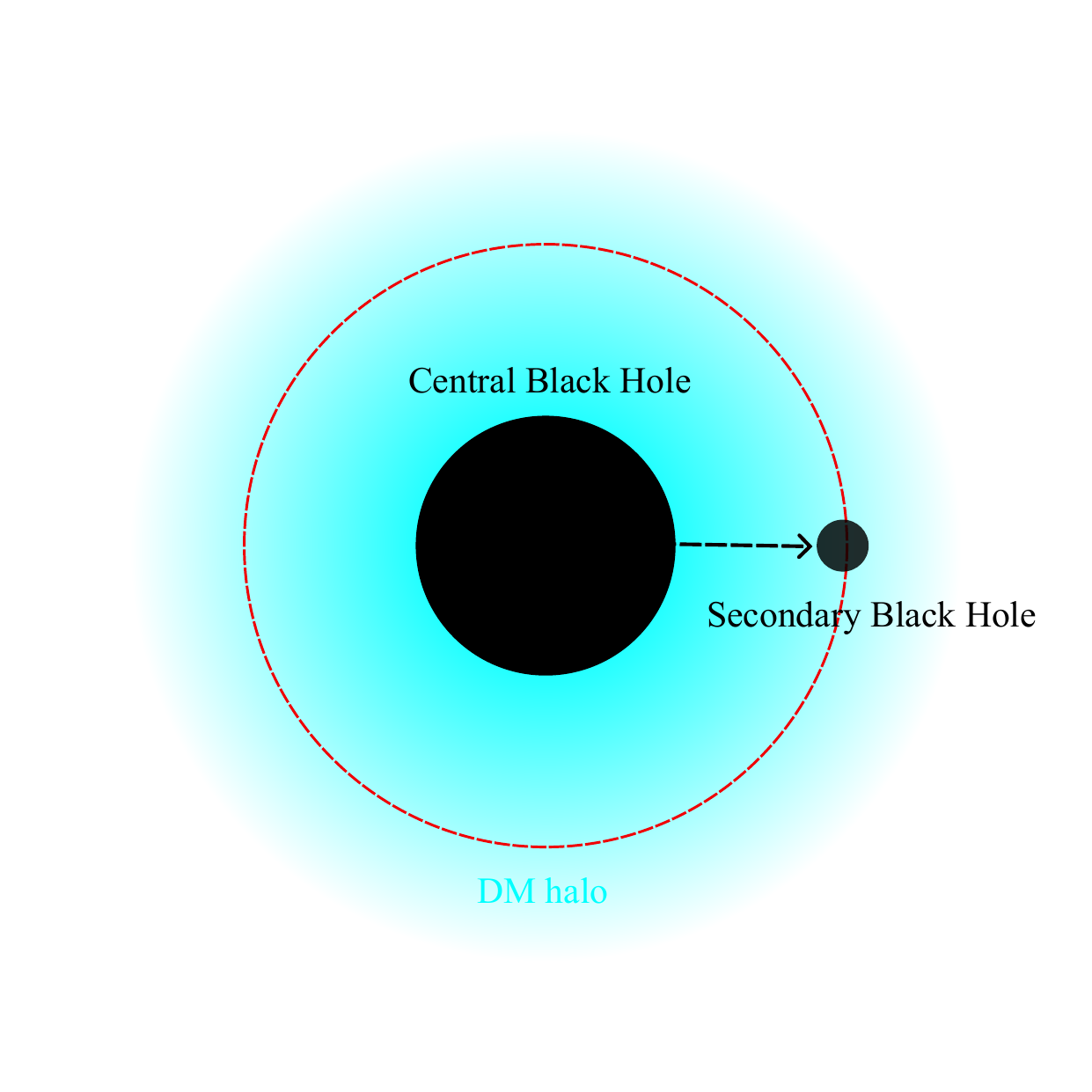}
    \caption{Schematic illustration of a BBH system, comprising a central massive BH embedded within a spherically symmetric DM halo, and a secondary BH undergoing an eccentric inspiral. For visual clarity, the secondary BH's orbit is depicted as circular.}
    \label{fig:DMHalo}
\end{figure}

In this section, we model a binary system embedded within a DM halo at the galactic center. This system comprises a central massive black hole (SMBH or IMBH) and a smaller secondary BH, forming a BBH with a mass ratio $q$. According to~\cite{2025ApJ...978..104G}, we choose the mass of the IMBH as $m_{\text{Pri}}=m_{\text{1}} =2.4\times10^4M_{\odot}$, and the mass of the secondary BH as $m_{\text{Sec}}=m_{\text{2}}$ in LMC; the mass of the SMBH as $m_{\text{Pri}} =m_{\text{1}}=1.4\times10^8M_{\odot}$, and the mass of the secondary BH as $m_{\text{Sec}}=m_{\text{2}}$ in M31. We assume that this BBH system undergoes a process of gradual inspiral and eventual merger driven by GWs. In Fig.~\ref{fig:DMHalo}, we depict such an inspiraling BBH where two celestial bodies move in the same plane, with the central BH surrounded by a spherically symmetric DM halo, and the secondary BH moving within the DM spike composed of the halo.

Following references~\cite{1996ApJ...462..563N,1997ApJ...490..493N}, we adopt the following DM distribution model:
\begin{equation}
    \rho_{\mathrm{NFW}}(r) = \frac{\rho_0}{(r/r_\mathrm{s})(1 + r/r_\mathrm{s})^2},
    \label{eqNFW}
\end{equation}
where $r$ is the distance from the test point to the central BH at the center of the galaxy. The NFW profile is an empirical formula discovered through cosmological simulations, which describes how the density $\rho$ of DM halo varies with radius $r$ from the galactic center to the outer regions under the cold dark matter cosmological model ($\Lambda\mathrm{CDM}$). $\rho_0$ denotes the typical scale density, and $r_\mathrm{s}$ represents the typical scale radius, defining the transition point between the inner core and outer regions of the halo.

In this study, we analytically derive the DM density profile by imposing specific macroscopic constraints on the galactic structure. We can define the virial mass $M$ as the total mass of the galaxy within a sphere centered on the galaxy's core with a radius equal to the virial radius $r_\mathrm{vir}$. Through references~\cite{2025ApJ...983..151C,2012A&A...546A...4T}, we have listed the relevant parameters for LMC and M31 in Table~\ref{tab:galaxy}.

\begin{table}
    \centering
    \begin{tabular}{cccc}
    \hline\hline
        galaxy & $M_\mathrm{vir}/M_{\odot}$ & $R_\mathrm{vir}$/kpc & $r_\mathrm{s}$/kpc\\
        \hline
        LMC & $1.8\times10^{11}$ & 120 & 13\\
        \hline
        M31 & $1.0\times10^{12}$ & 200 & 16\\
        \hline\hline
    \end{tabular}
    \caption{The virial mass $M_\mathrm{vir}$, virial radius $R_\mathrm{vir}$, and scale radius $r_\mathrm{s}$ for the LMC and M31 galaxies, assuming a spherically symmetric DM halo model.}
    \label{tab:galaxy}
\end{table}

\begin{figure}[htbp]
    \centering
    \includegraphics[width=0.4\textwidth]{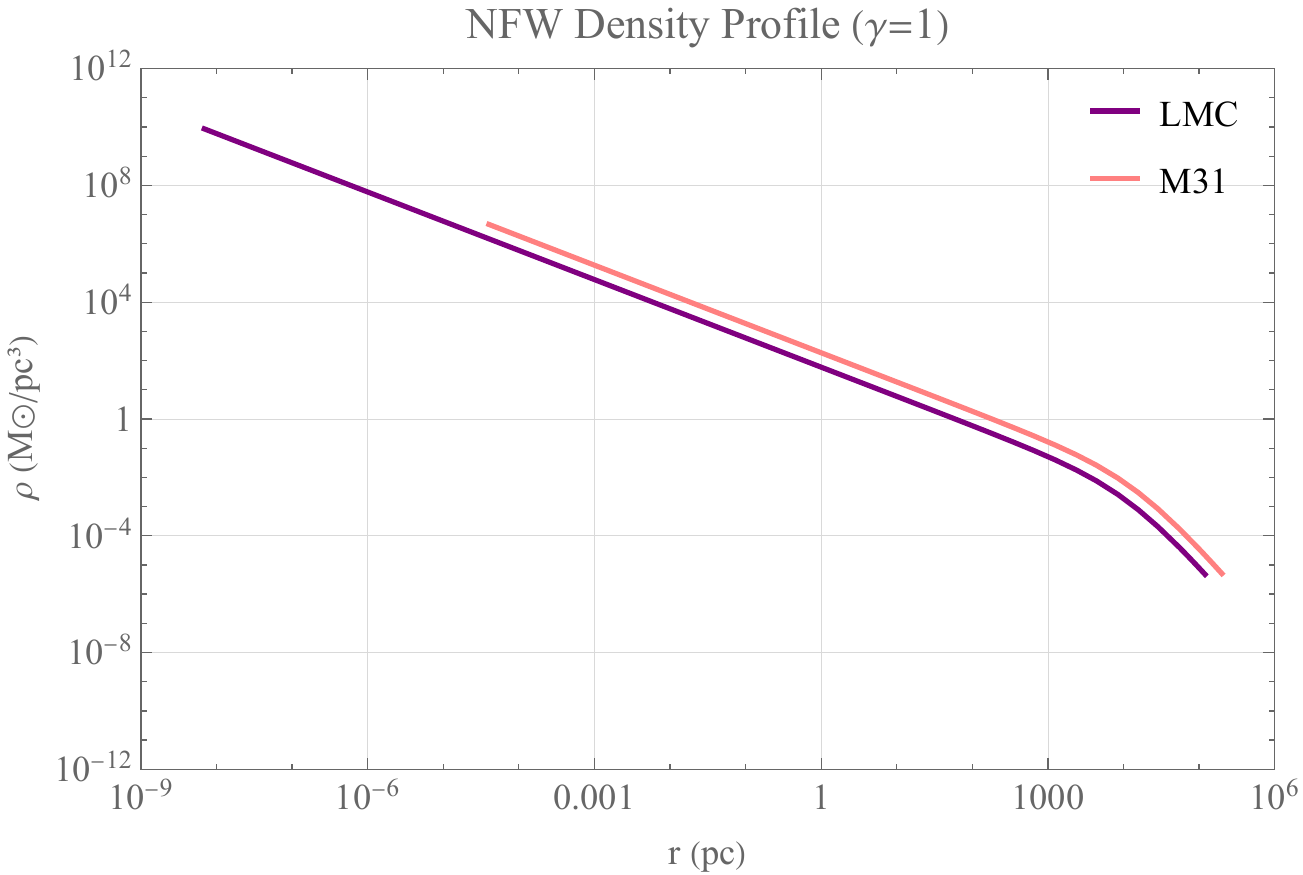}
    \caption{NFW density profiles for the LMC and M31, computed using the parameters detailed in Tables~\ref{tab:galaxy} and~\ref{tab:mpri}. The radial extent of each curve is bounded by  $r_{\mathrm{ISCO}}$ of the central BH and $R_\mathrm{vir}$ of the  galaxy. The larger $r_{\mathrm{ISCO}}$ cutoff for M31 reflects the greater mass of its central BH. The DM density is assumed to vanish in the region $r < r_{\mathrm{ISCO}}$.}
    \label{fig:NFW}
\end{figure}

Since the luminous matter of the LMC and M31 is negligible compared to the DM halo, the mass of the DM halo can be considered as the total mass of the galaxy~\cite{2025JCAP...02..067H}:
\begin{equation}
    \int_{r_{\text{ISCO}}}^{R_\mathrm{vir}} 4\pi\rho_{\text{gNFW}}(r)r^2 dr = M_\mathrm{vir},
    \label{eqMvir}
\end{equation}
here $r_{\text{ISCO}}$ represents the innermost stable circular orbit (ISCO) of the central BH~\cite{2015PhRvD..91d4045E}:
\begin{equation}
    r_{\mathrm{ISCO}} = 3R_{\mathrm{s}} = \frac{6Gm_{\text{1}}}{c^{2}},
\end{equation}
where $ m_{\text{1}} $ is the mass of central BH. $ R_{\mathrm{s}}=2Gm_{\text{1}}/c^2 $ is the Schwarzschild radius of the central BH. By combining~(\ref{eqNFW}) and~(\ref{eqMvir}) and plugging in the parameters of the central BH and the galaxy, we can obtain $\rho_0$ showed in Table~\ref{tab:mpri}. As illustrated in Fig.~\ref{fig:NFW}, the DM density increases significantly toward the galactic center.

\begin{table}
    \centering
    \begin{tabular}{ccc}
    \hline\hline
        galaxy & LMC & M31 \\
        \hline
        $r_{\mathrm{ISCO}}$[pc] & $6.89\times10^{-9}$  & $4.02\times10^{-5}$ \\
        \hline
        $R_{\mathrm{s}}$[pc] & $2.2967\times10^{-9}$ & $1.34\times10^{-5}$ \\
        \hline
        $m_{\text{1}}[M_\odot]$ & $2.4\times10^4$ &$1.4\times10^8$ \\
        \hline
        $\rho_0[M_{\odot}/{\rm pc}^3]$ & $0.00458125$ &$0.0115867$
        \\
        \hline\hline
    \end{tabular}
    \caption{Key parameters for the central BHs and the NFW DM profiles in the LMC and M31. The listed parameters include the ISCO $r_{\mathrm{ISCO}}$, the Schwarzschild radius $R_{s}$, the central BH mass $m_{\text{1}}$, and the NFW scale density $\rho_{0}$.}
    \label{tab:mpri}
\end{table}

Due to the immense gravity and adiabatic growth of the central BH, the surrounding DM will form a high-density region known as a DM spike within its gravitational influence radius $r_{sp} \approx 0.2r_h$~\cite{2004cbhg.symp..263M}. Here, $r_h$ is the radius of gravitational influence of the central BH, defined by the equation~\cite{2020PhRvD.102h3006K}
\begin{equation}
    \int_0^{r_h}4 \pi  \rho_{\mathrm{NFW}} \left( r \right) r^2 dr = 2 m_{\text{1}}.
    \label{eqrh}
\end{equation}

Consequently, we employ a piecewise function to model the DM halo, accounting for the density enhancement induced by the central BH~\cite{1995ApJ...440..554Q,PhysRevLett.83.1719,2015PhRvD..91d4045E,2022PhRvD.106f4003D}. We assume that at the outer regions of the central BH in the galaxy, the DM distribution is described by the NFW profile, while within the gravitational influence radius $r_{sp}$, it follows a spike profile $\rho_{\rm spike}(r)=\rho_{ sp}\left(1 - \frac{4R_{\rm s}}{r}\right)^3\left(\frac{r_{ sp}}{r}\right)^{\gamma_{\rm sp}}$. The final function describing the DM distribution is
\begin{equation}
    \rho(r) = \begin{cases}0 , & r \leq r_{\rm ISCO}, \\\rho_{ sp}\left(1 - \frac{4R_{\rm s}}{r}\right)^3\left(\frac{r_{ sp}}{r}\right)^{\gamma_{\rm sp}} , & r_{\rm ISCO} < r \leq r_{ sp}, \\\rho_{\rm NFW}(r)= \frac{\rho_0}{(r/r_\mathrm{s})(1 + r/r_\mathrm{s})^2} , & r_{ sp} < r.\end{cases}
    \label{eqfinal}
\end{equation}
Here, $r_{sp}$ is used to characterize the range of DM spike, specifically the maximum radius of the spike. $\rho_{sp}$ is the DM density at the distance $r_{sp}$. If the initial DM halo has an NFW profile with power law index $\gamma_{ini}=1$, after the adiabatic growth of the central BH the parameter $\gamma_{sp}=(9-2\gamma_{ini})/(4-2\gamma_{ini})=7/3$~\cite{2025JCAP...02..067H}. We require the DM density to vary continuously within the galaxy, therefore at $r=r_{sp}$, we have $\rho_{\rm spike}(r_{sp})=\rho_{\rm NFW}(r_{sp})$, and we can solve for $\rho_{sp}$:
\begin{equation}
    \rho_{\rm sp} = \frac{\rho_{\rm NFW}(r_{\rm sp})}{(1-4R_{\rm s}/r_{\rm sp})^3}.
    \label{eqsp}
\end{equation}

\begin{figure}[htbp]
    \centering
    \includegraphics[width=0.4\textwidth]{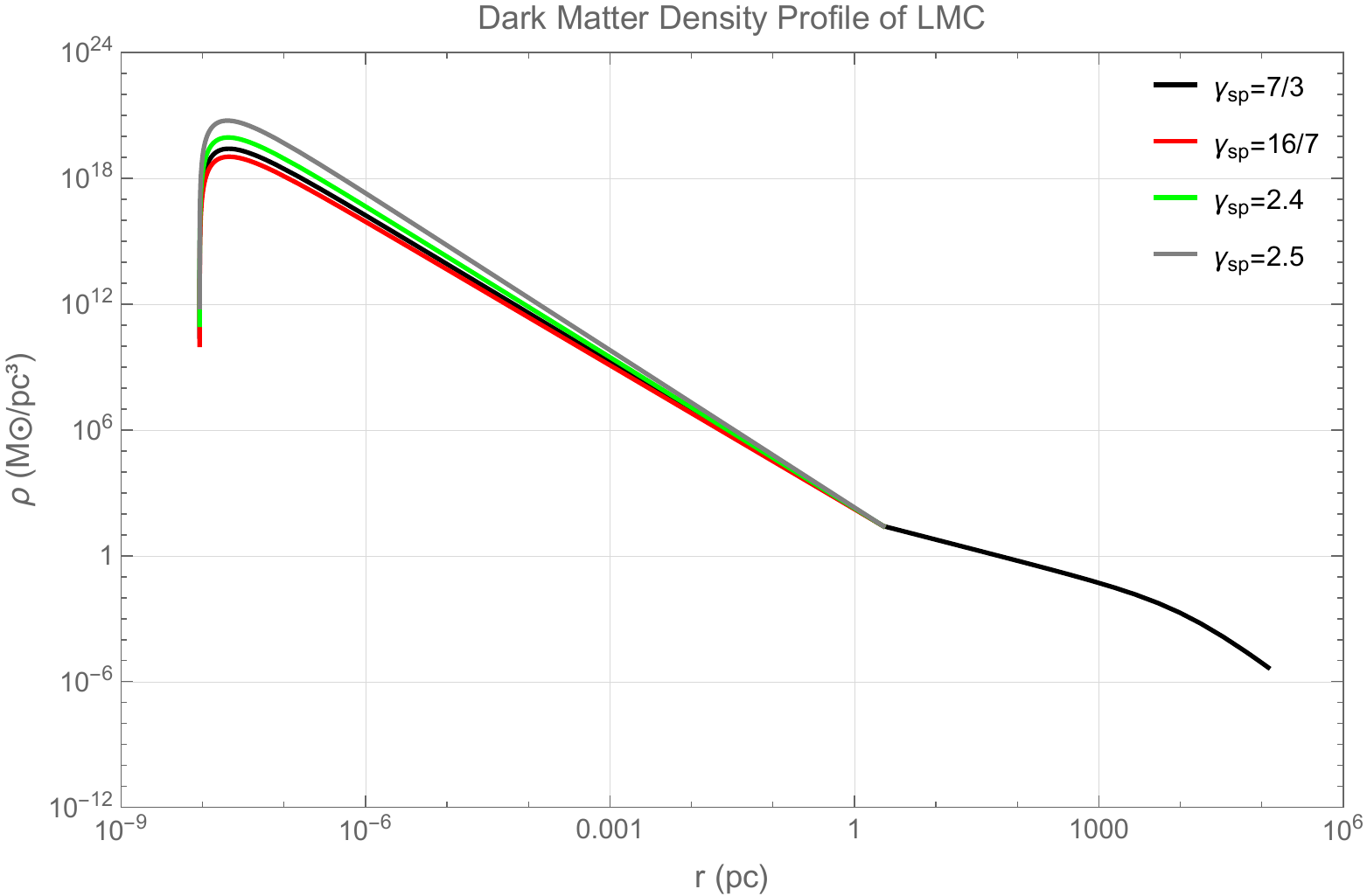}
    \\
    \includegraphics[width=0.4\textwidth]{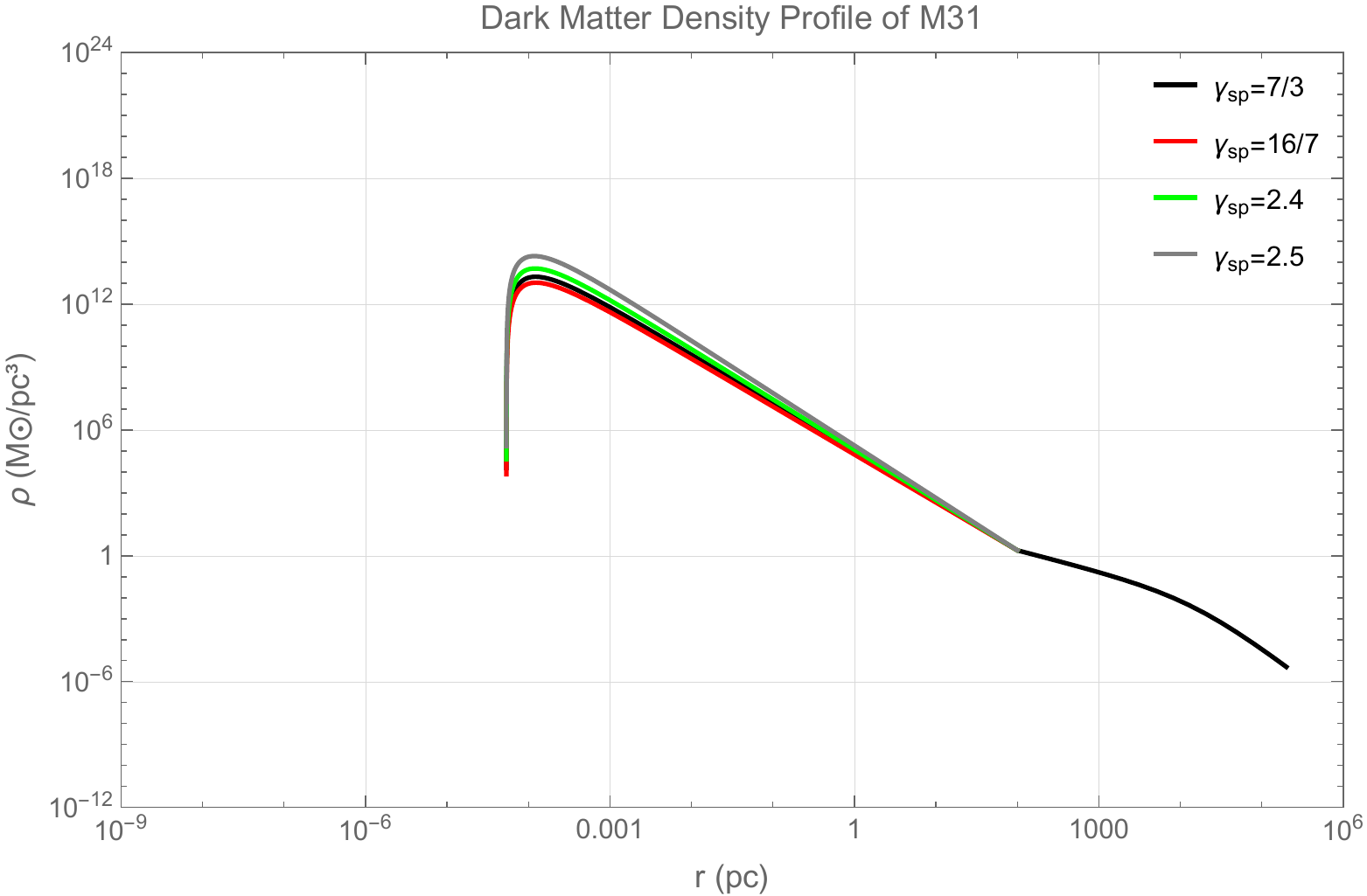}
    \caption{The initial DM distribution around the central BH follows the NFW profile. The strong gravitational potential and adiabatic growth of the central BH induce the formation of a dense DM spike extending up to a radius $r_{sp}$. Inward from $r_{sp}$, the DM density is substantially enhanced relative to the unperturbed NFW profile. Furthermore, a steeper spike index $\gamma_{sp}$ corresponds to a higher central DM density. The DM distribution is assumed to vanish within the ISCO.}
    \label{fig:DMSPIKE}
\end{figure}

By combining Eqs.~(\ref{eqNFW}) and~(\ref{eqrh}) and substituting the relevant parameters, we present the results in Table~\ref{tab:spike}. Then substituting the obtained parameters into Eq.~(\ref{eqfinal}), we obtain the final DM profile of the LMC and M31 showed in Fig.~\ref{fig:DMSPIKE}.

\begin{table}
    \centering
    \begin{tabular}{cccc}
    \hline\hline
         & $r_h$[$\mathrm{pc}$] & $r_{sp}$[$\mathrm{pc}$] & $\rho_{sp}$[$M_{\odot}/\mathrm{pc}^3$]\\          
         \hline
        LMC & 11.3323 & 2.26647 & 26.2679\\
        \hline
        M31 & 500.47 & 100.094 & 1.82917\\
        \hline\hline
    \end{tabular}
    \caption{Derived parameters for the DM spikes in the LMC and M31, including the gravitational influence radius $r_{h}$ of the central BH, the maximum spike radius $r_{sp}$, and the DM density $\rho_{sp}$ evaluated at $r_{sp}$.}
    \label{tab:spike}
\end{table}

\section{Dynamical equations}
\label{sec:equ}

In this section, we consider that the orbital evolution of the BBH is influenced by multiple dynamical factors, including the gravity of the central BH, the reaction of GWs, and the DF and accretion of the secondary BH caused by the DM spike. Reference~\cite{PhysRevD.106.064003} considered the orbital precession caused by the gravitational effect of the DM spike distribution, which we have ignored in this paper. We will discuss the effects of these factors one by one.

\subsection{Classical Keplerian motion}

\begin{figure}
    \centering
    \includegraphics[width=1\linewidth]{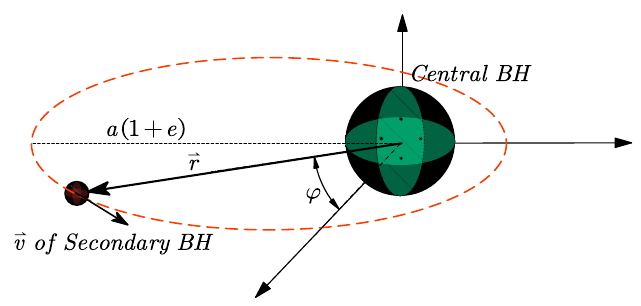}
    \caption{Orbital geometry of the BBH system, modeled as two Schwarzschild BHs restricted to the equatorial plane. The secondary BH follows a perturbed Keplerian orbit governed by the central BH's gravitational potential.}
    \label{fig:BBHs}
\end{figure}

Initially, we model the unperturbed system where the secondary BH undergoes classical Keplerian motion around the central BH. The schematic diagram of BBH is illustrated in Fig.~\ref{fig:BBHs}.

The Keplerian orbit is described by the following equation:
\begin{equation}
    r = \frac{p}{1 + e \cos(\varphi)},
    \label{eq8}
\end{equation}
where $r$ is the radial distance, $p$ is the semi-latus rectum, $e$ is the eccentricity, and $\varphi$ is the polar angle on the equatorial plane. Here, we set the total mass of BBH $m=m_{\text{1}}+m_{{\text{2}}}\approx m_{\text{1}}$ and the reduced mass $\mu=m_{\text{1}}m_{\text{2}}/m\approx m_{\text{2}}$. Based on the Newtonian Mechanics, the orbital angular momentum is
\begin{equation}
    L = \mu r ^ { 2 } \dot { \varphi },
    \label{eq9}
\end{equation}
and the total energy is
\begin{equation}
    E=\frac{1}{2}\mu(\dot{r}^2+r^2\dot{\varphi}^2)-\frac{G\mu m}{r}\\=\frac{1}{2}\mu\dot{r}^2+\frac{L^2}{2\mu r^2}-\frac{G\mu m}{r}.
    \label{eq10}
\end{equation}
Within the framework of classical mechanics, both orbital energy $E$ and angular momentum $L$ remain strictly conserved in the absence of dissipative forces.

It's convenient to describe Keplerian motion in terms of the semi-latus rectum $p$ and the eccentricity $e$ at any bounded equatorial orbit. Using the conclusions from the literature~\cite{PhysRevD.100.043013}, we can express $p$ and $e$ using $L$ and $E$:
\begin{equation}
    p = \frac{L^2}{Gm\mu^2},
    \label{eq11}
\end{equation}
\begin{equation}
    e^2 = 1 + \frac{2EL^2}{G^2m^2\mu^3}.
    \label{eq12}
\end{equation}
Differentiating the above two formulas, we obtain
\begin{equation}
    \frac{dp}{dt} = \frac { 2 L } { G m \mu ^ { 2 } } \dot { L } = 2 \sqrt { \frac { p } { G m \mu ^ { 2 } } } \dot { L },
    \label{eq13}
\end{equation}
\begin{equation}
    \frac{de}{dt}=\frac{p}{Gm\mu e}\dot{{E}}+\frac{(e^2-1)}{e\sqrt{Gm\mu^2 p}}\dot{{L}}.
    \label{eq14}
\end{equation}
Consequently, in a pure vacuum and absent GW emission, the orbital parameters remain static: $\frac{dp}{dt}=\frac{de}{dt}=0$.

\subsection{Reaction of GWs}

\begin{figure}[htbp]
    \centering
    \includegraphics[width=0.4\textwidth]{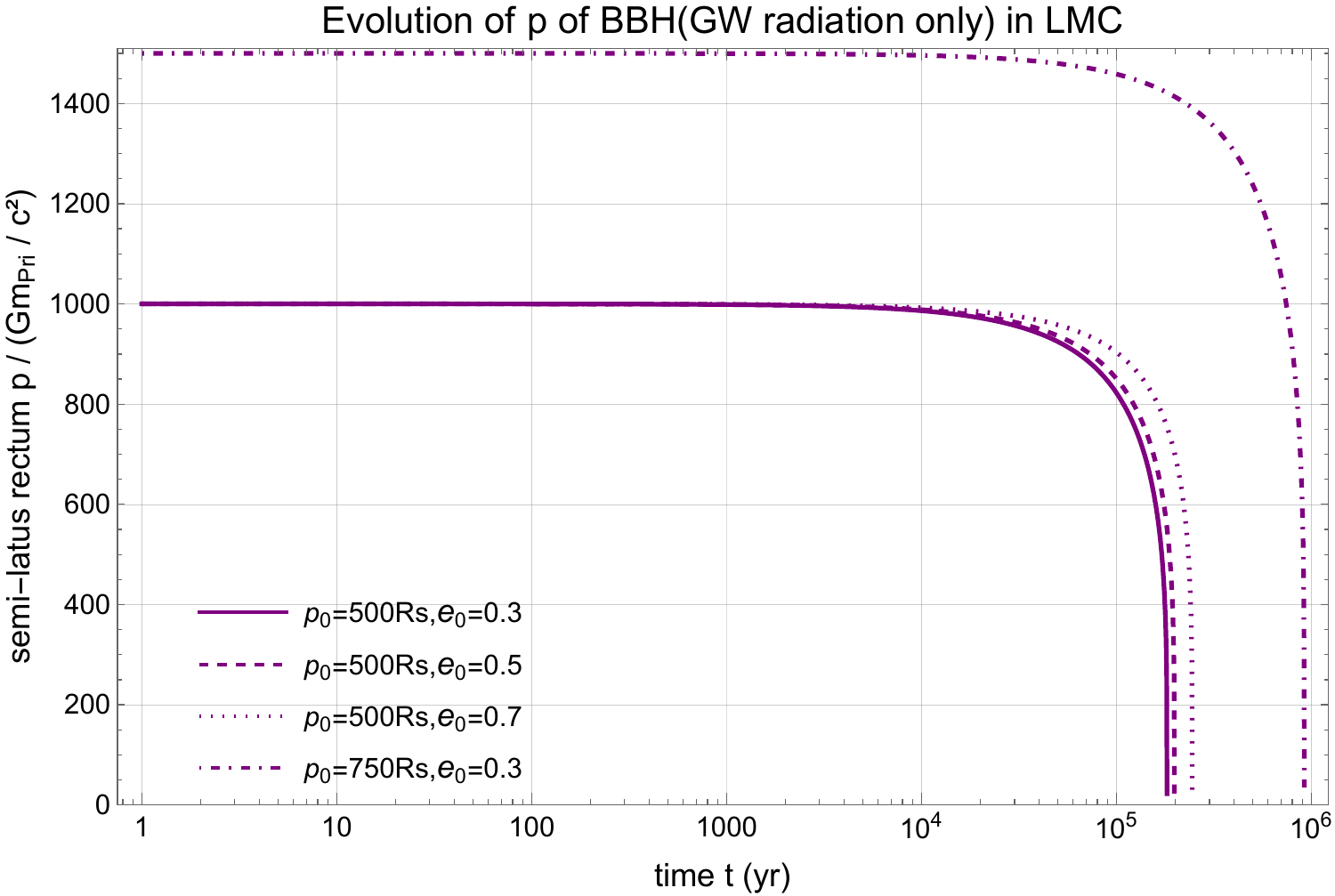}
    \\
    \includegraphics[width=0.4\textwidth]{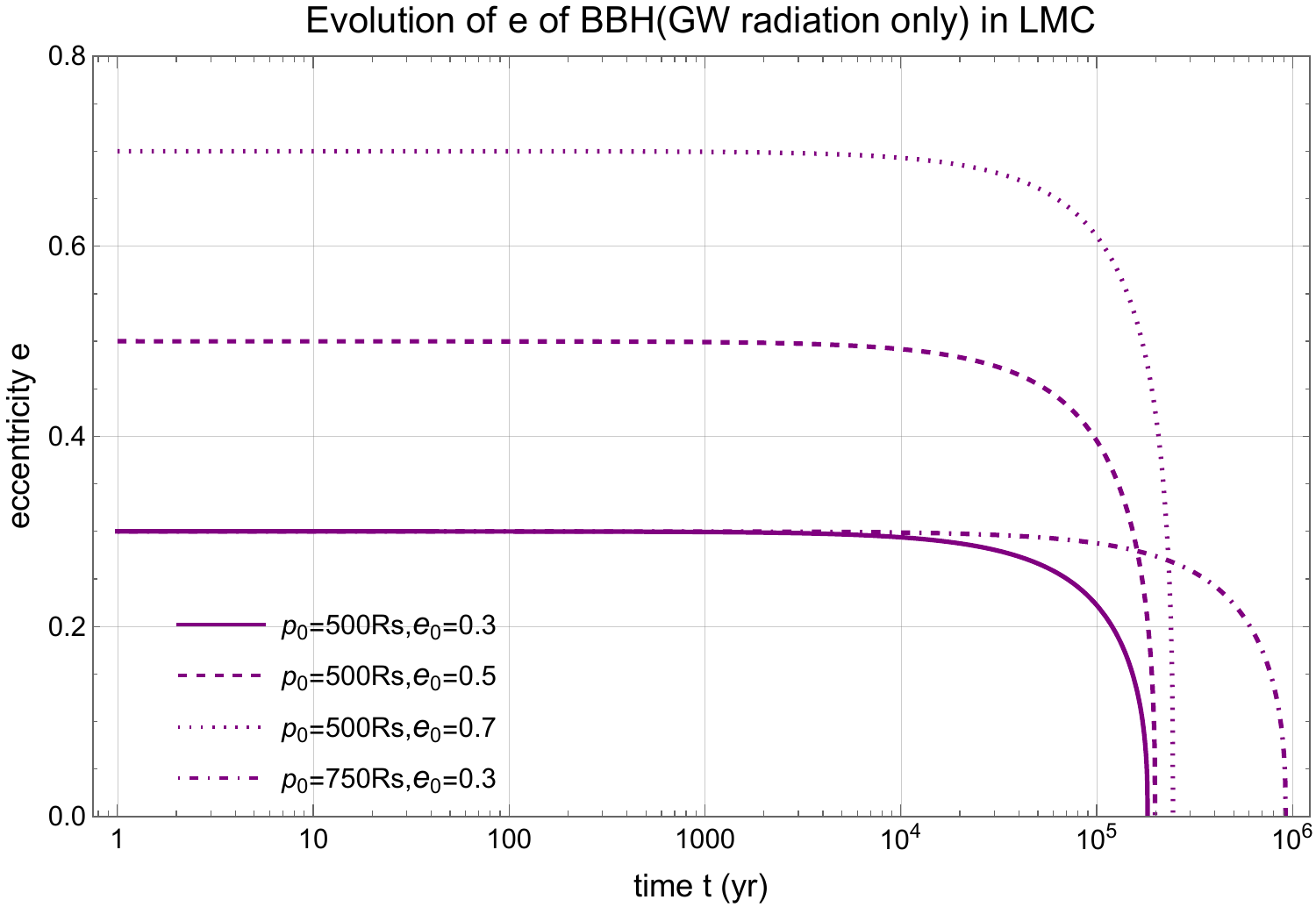}
     \caption{Evolution of the semi-latus rectum $p$ and eccentricity $e$ as a function of time $t$ for a BBH system at the center of the LMC, computed using Eqs.~(\ref{eq17}) and~(\ref{eq18}). GW radiation reaction drives the secular decay of both the orbital radius and eccentricity. For initial orbital parameters $p_0$ and $e_0$ shown in this figure, the inspiral times required for the secondary BH to reach the ISCO of the central BH are more than $10^{5}\mathrm{yr}$.}
    \label{fig:LMCgw}
\end{figure}

\begin{figure}[htbp]
    \centering
    \includegraphics[width=0.4\textwidth]{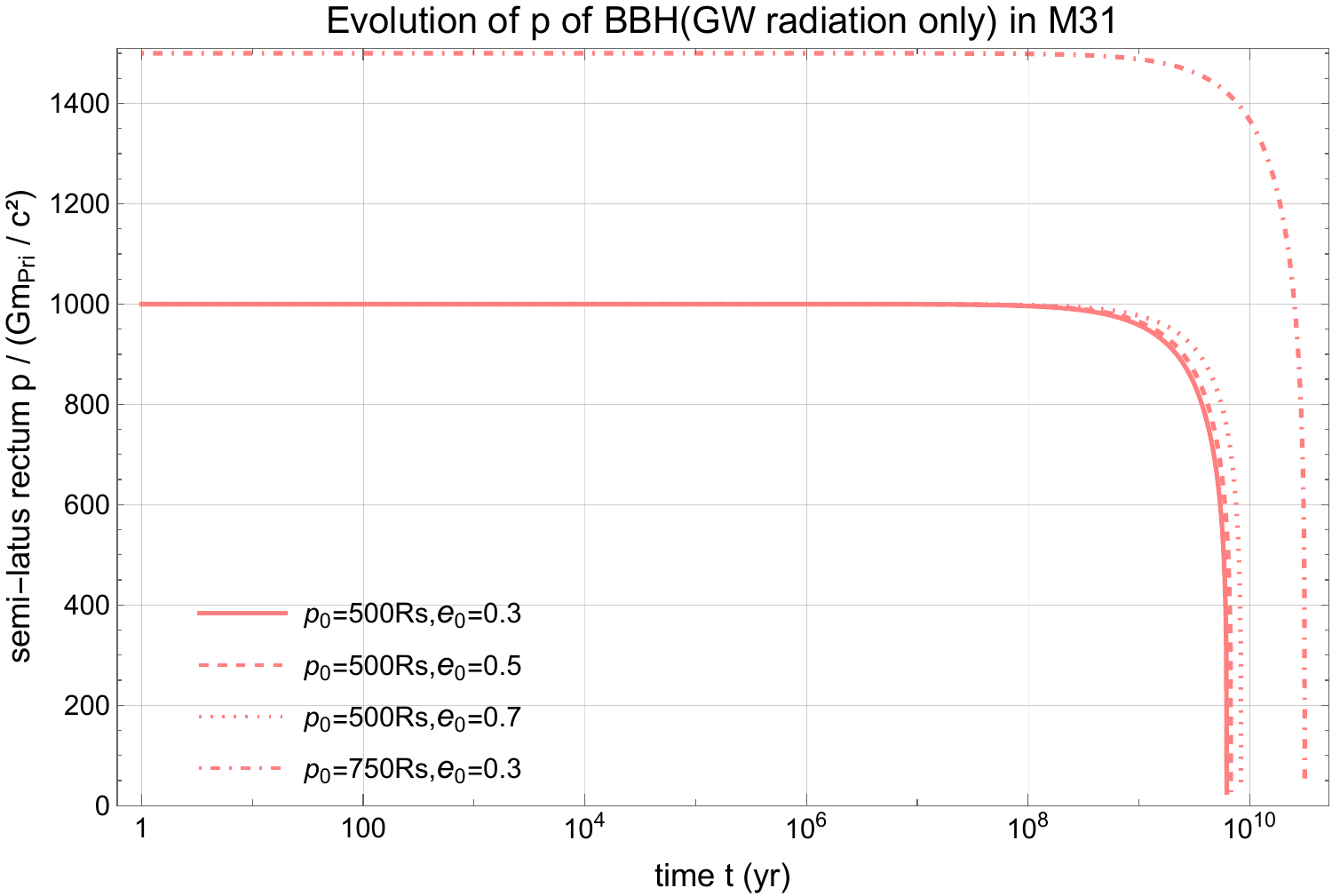}
    \\
    \includegraphics[width=0.4\textwidth]{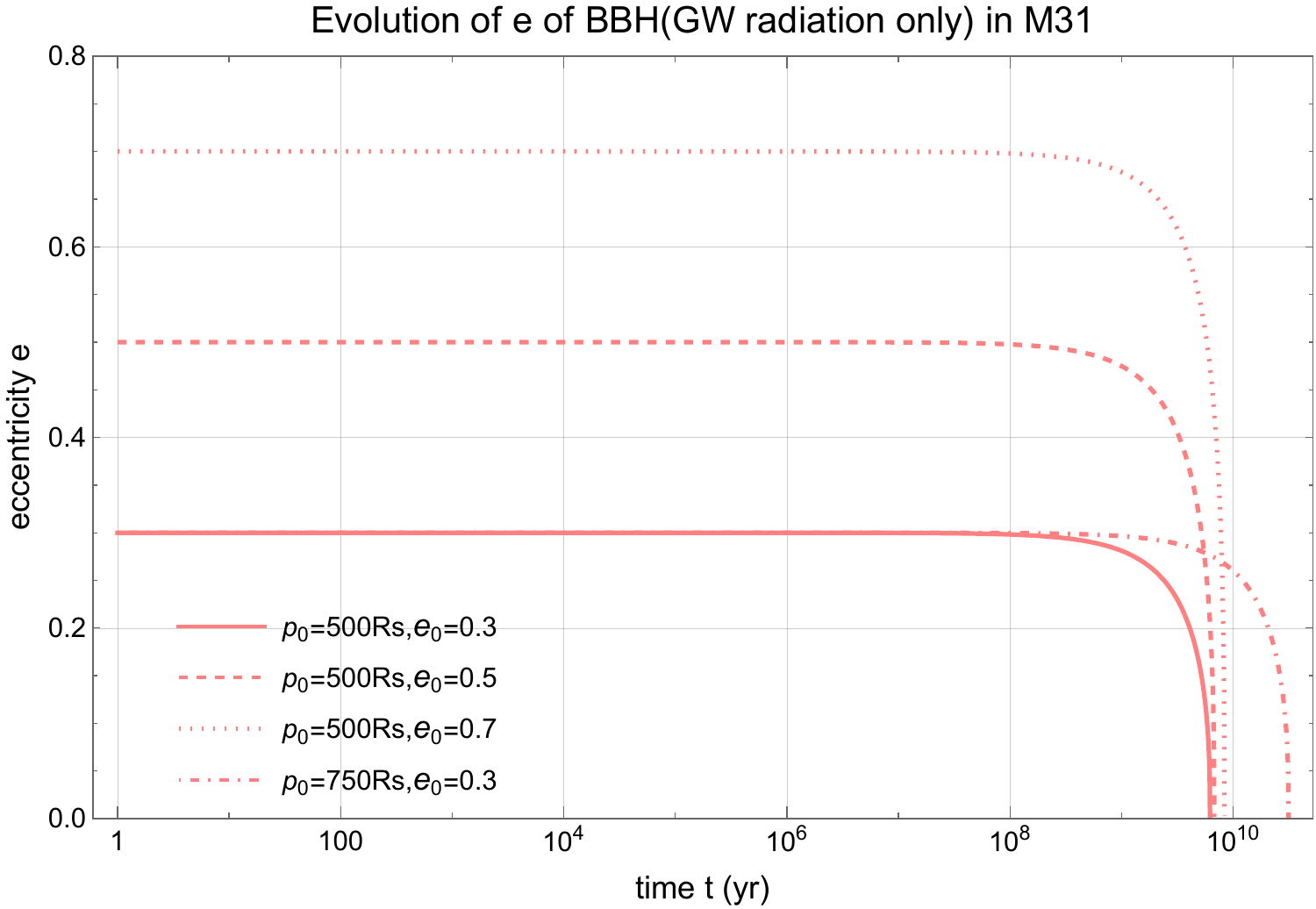}
    \caption{Temporal evolution of the semi-latus rectum $p$ and orbital eccentricity $e$ for a BBH system at the center of M31. GW radiation reaction drives the continuous decay of both orbital parameters. Compared to Fig.~\ref{fig:LMCgw}, the inspiral process extends over a significantly longer timescale, which is primarily attributed to the larger mass of the central BH and the higher mass ratio $q$. Assuming initial parameter configurations of $p_0$ and $e_0$, the secondary BH reaches $r_{\text{ISCO}}$ about $10^{10}\,\mathrm{yr}$. }
    \label{fig:M31gw}
\end{figure}

Incorporating the GW radiation reaction introduces secular dissipation into the BBH system, thereby breaking the conservation of $E$ and $L$. GWs carry away the orbital energy of the BBH, acting as a reaction force that reduces the orbital eccentricity and brings the orbits closer together. Therefore, the orbit of the secondary BH no longer follows purely Kepler's laws. According to the leading post-Newtonian order, the GW loss of energy and angular momentum can be expressed as
\begin{equation}
    \left\langle\frac{dE}{dt}\right\rangle_{GW}=-\frac{32}{5}\frac{G^4\mu^2m^3}{c^5p^5}(1-e^2)^{3/2}\left(1+\frac{73}{24}e^2+\frac{37}{96}e^4\right),
\end{equation}
\begin{equation}
    \left\langle\frac{dL}{dt}\right\rangle_{GW}=-\frac{32}{5}\frac{G^{7/2}\mu^2m^{5/2}}{c^5p^{7/2}}(1-e^2)^{3/2}\left(1+\frac{7}{8}e^2\right).
\end{equation}
Thus, the standard results yield the expressions for the secular changes in the semi-latus rectum $p$ and the eccentricity $e$ due to reaction of GWs~\cite{Poisson_Will_2014,PhysRevD.103.023015}:
\begin{equation}
    \left\langle\frac{dp}{dt}\right\rangle_{GW}=-\frac{8}{5}\eta\frac{G^3m^3}{c^5p^3}(1-e^2)^{3/2}\left(8+7e^2\right),
    \label{eq17}
\end{equation}
\begin{equation}
    \left\langle\frac{de}{dt}\right\rangle_{GW}=-\frac{8}{5}\eta\frac{G^3m^3}{c^5p^4}(1-e^2)^{3/2}\left(\frac{304}{24}e+\frac{121}{24}e^3\right),
    \label{eq18}
\end{equation}
where $\eta = m_{\text{1}}m_{\text{2}}/(m_{\text{1}} + m_{\text{2}})^2 $ is the symmetric mass ratio of BBH, and the subscript “GW” indicates that these effects arise from the reaction of GWs. It is evident that GW rediation leads to a gradual reduction in both the semi-latus rectum $p$ and the eccentricity $e$ of the binary system. Based on the above discussion, we successively substituted the relevant parameters of the LMC and M31 into Eqs.~(\ref{eq17}) and~(\ref{eq18}), and obtained the curves of $p$ and $e$ changing with time $t$, showed in Fig.~\ref{fig:LMCgw} and Fig.~\ref{fig:M31gw}.
 
\subsection{Effects of DM spike}

Next, we continue to introduce the effect of DM spike on BBH. Chandrasekhar proposed that moving celestial bodies experience a drag force from the gravitational pull of interstellar medium particles, which is known as gravitational drag or DF. The characteristics of DF depend on the velocity of the moving object, the density of the medium, and the sound speed~\cite{Kim_2007}. As the secondary BH moves through the DM spike around the central BH, it is subjected to the drag of DF from this DM spike, slowing down in its direction of motion and losing its kinetic energy and angular momentum. In this study, we focus on the supersonic regime, where the DF force can be expressed as
\begin{equation}
    \bm{f}_{DF} = - \frac{4 \pi G^2 \mu^2 \rho_{DM} I_v}{v^3} \bm{v},
\end{equation}
where $\bm{v}$ is the velocity of the secondary BH, $I_v$ is the Coulomb logarithm which depends on $\bm{v}$ and the sound speed of the DM spike. We can obtain the value of $v$ through following relationship:
\begin{equation}
    E = -\frac{Gm\mu}{r}+\frac{\mu v^2}{2},
\end{equation}
here is energy of the secondary BH in the gravitational field of the central BH. Then we can gain
\begin{align}
    v &= \sqrt{\frac{2E}{\mu} + \frac{2Gm}{r}} \\
      &= \sqrt{-\frac{Gm(1 - e^2)}{p} + \frac{2Gm}{p}(1 + e \cos \varphi)},
  \label{eq22}
\end{align}
where we have used Eqs.~(\ref{eq8}),~(\ref{eq11}) and~(\ref{eq12}) in the second step.

In this paper, we adopt $I_v = \ln \left( \sqrt{\frac{m_{\text{1}}}{m_{\text{2}}}} \right) $. According to~\cite{PhysRevD.100.043013,Poisson_Will_2014,Li:2025qtb}, we obtain the following equations with DF:
\begin{align}
    \left\langle\frac{dp}{dt}\right\rangle_{DF}&=-\int_{0}^{2\pi} d\varphi \frac{4G^{1/2}m_{\text{2}}\rho_{DM}I_vp^{5/2}}{(1-e^2)^{-3/2}m^{3/2}}\nonumber\\&\times\frac{1}{(1+e\cos{\varphi})^2(e^2+2e\cos{\varphi}+1)^{3/2}},
    \label{eq23}
\end{align}

\begin{align}
    \left\langle\frac{de}{dt}\right\rangle_{DF}
    &=-\int_{0}^{2\pi}d\varphi\frac{4G^{1/2}m_{\text{2}}\rho_{DM}I_vp^{3/2}}{(1-e^2)^{-3/2}m^{3/2}}\nonumber\\&\times \frac{e+\cos{\varphi}}{(1 + e \cos \varphi)^2 (e^2 + 2e \cos \varphi + 1)^{3/2}}.
    \label{eq24}
\end{align}
We obtain the dynamical equations that describe the evolution of $p$ and $e$ with respect to time $t$ for inspiraling BBHs under the effect of DF, which is manifested as a dissipative force. Under the influence of DF, the orbital radius of the system decreases while the eccentricity increases.

Additionally, we account for the accretion of DM by the secondary BH. In this paper, we assume that the radius of the secondary BH is greater than the mean free path of DM particles and consider only non-annihilating DM particles, ignoring all interactions except gravitational effects. The accretion process of the secondary BH is described by Bondi-Hoyle accretion and the mass flux at the horizon of the secondary BH is~\cite{PhysRevLett.126.101104}
\begin{equation}
    \dot{\mu}=4\pi G^2\lambda\frac{\mu^2\rho_{\rm DM}}{(v^2+c_s^2)^{3/2}}.
    \label{eq25}
\end{equation}
Here $c_s$ stands for the sound speed of the DM spike. We assume $v \gg c_s$ and the accretion term  $\dot{\mu}\bm{v}$ can be thought as a perturbation force:
\begin{equation}
    \bm{f}_{AC} = - \frac{4 \pi G^2 \mu^2 \rho_{DM} \lambda}{v^3} \bm{v},
\end{equation}
where the subscript ``AC" means that it is due to the effect of accretion. $\lambda$ is a dimensionless parameter of order unity, for simplicity, we adopt $\lambda = 1$ throughout this work. Similarly we obtain the following equations with AC:
\begin{align}
    \left\langle\frac{dp}{dt}\right\rangle_{AC} &=-\int_{0}^{2\pi}d\varphi\frac{4G^{1/2}m_{\text{2}}\rho_{DM}{\lambda}p^{5/2}}{(1-e^2)^{-3/2}m^{3/2}}\nonumber\\&\times
    \frac{1}{(1 + e \cos \varphi)^2 (e^2 + 2e \cos \varphi + 1)^{3/2}},
    \label{eq27}
\end{align}
\begin{align}
    \left\langle\frac{de}{dt}\right\rangle_{AC}&=-\int_{0}^{2\pi}d\varphi\frac{4G^{1/2}m_{\text{2}}\rho_{DM}{\lambda}p^{3/2}}{(1-e^2)^{-3/2}m^{3/2}}\nonumber\\&\times \frac{e+\cos{\varphi}}{(1 + e \cos\varphi)^2 (e^2 + 2e\cos \varphi + 1)^{3/2}}.
    \label{eq28}
\end{align}
For DF and accretion, their dynamic equations share the same form of differential equations. Through equations above, it is observed that DM spike reduce $p$ and accelerate the merger. In Appendix~\ref{onlyDM}, one can observe the impact on orbits when only DM is present.

\subsection{Total effects}

\begin{figure}[htbp]
    \centering
    \includegraphics[width=0.4\textwidth]{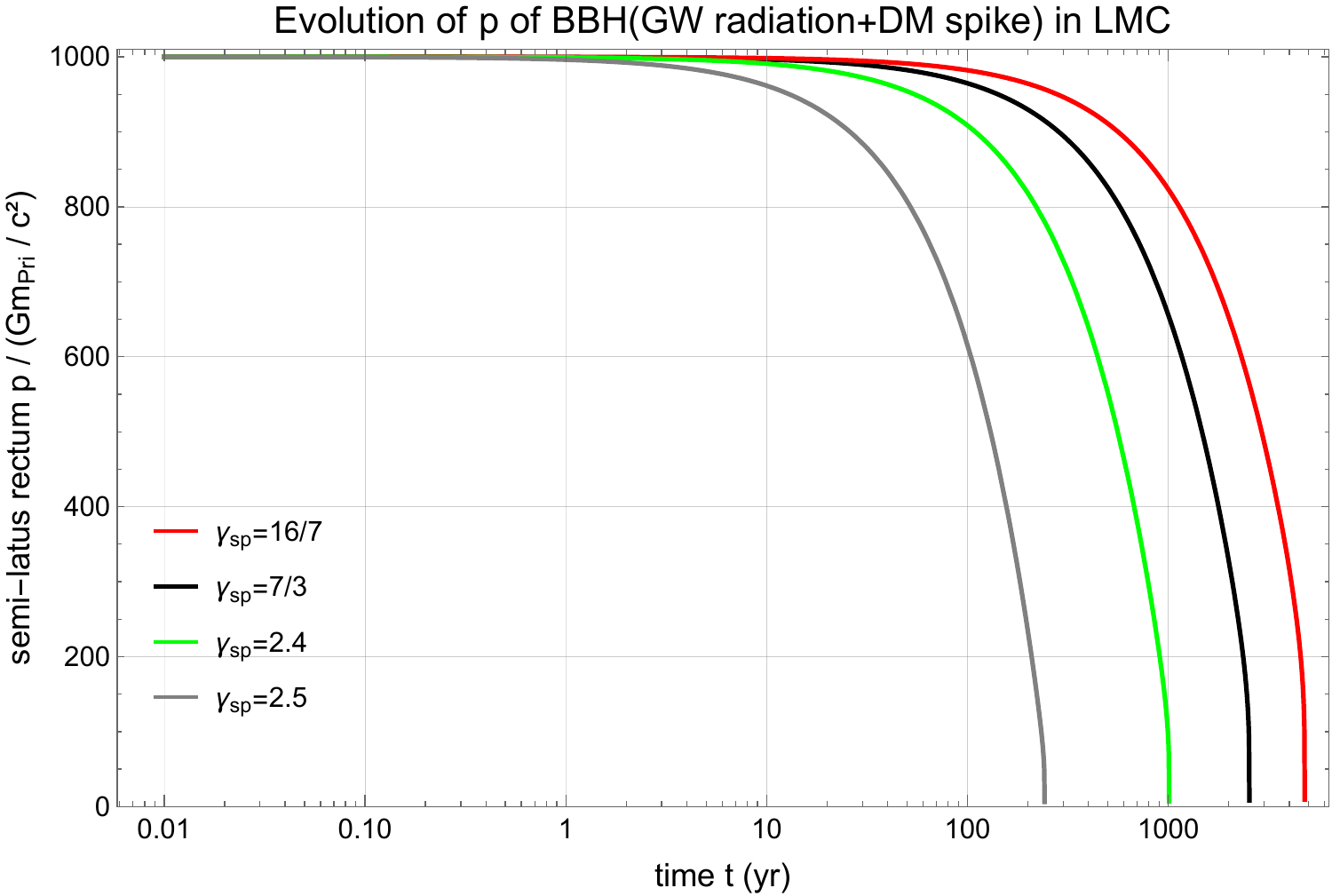}
    \\
    \includegraphics[width=0.4\textwidth]{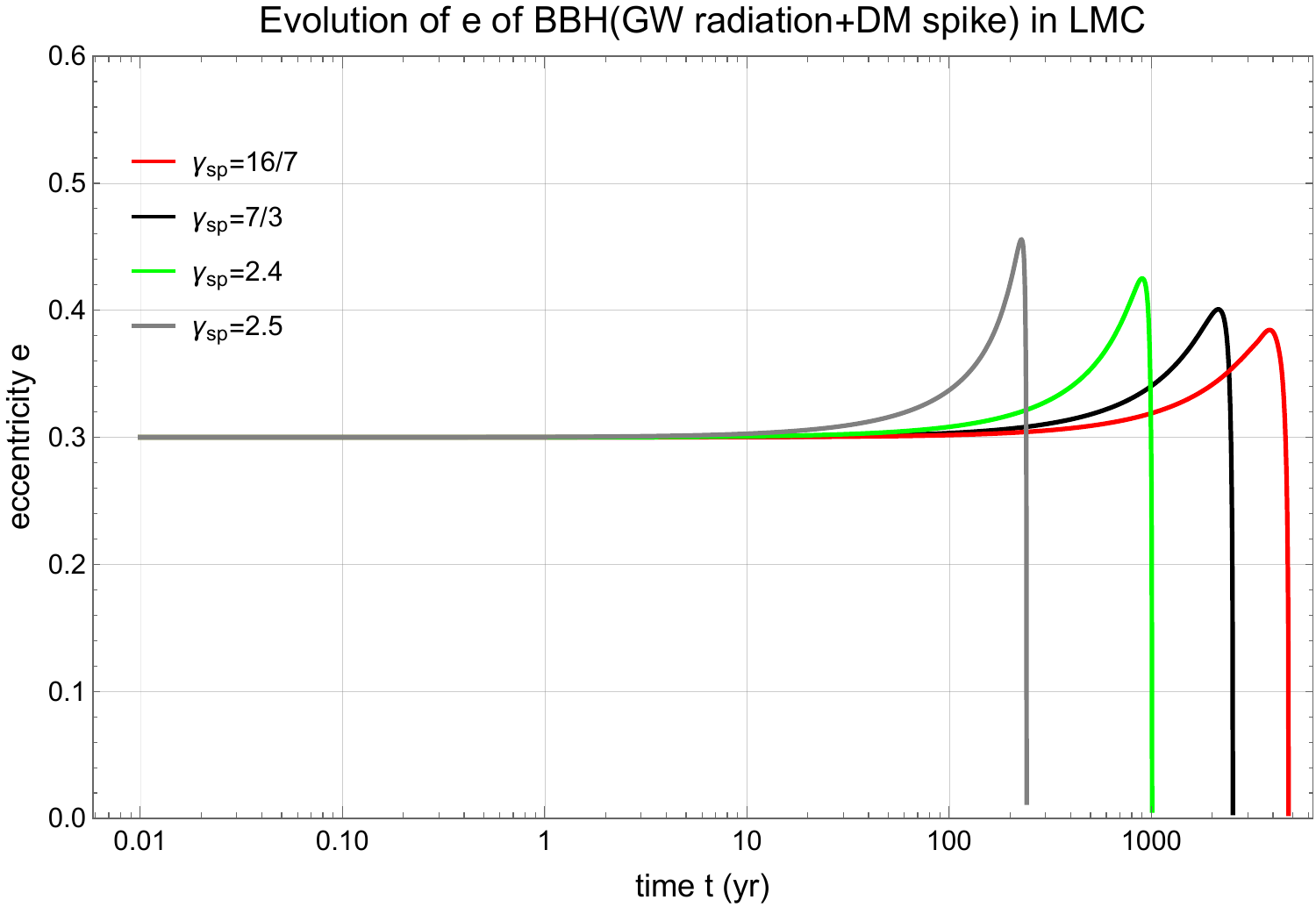}
    \\
    \includegraphics[width=0.4\textwidth]{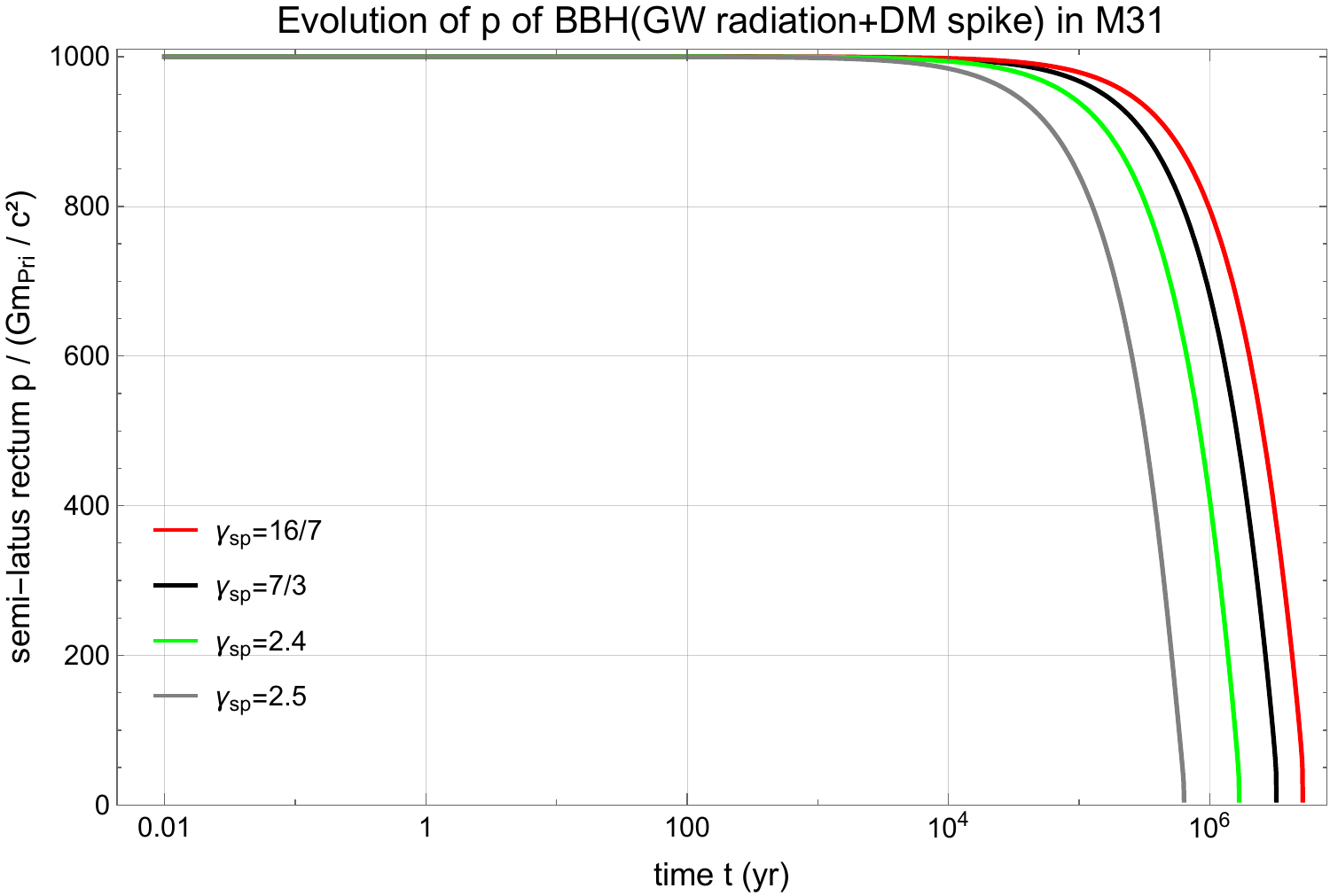}
    \\
    \includegraphics[width=0.4\textwidth]{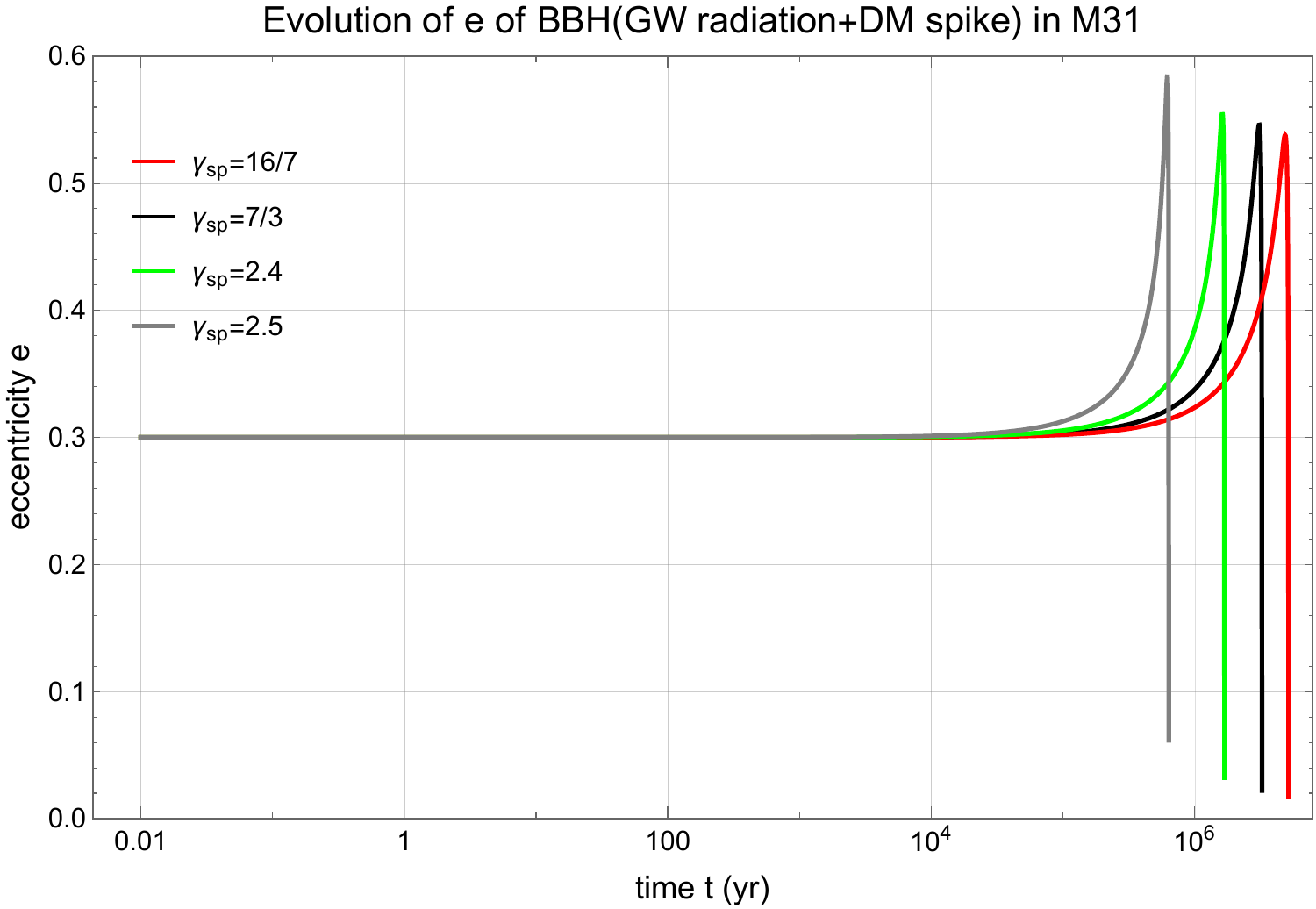}
    \caption{The combined effects of GW radiation and a DM spike on the orbital evolution of the BBH system. The semi-latus rectum $p$ decays more rapidly compared to the pure GW emission scenario. The orbital eccentricity $e$ initially undergoes an excitation to a maximum value, followed by a rapid decay during the late inspiral phase. The orbital evolution is terminated once the orbital radius satisfies $r \le r_{\text{ISCO}}$. }
    \label{fig:FINALDM}
\end{figure}

Based on all the factors above, we found that reaction of GWs causes a decrease in both orbital radius and eccentricity, while DF and accretion by secondary BHs with DM spike lead to a reduction in orbital radius and an increase in eccentricity. The final expressions for the semi-latus rectum $p$ and eccentricity $e$ are as follows
\begin{equation}
    \left\langle\frac{dp}{dt}\right\rangle_{Total}=\left\langle\frac{dp}{dt}\right\rangle_{GW}+\left\langle\frac{dp}{dt}\right\rangle_{DF}+\left\langle\frac{dp}{dt}\right\rangle_{AC},
    \label{eq29}
\end{equation}
\begin{equation}
    \left\langle\frac{de}{dt}\right\rangle_{Total}=\left\langle\frac{de}{dt}\right\rangle_{GW}+\left\langle\frac{de}{dt}\right\rangle_{DF}+\left\langle\frac{de}{dt}\right\rangle_{AC}
    \label{eq30}.
\end{equation}
Taking into account the total effects of GW's reaction, DF and accretion, we have plotted the changes with time $t$ in semi-latus rectum $p$ and eccentricity $e$ of the inspiraling BBHs in Fig.~\ref{fig:FINALDM} with initial semi-latus rectum $p_0 = 1000 Gm_{\text{1}}/c^2=500R_s$ and $e_0=0.3$. The corresponding numerical results are presented in Tables~\ref{tab:44}. We see orbital eccentricity $e$ may increase under the influence of DF and accretion due to surrounding DM, and decrease through GW radiation. 

\begin{table}[htbp]
    \centering
    \begin{subtable}{0.5\textwidth}
        \centering
        \begin{tabular}{cccc}
            \hline\hline
            spike index $\gamma_{sp}$  & final time[$\mathrm{yr}$] & $p_{final}$ & $e_{final}$\\
            \hline
            $16/7$ &$4476$  & $3.01R_s$ & 0.00173\\
            \hline
            $7/3$ & $2532$ & $3.0075R_s$ & 0.00262\\
            \hline
            $2.4$ &1009 &3.0107$R_s$ & 0.00473\\
            \hline
            $2.5$ & 242 &3.0345$R_s$ &0.0116\\
            \hline\hline
        \end{tabular}
        \caption{Final numerical results of BBH in LMC.}
        \label{tab:4a}
    \end{subtable}
    \begin{subtable}{0.5\textwidth}
        \centering
        \begin{tabular}{cccc}
            \hline\hline
            spike index $\gamma_{sp}$ & final time[$\mathrm{yr}$] & $p_{final}$ & $e_{final}$\\
            \hline
            $16/7$ &5.161$\times10^6$  & $3.0472R_s$ & 0.0157\\
            \hline
            $7/3$ & $3.243\times10^6$ & $3.063R_s$ & 0.02092\\
            \hline
            $2.4$ &1.682$\times10^6$ &3.095$R_s$ & 0.0315\\
            \hline
            $2.5$ & 6.362$\times10^5$&3.184$R_s$ &0.061\\
            \hline\hline
        \end{tabular}
        \caption{Final numerical results of BBH in M31.}
        \label{tab:4b}
    \end{subtable}
    \caption{Final numerical results of the inspiraling BBH under the combined effects of GW radiation and a DM spike in the LMC (a) and M31 (b). The table presents the total evolution time, final semi-latus rectum $p_{{final}}$, and final eccentricity $e_{{final}}$ for various DM spike indices $\gamma_{sp}$. The initial orbital parameters are set to $p_0 = 500R_s$ and $e_0 = 0.3$. The orbital evolution is considered terminated when the secondary BH reaches the ISCO.}
    \label{tab:44}
\end{table}

\section{Accretion disk}
\label{sec:acc}

Unlike DM spikes, whose existence around central massive BHs is primarily theoretically motivated, the presence of baryonic accretion disks is well-established by multi-wavelength observations~\cite{Padovani_2017}. Such gaseous environments introduce additional hydrodynamical drag that, alongside the DM spike, further modifies the orbital dynamics and GW signatures of the embedded BBH system.

\subsection{Accretion disk distribution}

\begin{figure}
    \centering
    \includegraphics[width=1\linewidth]{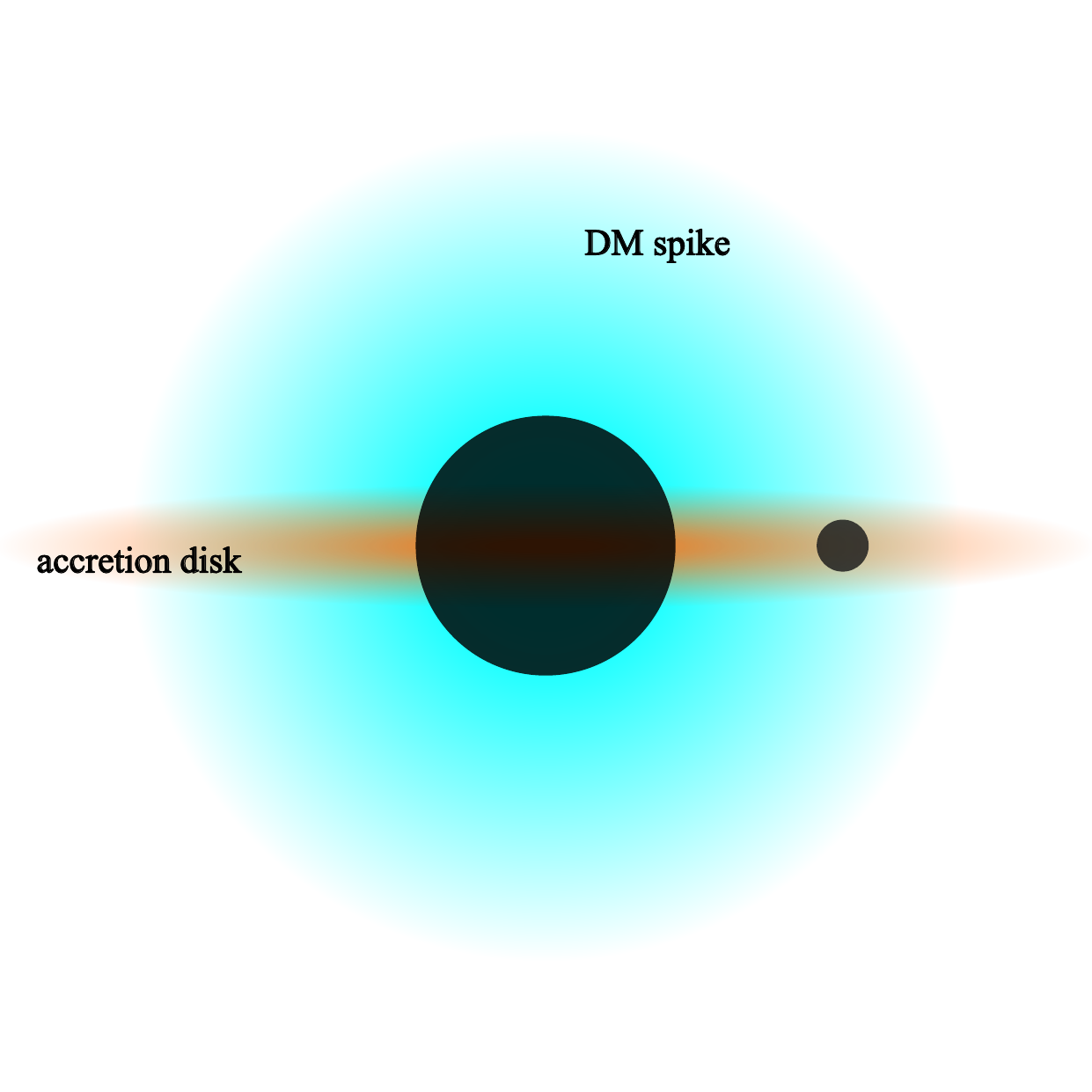}
    \caption{Edge-on schematic view of a BBH system embedded within both a DM spike and an axisymmetric baryonic accretion disk. The secondary BH moves entirely within the disk.}
    \label{fig:acdBBH}
\end{figure}

This section shows the theoretical modeling of the accretion disk surrounding the central massive BH. We assume a stationary accretion disk that is geometrically thin, optically thick, and highly radiative efficient, which is known as the standard thin disk model ($H\ll R$)~\cite{1973A&A....24..337S}, as detailed in reference~\cite{Speri_2023,Abramowicz_2013,PhysRevD.84.024032,Frank_King_Raine_2002,Becker:2022wlo,Sánchez-Salcedo_2020,Liu_2024}. We assume an axisymmetric accretion disk and consider a scenario where the secondary BH orbits entirely within the disk plane. To construct a physically reasonable and complete density distribution, we divide the disk into an inner region ($r < R_Q = 200 r_{\rm{ISCO}}$) and an outer region ($R_Q < r $)~\cite{Cheng:2024mgl}. The schematic diagram is shown in Fig.~\ref{fig:acdBBH}. 

In the inner region of the accretion disk, the $\alpha$ disk and $\beta$ disk are used. According to~\cite{Speri_2023,PhysRevD.86.049907,1981ApJ...247...19S}, these standard disk models remain physically valid within the regime $r<R_Q$ and we can parameterize the surface density $\Sigma$ and scale height $H$. The corresponding volume density of the accretion disk is $\rho_b$ = $\Sigma/2H$ ($b=\alpha, \beta$). In contrast to~\cite{Speri_2023,1981ApJ...247...19S}, we use the International System of Units:

\begin{align}
    \Sigma_{\alpha} \left[ \frac{\text{kg}}{\text{m}^2} \right] \approx 
    & 5.4 \times 10^3 \left( \frac{\alpha}{0.1} \right)^{-1} \left( \frac{f_{\text{Edd}}}{0.1} \frac{0.1}{\epsilon} \right)^{-1} \nonumber \\
    & \times\left( \frac{c^2 r}{10 Gm_{\text{1}}} \right)^{3/2},
\end{align}

\begin{align}
    \Sigma_{\beta} \left[ \frac{\text{kg}}{\text{m}^2} \right] \approx 
    & 2.1 \times 10^7 \left( \frac{\alpha}{0.1} \right)^{-4/5} \left( \frac{f_{\text{Edd}}}{0.1} \frac{0.1}{\epsilon} \right)^{3/5} \nonumber \\
    & \times \left( \frac{m_{\text{1}}}{10^6 M_\odot} \right)^{1/5} \left( \frac{c^2 r}{10 Gm_{\text{1}}} \right)^{-3/5},
\end{align}

\begin{equation}
    H[m] = 1.5 \left( \frac{f_{\text{Edd}}}{0.1} \frac{0.1}{\epsilon} \right) \left( \frac{Gm_{\text{1}}}{c^2} \right).
\end{equation}
Here we describe the inner region with a constant scale height $H[m]$.\footnote{Note that the subscript $\alpha$ on the left side of the equation refers to the $\alpha$-disk model, while the $\alpha$ on the right side represents the viscosity parameter, with an estimated range of 0.01--0.1~\cite{10.1111/j.1365-2966.2007.11556.x}. The brackets on the left side denote the dimensions of surface density $\Sigma$ and scale height $H$.} 

\begin{figure}[htbp]
    \centering
    \includegraphics[width=0.4\textwidth]{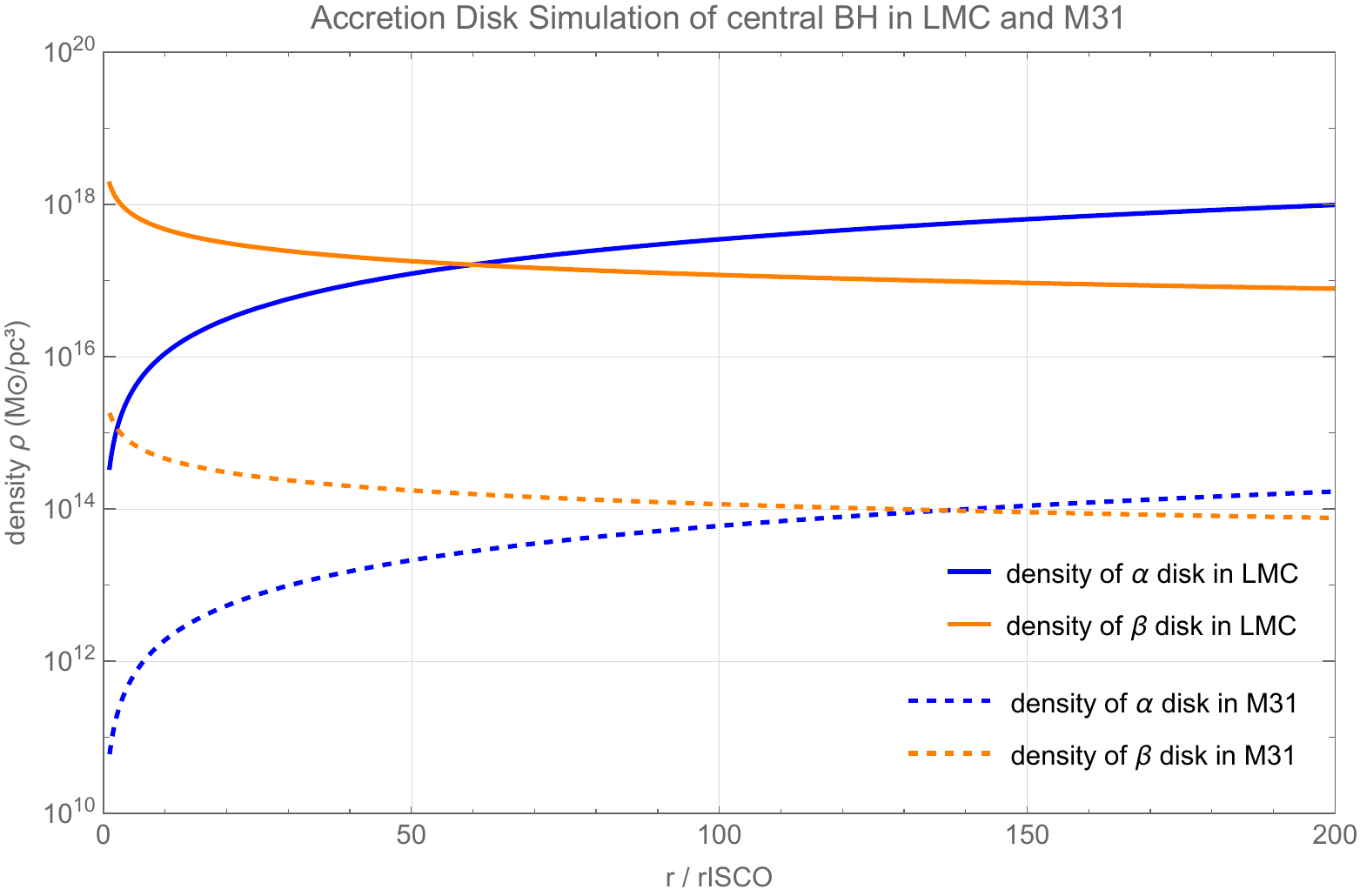}
    \caption{Radial volume density profiles of the $\alpha$-disk and $\beta$-disk accretion models for the LMC and M31. The blue line indicates the $\alpha$ accretion disk model, and the orange line indicates the $\beta$ accretion disk model.}
    \label{fig:LMCM31disk}
\end{figure}

In this paper, for simplicity, we adopt \( f_{\mathrm{Edd}} = \epsilon = 0.1 \), where \( f_{\mathrm{Edd}} \) is the central BH's Eddington accretion rate fraction, and \( \epsilon \) stands for the disk's efficiency of mass-energy conversion into luminosity. In Fig.~\ref{fig:LMCM31disk}, we respectively plot the variation of accretion disk density of the central BHs in LMC and M31 as a function of radial distance.

In the outer region of the accretion disk, the self-gravitating disk model to describe the outer  region~\cite{Goodman_2004,10.1046/j.1365-8711.2003.06241.x}, with the density $\rho_b(r)=\rho_0(\frac{R_Q}{r})^3\exp(-\frac{r^2}{2H^2})$.

\subsection{Effects of accretion disk}

The dynamical imprints of accretion disks and DM spikes are physically distinguishable, as they dominate at different orbital stages and GW frequency bands. Specifically, at larger orbital separations during the early inspiral phase, the gaseous interaction from the accretion disk plays a dominant role. Conversely, as the binary separation shrinks in the late inspiral stages, the environmental influence of the DM spike becomes increasingly significant~\cite{Becker:2022wlo}. In our study, we comprehensively consider the net effect of the DM spike and the accretion disk on the BBH system. As the secondary BH traverses the accretion disk, it is subjected to gaseous friction, which is conventionally described by the Ostriker formalism~\cite{Ostriker_1999,10.1093/mnras/stac1294,Zwick:2025ine}:

\begin{equation}
    \bm{F}_{Ostriker} = - \frac{4 \pi G^2 m_{\text{2}}^2 \rho_{b}(r) I_b}{v_{rel}^2} \frac{\bm{v}}{v},
    \label{eq34}
\end{equation}
with
\begin{equation}
    I_b = \frac{1}{2} 
        \begin{cases} 
            \ln \dfrac{1 - v_{\text{rel}}/c_b}{1 + v/c_b} - v_{\text{rel}}/c_b & \text{subsonic}, \\ 
            \ln \left( 1 - (v_{\text{rel}}/c_b)^{-2} \right) + \ln \Lambda & \text{supersonic},
        \end{cases}
\end{equation}
where $v_{rel}$ stands for the relative velocity between the secodary BH and the disk. Here $c_b$ is the sound speed of the gas in the disk. $\ln \Lambda$ is regarded as the Coulomb logarithm of the Ostriker model in accretion disk~\cite{Trani:2025edb,Li:2024gfi}.

Similarly, we need to consider the accretion of gas in the accretion disk by the secondary BH. For simplicity, we still use Bondi-Hoyle accretion\footnote{Note that while the Bondi-Hoyle formalism in Eq.~(\ref{eq36}) provides a straightforward analytical treatment for the accretion rate, it may overestimate the actual mass capture in a disk environment due to the high specific angular momentum of the gas. In a realistic accretion disk, the centrifugal barrier and local vorticity can significantly impede the direct inflow of material compared to the classical Bondi-Hoyle case~\cite{Krumholz:2004vj,Jiao:2025frj}. Therefore, the results derived from this model should be interpreted as an upper-bound approximation for the environmental impact, representing the maximum potential effect of disk-driven accretion on the binary evolution.}:
\begin{equation}
    \dot{\mu}=4\pi G^2\lambda_b\frac{\mu^2\rho_{\rm b}}{(v^2+c_b^2)^{3/2}}.
    \label{eq36}
\end{equation}

In Appendix~\ref{A}, we present the detailed calculations, and only the key conclusions are shown here:

\begin{align}
    \left\langle\frac{dE}{dt}\right\rangle_{AC  disk} &=-\int_{0}^{2\pi}\frac{2G^{3/2}m_{\text{2}}^2(I_b+\lambda_b)\rho_{b}(r)p^{1/2}}{(1-e^2)^{-3/2}m^{1/2}}\nonumber\\&\times
    \frac{1}{(1 + e \cos \varphi)^2 (e^2 + 2e \cos \varphi + 1)^{1/2}}d\varphi,
    \label{eq37}
\end{align}

\begin{align}
    \left\langle\frac{dL}{dt}\right\rangle_{AC  disk} &=-\int_{0}^{2\pi}\frac{2Gm_{\text{2}}^2(I_b+\lambda_b)\rho_{b}(r)p^{2}}{(1-e^2)^{-3/2}m}\nonumber\\&\times
    \frac{1}{(1 + e \cos \varphi)^2 (e^2 + 2e \cos \varphi + 1)^{3/2}}d\varphi,
    \label{eq38}
\end{align}

\begin{align}
    \left\langle\frac{dp}{dt}\right\rangle_{AC   disk} &=-\int_{0}^{2\pi}\frac{4G^{1/2}m_{\text{2}}(I_b+\lambda_b)\rho_{b}(r)p^{5/2}}{(1-e^2)^{-3/2}m^{3/2}}\nonumber\\&\times
    \frac{1}{(1 + e \cos \varphi)^2 (e^2 + 2e \cos \varphi + 1)^{3/2}}d\varphi,
    \label{eq39}
\end{align}

\begin{align}
    \left\langle\frac{de}{dt}\right\rangle_{AC  disk} &=-\int_{0}^{2\pi}\frac{4G^{1/2}m_{\text{2}}(I_b+\lambda_b)\rho_{b}(r)p^{3/2}}{(1-e^2)^{-3/2}m^{3/2}}\nonumber\\&\times
    \frac{e+cos\varphi}{(1 + e \cos \varphi)^2 (e^2 + 2e \cos \varphi + 1)^{3/2}}d\varphi.
    \label{eq40}
\end{align}

\begin{figure}[htbp]
    \centering
    \includegraphics[width=0.4\textwidth]{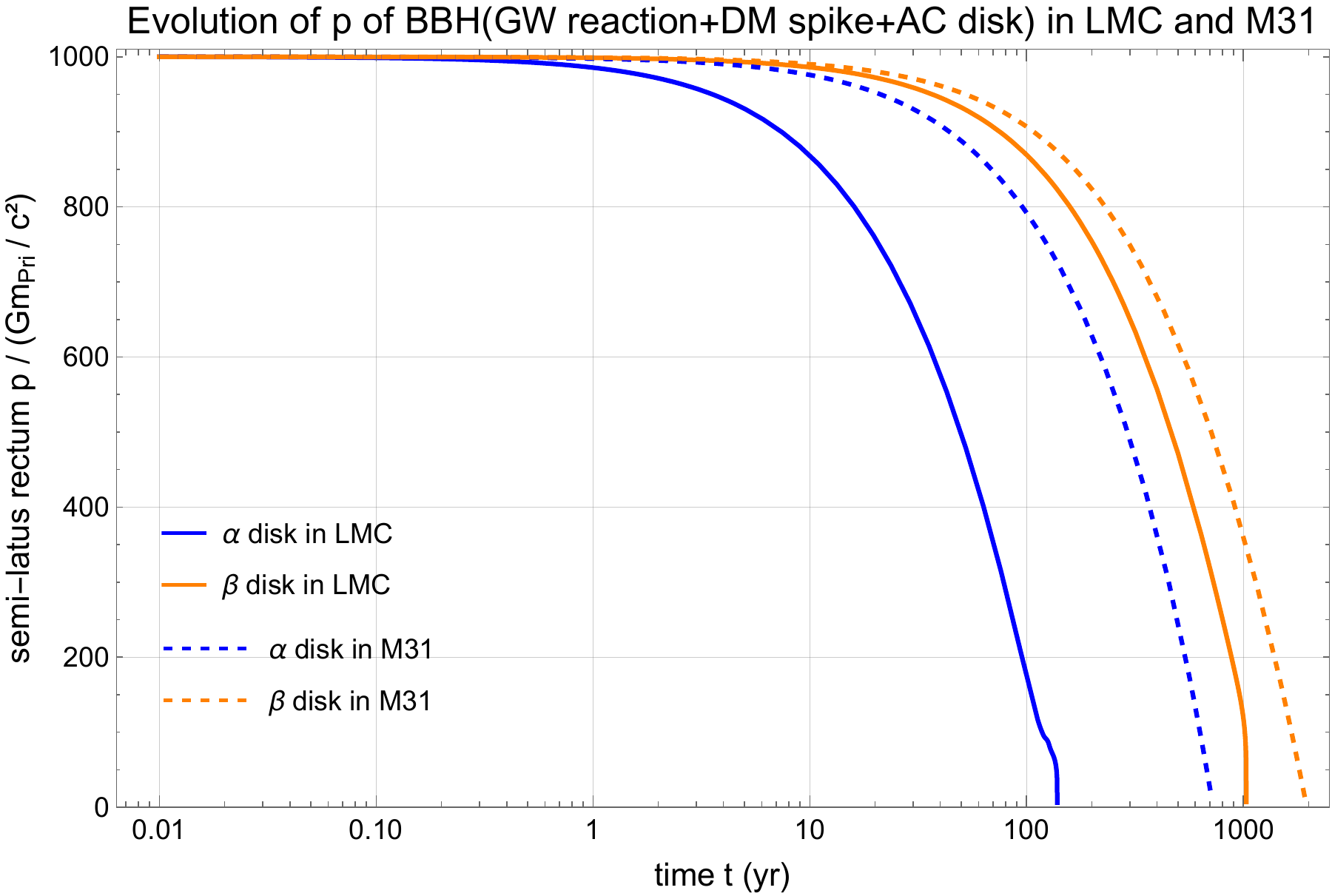}
    \\
    \includegraphics[width=0.4\textwidth]{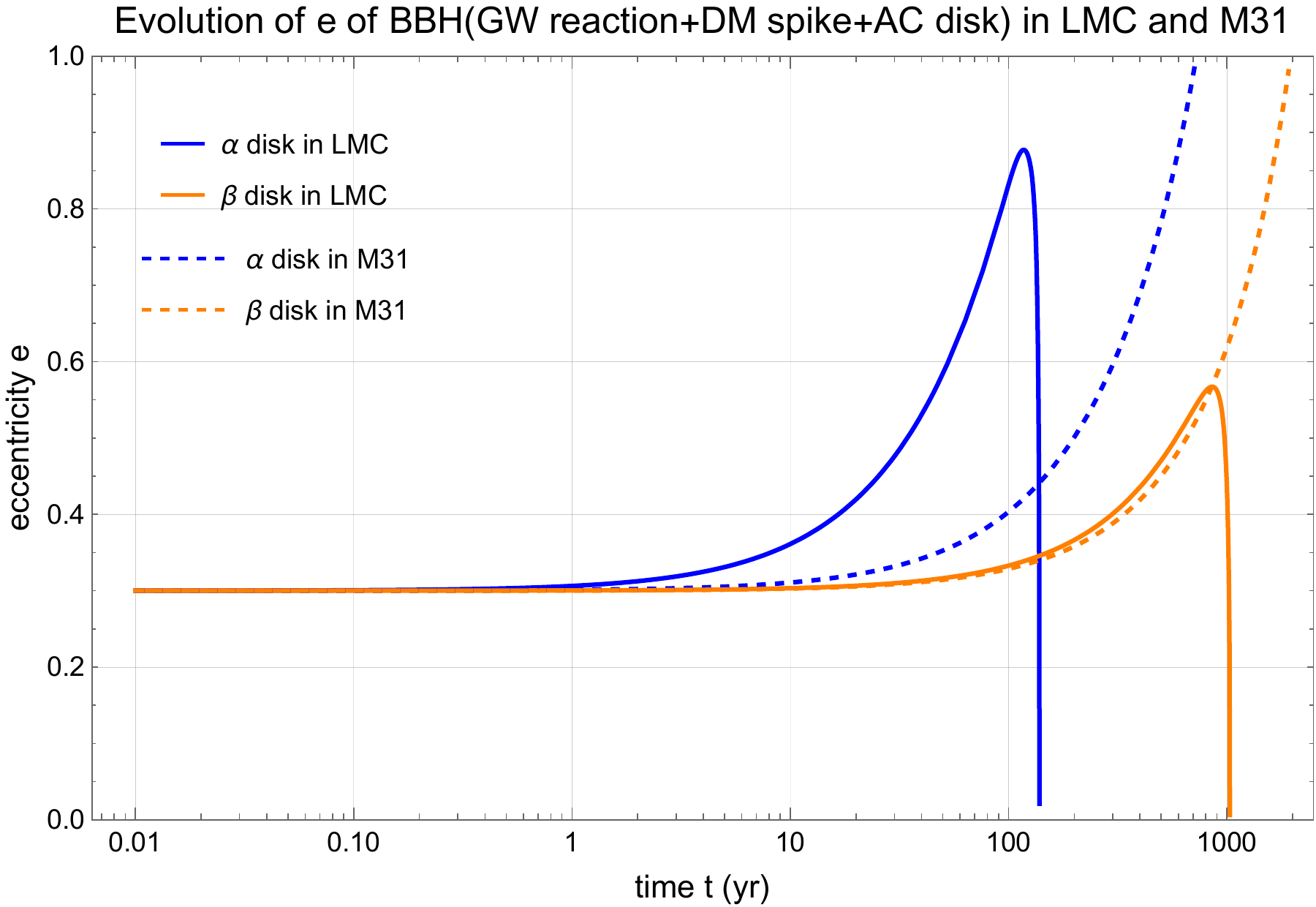}
    \caption{Evolution of orbital elements ($p$ and $e$) under the joint influence of GW radiation, a DM spike ($\gamma_{sp}=7/3$), and an accretion disk ($\alpha$-disk in blue, $\beta$-disk in orange). The orbital evolution terminates at $r \le  r_{\text{ISCO}}$.The initial semi-latus rectum is $p_0 = 1000Gm_{\text{1}}/c^2=500R_s$ and the initial eccentricity is $e_0=0.3$. }
    \label{fig:acdfianl}
\end{figure}

We now consider the model of an inspiraling BBH system that includes both an accretion disk and a DM spike. Then Eqs.~(\ref{eq29}) and~(\ref{eq30}) can be modified to as all terms are considered independent minor perturbations to the Keplerian orbit:

\begin{equation}
    \left\langle\frac{dp}{dt}\right\rangle_{Total}=\left\langle\frac{dp}{dt}\right\rangle_{GW}+\left\langle\frac{dp}{dt}\right\rangle_{DM spike}+\left\langle\frac{dp}{dt}\right\rangle_{AC disk},
    \label{eq41}
\end{equation}
\begin{equation}
    \left\langle\frac{de}{dt}\right\rangle_{Total}=\left\langle\frac{de}{dt}\right\rangle_{GW}+\left\langle\frac{de}{dt}\right\rangle_{DM spike}+\left\langle\frac{de}{dt}\right\rangle_{AC disk}.
    \label{eq42}
\end{equation}
We calculate their total effects on the inspiral process shown in Fig.~\ref{fig:acdfianl}. The initial semi-latus rectum is $p_0 = 1000Gm_{\text{1}}/c^2=500R_s$ and the initial eccentricity is $e_0=0.3$. As can be seen from Fig.~\ref{fig:acdfianl}, BBHs merge more rapidly in $\alpha$ accretion disk. 

By comparing Figs.~\ref{fig:LMCgw},~\ref{fig:M31gw},~\ref{fig:FINALDM}, and~\ref{fig:acdfianl}, it is obvious that the orbital decay is slow when only GW radiation is considered. With the introduction of a DM spike, the orbital contraction is significantly accelerated, and the decay rate strongly depends on the density distribution index $\gamma_{sp}$ of the spike. Furthermore, when the accretion disk is included, the orbital evolution time is further shortened, highlighting the decisive influence of environmental factors on the merger timescale of BBHs. We have calculated the numerical results for Fig.~\ref{fig:acdfianl} in Table~\ref{tab55}. Appendix~\ref{GC} shows the evolution of an inspiraling BBH at the GC under environmental influences including DM spike and accretion disk.

\begin{table}[htbp]
    \centering
    \begin{subtable}{\linewidth}
        \centering
        \begin{tabular}{cccc}
            \hline\hline
            AC disk  & final time[$\mathrm{yr}$] & $p_{final}$ & $e_{final}$\\
            \hline
            $\alpha$ disk &$138.6$  & $3.06R_s$ & 0.0202\\
            \hline
            $\beta$ disk & $1032.7$ & $3.014R_s$ & 0.00477\\
            \hline\hline
        \end{tabular}
        \caption{Final numerical results of BBH in LMC.}
        \label{tab:6}
    \end{subtable}
    
    \vspace{1.5em} 
    
    \begin{subtable}{\linewidth}
        \centering
        \begin{tabular}{cccc}
            \hline\hline
            AC disk  & final time[$\mathrm{yr}$] & $p_{final}$ & $e_{final}$\\
            \hline
            $\alpha$ disk &$714$  & $5.9655R_s$ & 0.9883\\
            \hline
            $\beta$ disk & $1924.8$ & $5.9495R_s$ & 0.9831\\
            \hline\hline
        \end{tabular}
        \caption{Final numerical results of BBH in M31.}
        \label{tab:7}
    \end{subtable}
    \caption{Final numerical results for the inspiraling BBH systems under the combined influence of GW radiation, a DM spike, and an accretion disk ($\alpha$ or $\beta$ model) in the LMC (a) and M31 (b). The initial orbital parameters remain $p_{0}=500R_{s}$ and $e_{0}=0.3$.}\label{tab55}
\end{table}

\section{GW waveform and Detectability}
\label{sec:mismatch}

In the preceding sections, we investigated the temporal dynamical evolution of the semi-latus rectum $p$ and orbital eccentricity $e$ for BBH systems situated in the LMC and M31. In this section, we use the conclusions obtained earlier to investigate the impact of the DM spike and accretion disk on the GW waveform over relatively long time scales. Subsequently, we will calculate the SNR after considering the total effects.

Consistent with our prior assumptions, the binary components and the accretion disk are coplanar, defined herein as the equatorial $(x, y)$ plane. Following the formalisms presented in~\cite{10.1093/acprof:oso/9780198570745.001.0001,Jaranowski_Krolak_2009}, the non-zero components of the mass quadrupole moment tensor in the center-of-mass frame are given by

\begin{equation}
    M_{ab} = \mu(t) r(t)^2 \begin{pmatrix} \cos^2 \varphi(t) & \sin \varphi(t) \cos \varphi(t) \\ \sin \varphi(t) \cos \varphi(t) & \sin^2 \varphi(t) \end{pmatrix}_{ab},
    \label{eq43}
\end{equation}
where $a,b = 1,2$ are indices in the $(x,y)$ plane. 

The plus and cross modes of GWs are as follows
\begin{align}
    h_+(t; \theta, \varphi) = \frac{1}{R} \frac{G}{c^4} \big[   & \ddot{M}_{11}(\cos^2 \varphi - \sin^2 \varphi \cos^2 \theta) \nonumber \\
    & + \ddot{M}_{22}(\sin^2 \varphi - \cos^2 \varphi \cos^2 \theta) \nonumber \\
    & - \ddot{M}_{33}\sin^2 \theta \nonumber \\
    & - \ddot{M}_{12}\sin 2\varphi (1 + \cos^2 \theta) \nonumber \\
    & + \ddot{M}_{13}\sin \varphi \sin 2\theta \nonumber \\
    & +  \ddot{M}_{23}\cos \varphi \sin 2\theta \big],
    \label{eq47}
\end{align}

\begin{align}
    h_{\times}(t; \theta, \varphi) = \frac{1}{R} \frac{G}{c^4} \big[   & (\ddot{M}_{11} - \ddot{M}_{22})\sin 2\varphi \cos \theta \nonumber \\
    &  + 2\ddot{M}_{12}\cos 2\varphi  \cos \theta \nonumber \\
    & - 2\ddot{M}_{13}\cos \varphi \sin \theta \nonumber \\
    & +  2 \ddot{M}_{23}\sin \varphi \sin \theta \big].
    \label{eq48}
\end{align}
In this work, we set $\theta$ to 0. Finally, we organize Eqs.~(\ref{eq47}) and~(\ref{eq48}) as follows

\begin{equation}
    h_{+}(t)= \frac{G}{Rc^4}  \bigg[(\ddot{M}_{11}-\ddot{M}_{22})\cos\left(2\varphi\right)-2\ddot{M}_{12}\sin(2\varphi) \bigg],
    \label{eq49}
\end{equation}

\begin{equation}
    h_{\times}(t)= \frac{G}{Rc^4}  \bigg[(\ddot{M}_{11}-\ddot{M}_{22})\sin\left(2\varphi\right)+2\ddot{M}_{12}\cos(2\varphi) \bigg],
    \label{eq50}
\end{equation}
where $m$ is the total mass of the BBH system and $\mu$ denotes the reduced mass. The parameter $R$ represents the distance from the BBH's center of mass to the observer according to~\cite{2025ApJ...978..104G}. $\varphi$ can be gained through solve with $ \frac{d\varphi}{dt}=\sqrt{\frac{Gm}{p^3}}(1+e\cos\varphi)^{2} $.

\begin{figure}
    \centering
    \includegraphics[height=0.2\textheight]{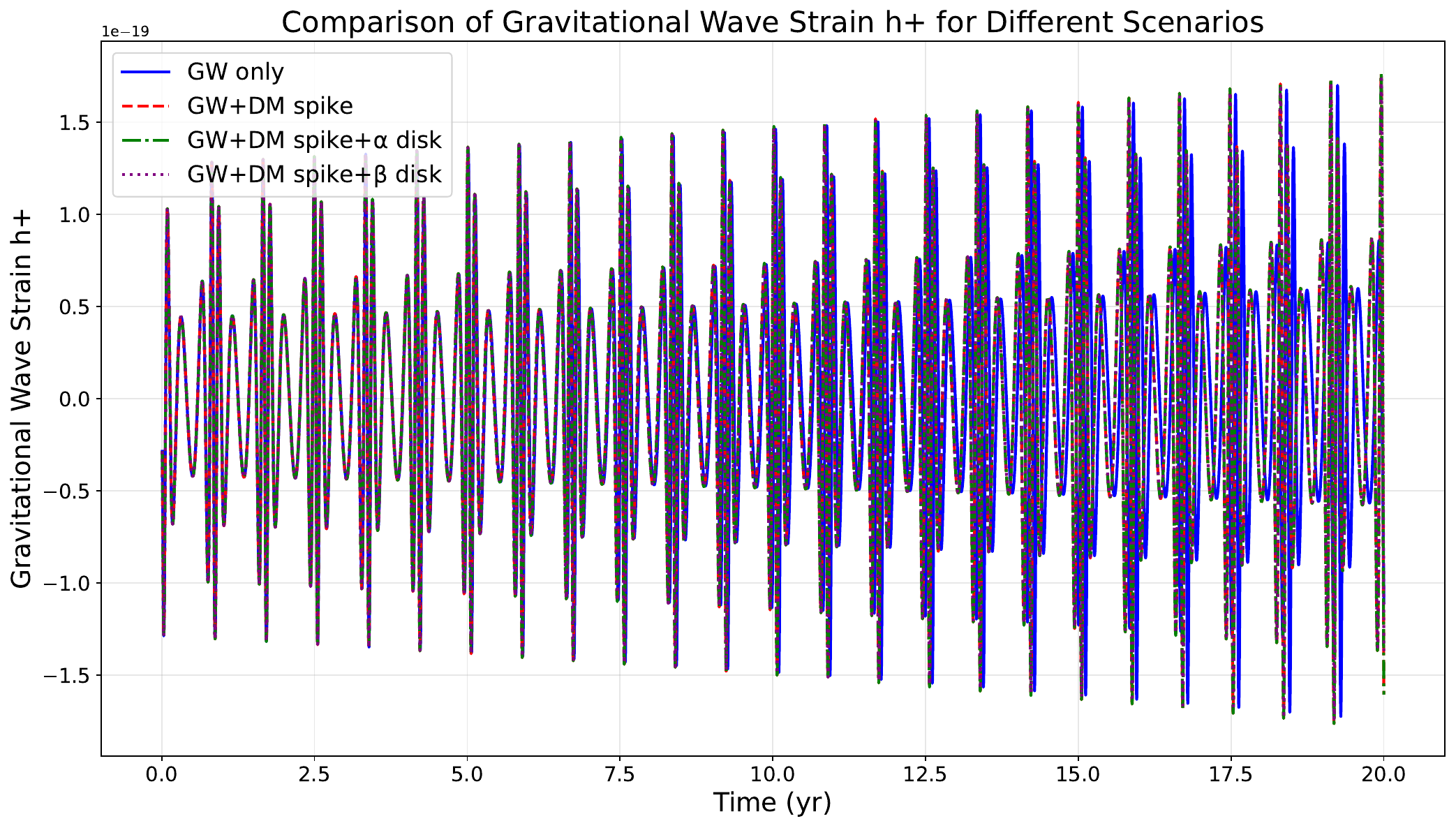}
    \\
    \includegraphics[height=0.2\textheight]{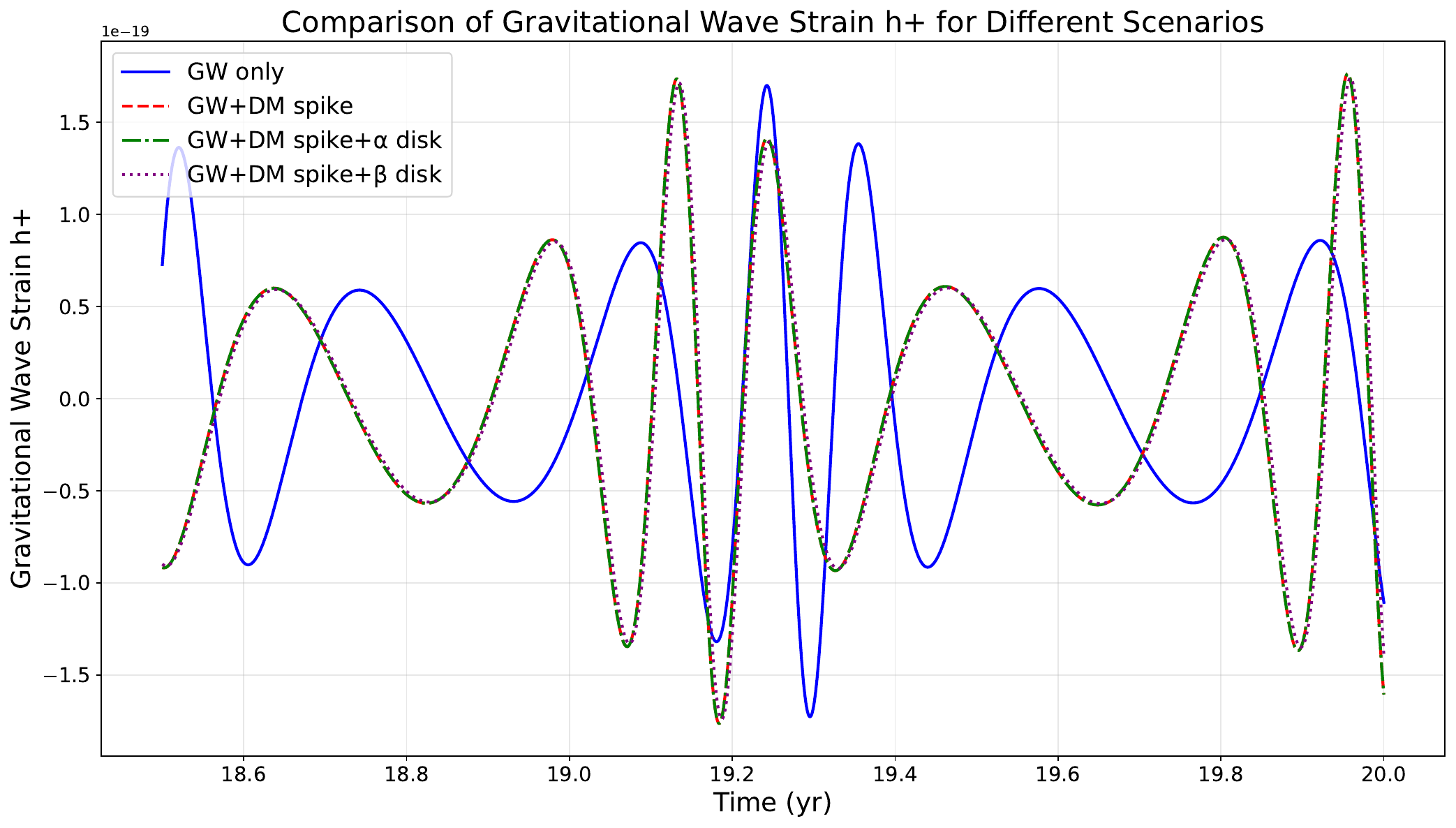}
    \\
    \includegraphics[height=0.2\textheight]{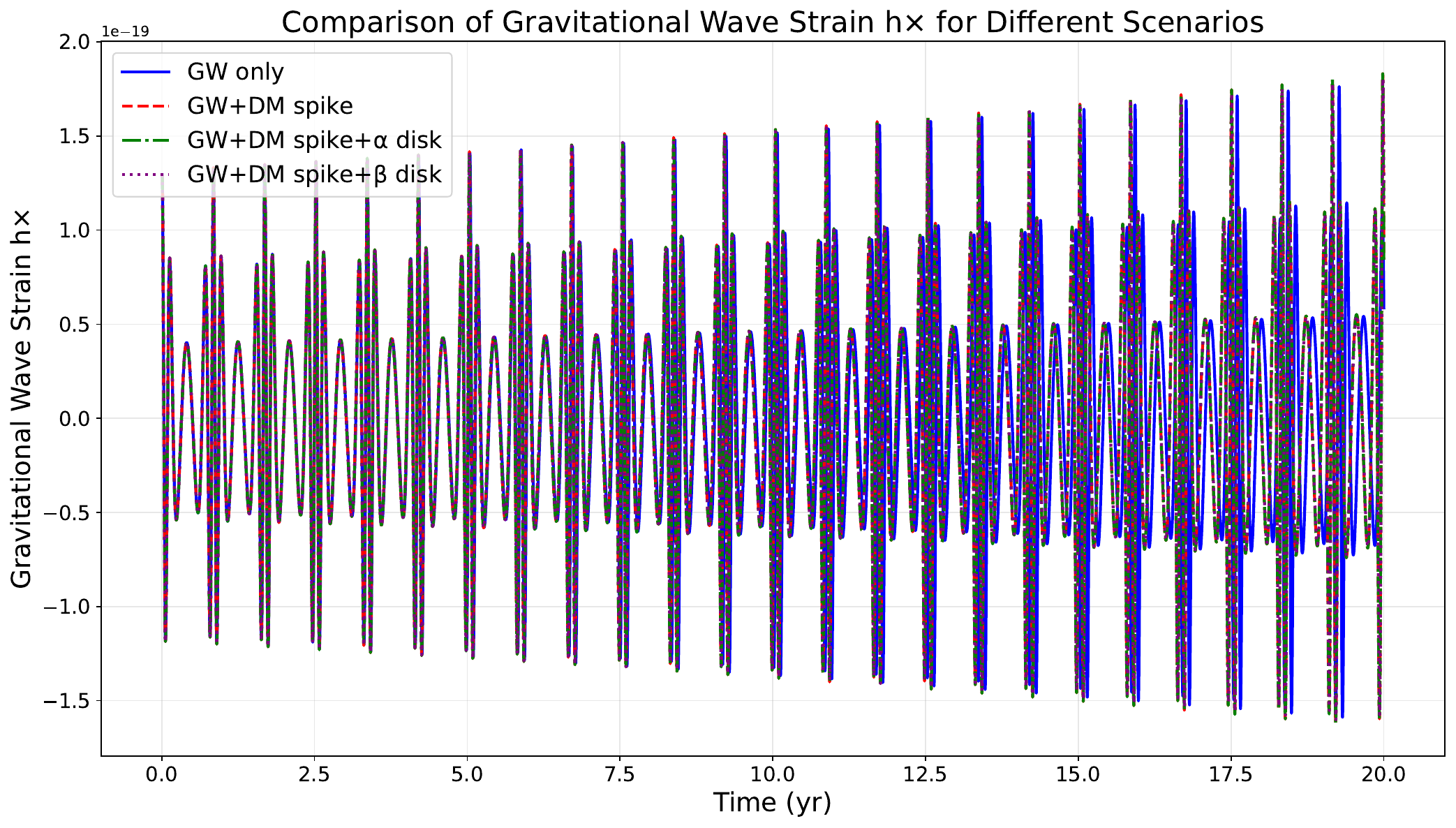}
    \\
    \includegraphics[height=0.2\textheight]{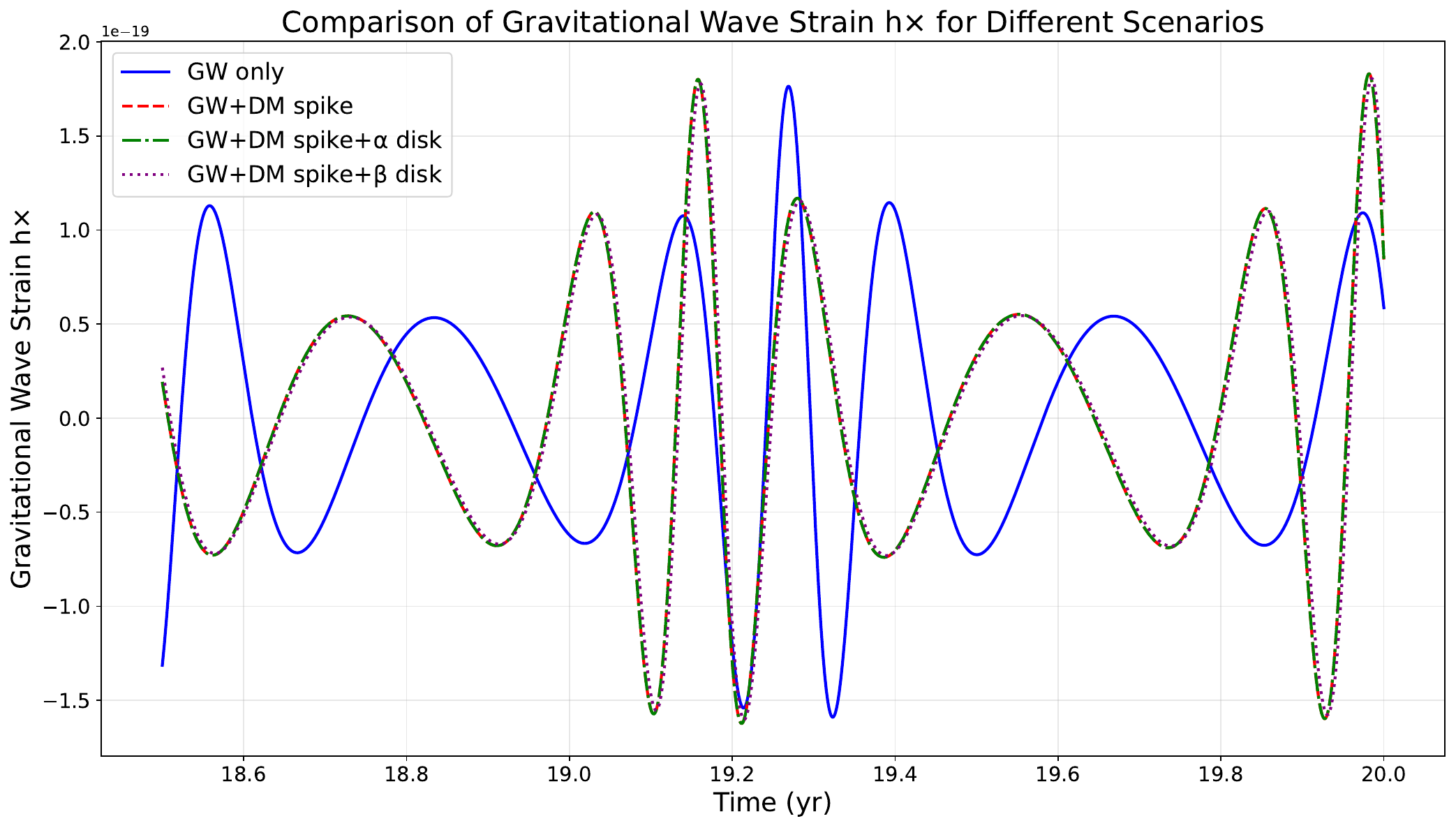}
    \caption{Time evolution of the GW strains, $h_+$ and $h_{\times}$, generated by a secondary BH ($m_{\text{2}}=1000M_{\odot}$) orbiting a central IMBH ($m_{\text{1}} =2.4\times10^4M_{\odot}$) within the LMC. The orbital dynamics are evaluated under four distinct astrophysical scenarios: pure GW radiation (vacuum), GW radiation coupled with a DM spike, and GW radiation combined with a DM spike and either an $\alpha$-disk or a $\beta$-disk. The initial orbital parameters are set to a semi-latus rectum of $p_0 =50000R_s$ and an eccentricity of $e_0=0.3$. }
    \label{fig:GWLMC}
\end{figure}

Fig.~\ref{fig:GWLMC} illustrates the GW waveforms of a BBH system in the LMC, initialized with an eccentricity of $e_0=0.3$ and a semi-latus rectum of $p_0=10^5Gm_{\text{1}}^{LMC}/c^2=50000R_s$. The orbital evolution is evaluated under four distinct scenarios: pure GW radiation (blue line), GW radiation coupled with a DM spike (red line), and GW radiation combined with a DM spike alongside either an $\alpha$-disk (green line) or a $\beta$-disk (purple line). Adopting the BBH and environmental parameters from~\cite{2025ApJ...978..104G}, we track the waveform evolution over a 20-year observational window. During the early inspiral phase, the waveforms are nearly indistinguishable, indicating that the secondary BH experiences negligible environmental perturbations. However, as the inspiral progresses, significant phase deviations emerge. For these simulations, we set the DM spike index to $\gamma_{\mathrm{sp}}=7/3$. Furthermore, because the secondary BH orbits within the outer region of the accretion disk, the local gas density is characterized by $\rho_b(r)=\rho_0(\frac{R_Q}{r})^3\exp(-\frac{r^2}{2H^2})$.

\begin{figure}
    \centering
    \includegraphics[height=0.2\textheight]{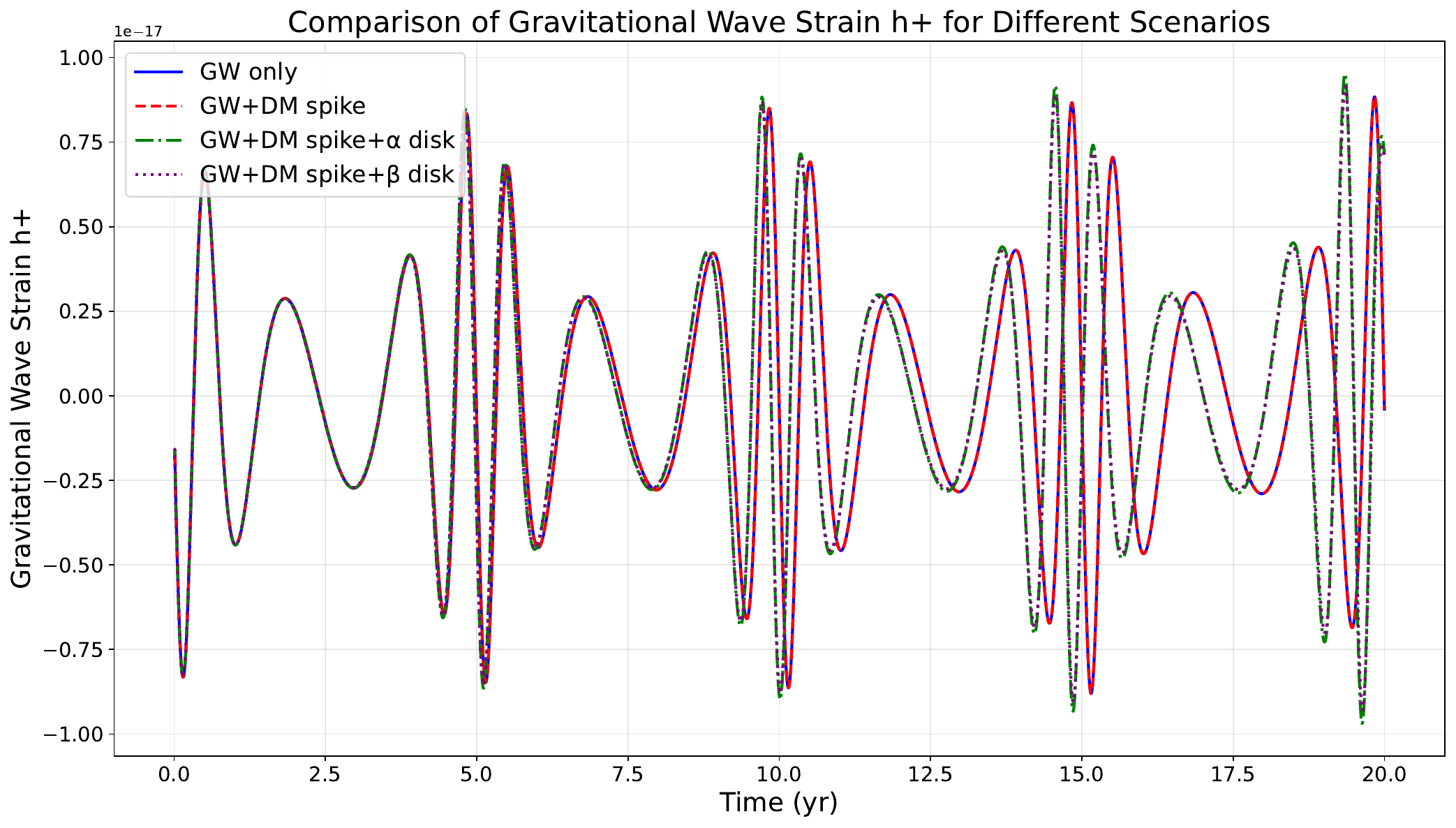}
    \\
    \includegraphics[height=0.2\textheight]{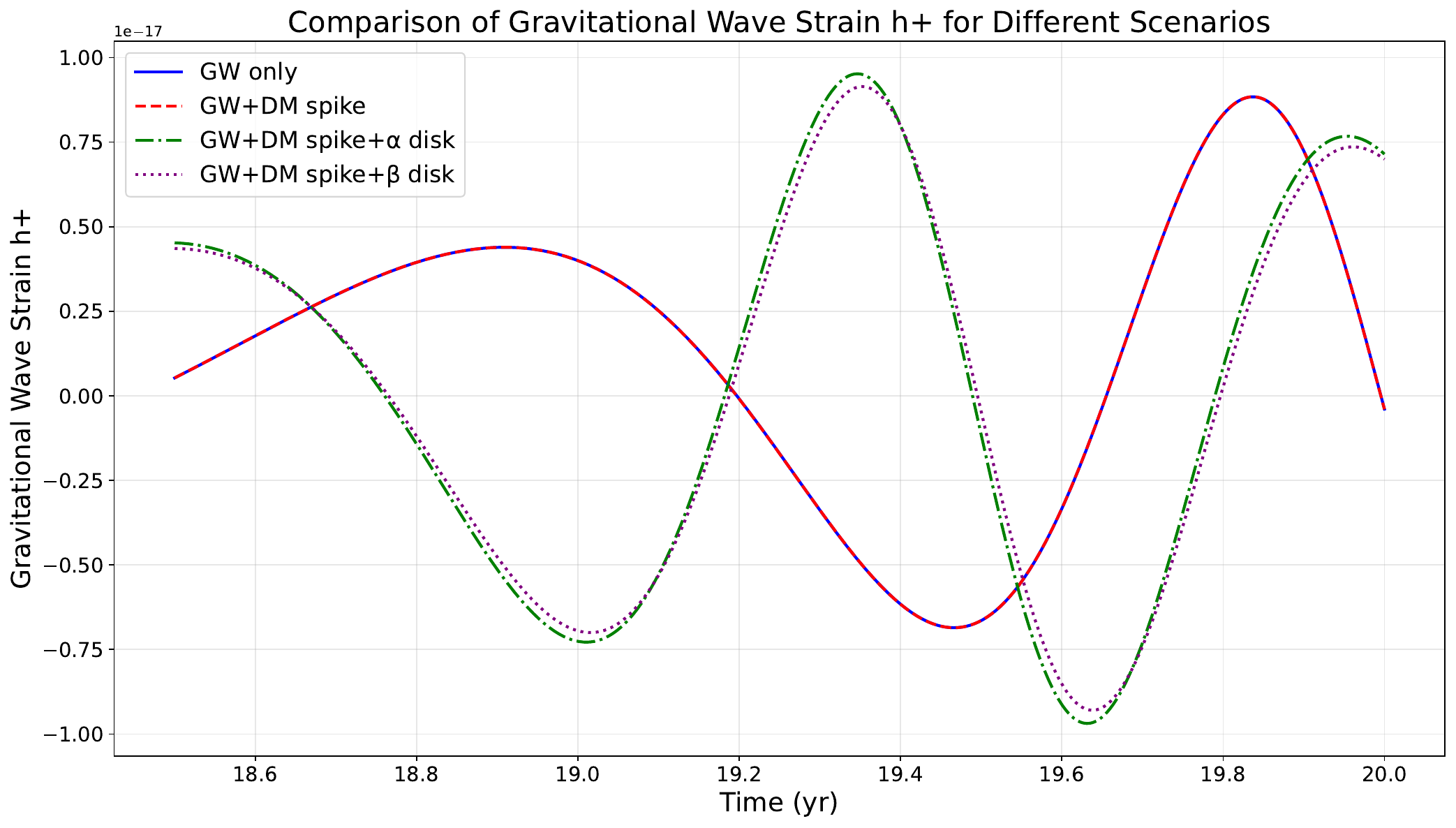}
    \\
    \includegraphics[height=0.2\textheight]{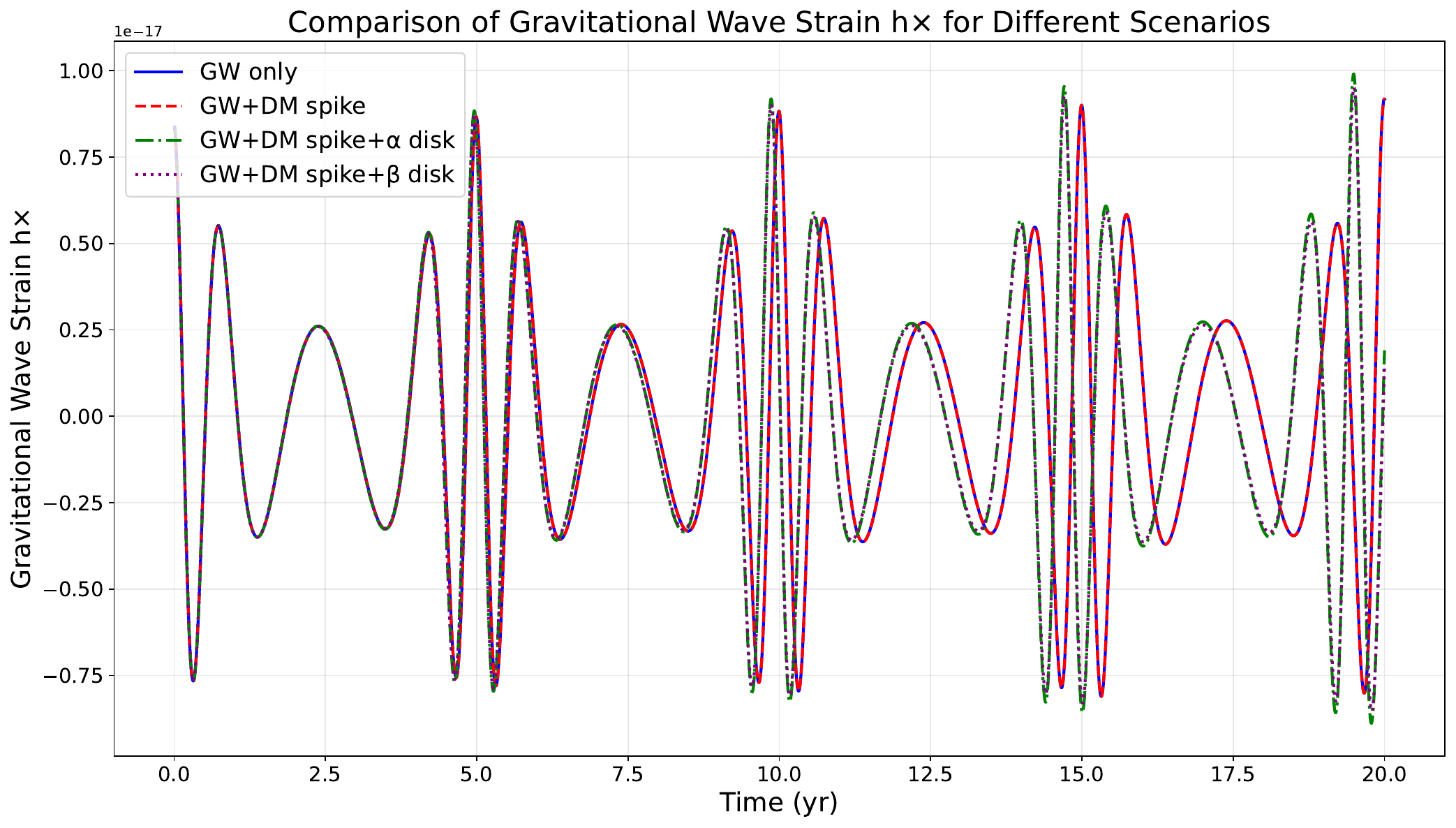}
    \\
    \includegraphics[height=0.2\textheight]{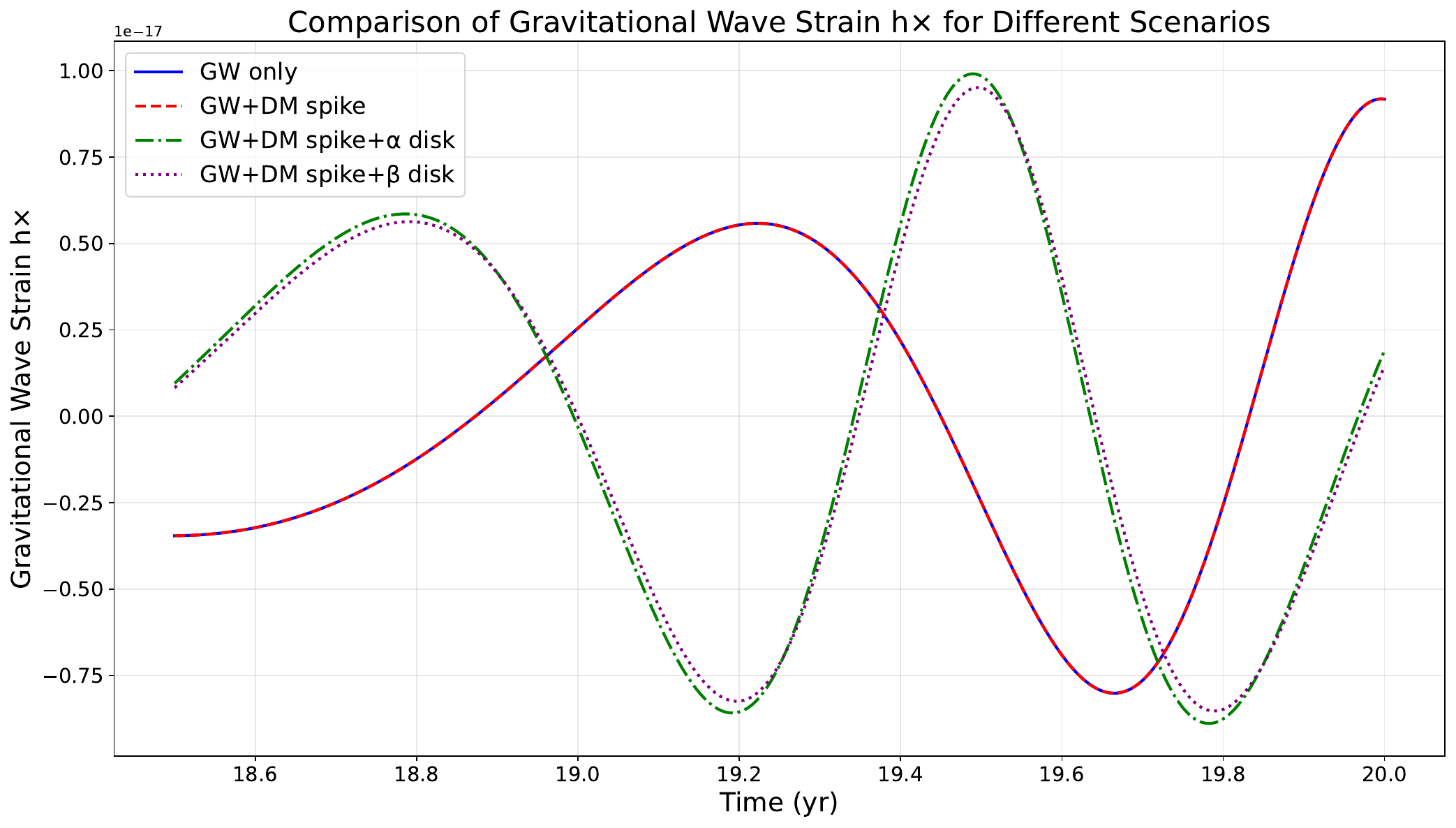}
    \caption{GW strains, $h_+$ and $h_{\times}$, generated by the inspiral of a secondary BH ($m_{\text{2}}=10000M_{\odot}$) orbiting a central SMBH ($m_{\text{1}}=1.4\times10^8M_{\odot}$) in M31. The orbital evolution is evaluated under four distinct dynamical scenarios: pure GW radiation, GW emission coupled with a DM spike, and the combined effects of GW emission, a DM spike, and either the $\alpha$ or $\beta$ accretion disk. Initial orbital parameters are set to a semi-latus rectum of $p_0 = 1000Gm_{\text{1}}/c^2=500R_s$ and an eccentricity of $e_0=0.3$. }
    \label{fig:GWM31}
\end{figure}

Fig.~\ref{fig:GWM31} shows the GW waveform of a BBH system in the M31, with an initial eccentricity  of $e_0=0.3$ and semi-latus rectum of $p_0=1000 Gm_{\text{1}}^{M31}/c^2=500R_s$. Given the substantial masses of the selected central BHs~\cite{2025ApJ...978..104G}, the corresponding orbital periods of the secondary BHs are significantly extended. Similar to the case in LMC, the GW waveforms of the BBH system in M31 show no significant difference under four different scenario at the very beginning. As time progresses, the environment leads to a mismatch in the waveforms. Unlike the case of LMC, the evolution of the waveforms in M31 requires a relatively long time due to the enormous mass of the BBH. Here, we select the DM spike index as $\gamma_{\mathrm{sp}} = 7/3$.

Compared to the scenario without DM or accretion disk, GW signals from BBH may exhibit detectable differences due to environmental effects. In Fig.~\ref{fig:GWLMC}, if there is a DM spike around the central BH, the waveform changes significantly for the blue line compared to the red line due to the large $\rho_{sp}$. Since the density in the outer region of the accretion disk is very small, after introducing the $\alpha$ disk and $\beta$ disk, the distinctions among the red, green, and purple lines are not obvious. In Fig.~\ref{fig:GWM31}, due to the particularly small value of $\rho_{sp}$, the blue line and the red line essentially overlap over a period of 20 years. Given the high density in the inner region of the accretion disk, after considering the $\alpha$ disk and $\beta$ disk, the red line, green line, and purple line become distinguishable.

Now we calculate the SNR for GW radiation from BBH with and without environmental effects and the Mismatch between these two GWs. We can obtain the GW waveform in the frequency domain through the Fourier transform:
\begin{equation}
     \tilde{h}_{+,\times}(f) = \int_{-\infty}^{\infty} h_{+,\times}(t) \, e^{-i 2\pi f t} \, dt.
\end{equation}
For two signals $h_1(t)$ and $h_2(t)$ given, we can define the inner product as
\begin{equation}
    (h_1|h_2) \equiv 4{\rm Re}\int_{f_{\min}}^{f_{\max}} \frac{\tilde{h}_1(f)\tilde{h}_2^*(f)}{S_n(f)}  df,
\end{equation}
where $\rm{Re}$ stands for the real part. $\tilde{h}(f)$ is the Fourier transformation of the time series $h(t)$, $h^*(f)$ denotes the complex conjugation and $S_n(f)$ is the one-sided noise power spectral density(PSD)~\cite{2022PhRvD.106f4003D}. The GW frequency range probed by the PTA is limited by the cadence $(\Delta t)$ and the total observation period $(T_{\rm{obs}})$, i.e., $1/T_{\mathrm{obs}}=f_{min}\lesssim f \lesssim 1/\Delta t = f_{max}$. From literature~\cite{2025ApJ...978..104G}, we can learn that the observation of current PTAs is normally set at a cadence of 1-2 weeks($\Delta t \sim 0.02\mathrm{yr} - 0.04\mathrm{yr}$), and the total observation duration that has been continuously operated is about 20 years($T_{\rm{obs}} \sim 20\mathrm{yr}$).

Paper~\cite{Gourgoulhon_2019} gave the effective SNR $\varrho$ by  
\begin{equation}
    \mathcal{\varrho}^2 =(h|h)=  4 \int_{f_{\min}}^{f_{\max}} df \frac{|\tilde{h}_+(f)|^2 + |\tilde{h}_{\times}(f)|^2}{S_n(f)}.
\end{equation}
For a PTA with $N_p (\ge 3)$ millisecond pulsars (MSPs) all located at a distance $r_p$ from the galactic center, our study employs the matched filtering approach for data processing, and the corresponding SNR expression is as follows~\cite{Guo_2022}
\begin{equation}
    \mathcal{\varrho}^2=  \sum_{i=1}^{N_p} 4\chi_i^2 \int_{f_{\min}}^{f_{\max}} df \frac{|\tilde{h}_+(f)|^2 + |\tilde{h}_{\times}(f)|^2}{S_{n,i}(f)},
\end{equation}
here $\chi$ is the geometric factor which equals 0.365 under the far-field approximation. For the convenience of theoretical analysis, assuming that all MSPs contribute equally to the SNR~\cite{Moore_2015}, the total SNR used can be approximately expressed as
\begin{equation}
    \mathcal{\varrho}^2=   4N_p \chi^2 \int_{f_{\min}}^{f_{\max}} df \frac{|\tilde{h}_+(f)|^2 + |\tilde{h}_{\times}(f)|^2}{S_n(f)}.
\end{equation}

In the above, $S_n(f) = S_{n,s} + \frac{h_b^2}{f}$ is the total noise for individual PTA sources. References~\cite{10.1093/mnras/stv1098,10.1093/mnras/stz420,Chen_2020} suggest that the noise affecting the detection of individual GW sources by PTA is primarily composed of two parts: shot noise and confusion noise with the gravitational wave background (GWB). In our calculations, we treat the GWB as a noise component. The PSD of the GWB strain originating from shot noise is typically characterized as~\cite{doi:https://doi.org/10.1002/9783527636037.ch6}
\begin{equation}
    S_{n,s}(f) = 8\pi^2\sigma_t^2 f^2 \Delta t.
\end{equation}
The strain of the GWB due to GW radiation from numerous distant inspiraling BBHs can be described as~\cite{Chen_2020}
\begin{equation}
    h_b = \mathcal{A} \frac{(f/1\text{yr}^{-1})^{-2/3}}{[1 + (f_{\text{bend}} / f)^{\kappa_{\text{gw}} \gamma_{\text{gw}}}]^{1/(2\gamma_{\text{gw}})}}.
\end{equation}
The parameters above are the median values for the GWB predictions, which can be found in~\cite{Guo_2022,Chen_2020}.

The Mismatch between two signals is defined as
\begin{equation}
    \text{Mismatch} = 1 - \mathcal{O}(h_1, h_2),
\end{equation}
here $\mathcal{O}(h_1, h_2)$ is the overlap between two GW signals quantified as the ``Fitting Factor"(FF)~\cite{PhysRevD.49.6274,PhysRevD.78.124020}:
\begin{equation}
    \mathcal{O}(h_1, h_2) =\text{FF}= \frac{( h_1 | h_2)}{\sqrt{( h_1 |h_1) ( h_2 | h_2 )}}.
\end{equation}
Obviously, the Mismatch is zero if two signals are identical. The two waveforms are indistinguishable if $(\delta h|\delta h)=(h_1-h_2|h_1-h_2)<1$~\cite{Fang_2019}.

\begin{table}
    \centering
    \begin{subtable}{0.5\textwidth}
        \centering
        \begin{tabular}{c|c|ccc}
            \hline\hline
              \multirow{2}{*}{LMCC-PTA} &  \multirow{2}{*}{$m_{\text{2}}/M_\odot$} & \multicolumn{3}{c}{$a_0$/AU}\\
              \cline{3-5}
             & & 12 & 24 & 144\\
            \hline
         GW+DM+$\alpha$ disk   &\multirow{2}{*}{500} & 3.54 & 4.56 & 2.45\\
            \cline{3-5}
           GW+DM+$\beta$ disk   &  & 3.43 & 4.56 & 2.45 \\
            \hline
         GW+DM+$\alpha$ disk   &\multirow{2}{*}{1000} & 7.25 & 10.6 & 5.02\\
            \cline{3-5}
           GW+DM+$\beta$ disk   &  & 6.8 & 10.6 & 5.02\\
            \hline
         GW+DM+$\alpha$ disk   &\multirow{2}{*}{2000} & 15.2 & 18.4 & 10.6\\
            \cline{3-5}
           GW+DM+$\beta$ disk   &  & 13.2 & 18.4 & 10.6\\
            \hline\hline
        \end{tabular}
        \caption{SNRs detected by LMCC-PTA for BBH in LMC.} 
        \label{tab:sub_a}
    \end{subtable}
    
    \bigskip

    \begin{subtable}{0.5\textwidth}
        \centering
        \begin{tabular}{c|c|ccc}
            \hline\hline
              \multirow{2}{*}{M31C-PTA} &  \multirow{2}{*}{$m_{\text{2}}/M_\odot$} & \multicolumn{3}{c}{$a_0$/AU}\\
              \cline{3-5}
             & & 138 & 1380 & 2760\\
            \hline
         GW+DM+$\alpha$ disk   &\multirow{2}{*}{20} & 3.32 & 8.62 & 3.05\\
            \cline{3-5}
           GW+DM+$\beta$ disk   &  & 3.32 & 8.62 & 3.05 \\
            \hline
         GW+DM+$\alpha$ disk   &\multirow{2}{*}{100} & 16.6 & 43.1 & 15.3\\
            \cline{3-5}
           GW+DM+$\beta$ disk   &  & 16.6 & 43.1 & 15.3 \\
            \hline
         GW+DM+$\alpha$ disk   &\multirow{2}{*}{1000} & 166 & 435 & 162\\
            \cline{3-5}
           GW+DM+$\beta$ disk   &  & 166 & 434 & 157 \\
            \hline
         GW+DM+$\alpha$ disk   &\multirow{2}{*}{10000} & 1650 & 3850 & 2720\\
            \cline{3-5}
           GW+DM+$\beta$ disk   &  & 1660 & 3780 & 1930 \\
            \hline\hline
              \multirow{2}{*}{SKA-PTA} &  \multirow{2}{*}{$m_{\text{2}}/M_\odot$} & \multicolumn{3}{c}{$a_0$/AU}\\
              \cline{3-5}
             & & 138 & 276 & 552\\
            \hline
         GW+DM+$\alpha$ disk   &\multirow{2}{*}{$10^5$} & 1.82 & 3.42 & 4.30\\
            \cline{3-5}
           GW+DM+$\beta$ disk   &  & 1.86 & 3.54 & 4.50 \\
            \hline
         GW+DM+$\alpha$ disk   &\multirow{2}{*}{$10^6$} & 18.4 & 33.5 & 54.8\\
            \cline{3-5}
           GW+DM+$\beta$ disk   &  & 25.2 & 66.6 & 48.5 \\
            \hline
         GW+DM+$\alpha$ disk   &\multirow{2}{*}{$2\times10^6$} & 37.7 & 68.2 & 258\\
            \cline{3-5}
           GW+DM+$\beta$ disk   &  & 84.2 & 68.6 & 117 \\
            \hline\hline
        \end{tabular}
        \caption{SNRs detected by M31C-PTA and SKA-PTA for BBH in M31.}
        \label{tab:sub_b}
    \end{subtable}
    \caption{SNRs detected by different PTAs (see~\cite{2025ApJ...978..104G}) for BBH with different mass $m_2$ and initial semimajor axis $a_0$ in LMC (a)  and M31 (b)  in the case of GW radiation + DM spike + $\alpha$ (first row for each mass value)  or $\beta$ disk (second row for each mass value). Here, we set the eccentricity $e_0=0.3$. }
    \label{tab:SSNNRR}
\end{table}

As presented in Table~\ref{tab:SSNNRR}, assuming a rigorous detection threshold of SNR $\ge 8$ to define the theoretical horizon, the detectability of the GW signals is highly dependent on the adopted PTA and the specific parameters of the binary system. 
For the LMC scenarios observed by LMCC-PTA (parameters of Fig.~\ref{fig:GWLMC}), the secondary BH must possess a mass of at least $m_2 \gtrsim 10^3 M_\odot$ to be detectable. Even at this mass, the binary is restricted to specific intermediate orbits, such as $a_0 \approx 24$ AU, since tighter ($12$ AU) or broader ($144$ AU) orbits yield SNRs below 8. To guarantee robust detection across a wider spatial range ($a_0 \in [12, 144]$ AU), $m_2$ must exceed $2 \times 10^3 M_\odot$. 
Conversely, in the M31 scenarios (parameters of Fig.~\ref{fig:GWM31}), the enormous mass of the primary BH substantially amplifies the GW strain. When utilizing M31C-PTA, a relatively light secondary BH of $m_2 \approx 20 M_\odot$ can just reach the detection threshold, but it is strictly limited to orbits around $a_0 \approx 1380$ AU. To safely exceed the SNR $\ge 8$ threshold across a vast semi-major axis span of $a_0 \in [138, 2760]$ AU, a mass of $m_2 \gtrsim 100 M_\odot$ is required. 
Interestingly, when assessing the M31 center with SKA-PTA, the parameter space for detectability shifts significantly. To surpass the SNR $\ge 8$ threshold, the secondary BH must be substantially more massive, requiring $m_2 \gtrsim 10^6 M_\odot$, where signals remain strong across $a_0 \in [138, 552]$ AU. At lower masses, such as $m_2 = 10^5 M_\odot$, the maximum SNR drops to roughly $4.5$, falling completely short of the threshold regardless of the orbital separation. 
These results demonstrate that the GW signals from such BBH systems we set are well within the detectable threshold of future PTA observations, thereby confirming the physical viability of the parameters adopted in our model~\cite{2025ApJ...978..104G}. The other SNR values, which are deeply modulated by the distance between the two BHs and the mass of the secondary BH, are also presented in Table~\ref{tab:SSNNRR}.

\begin{figure}
    \centering
    \includegraphics[height=0.2\textheight]{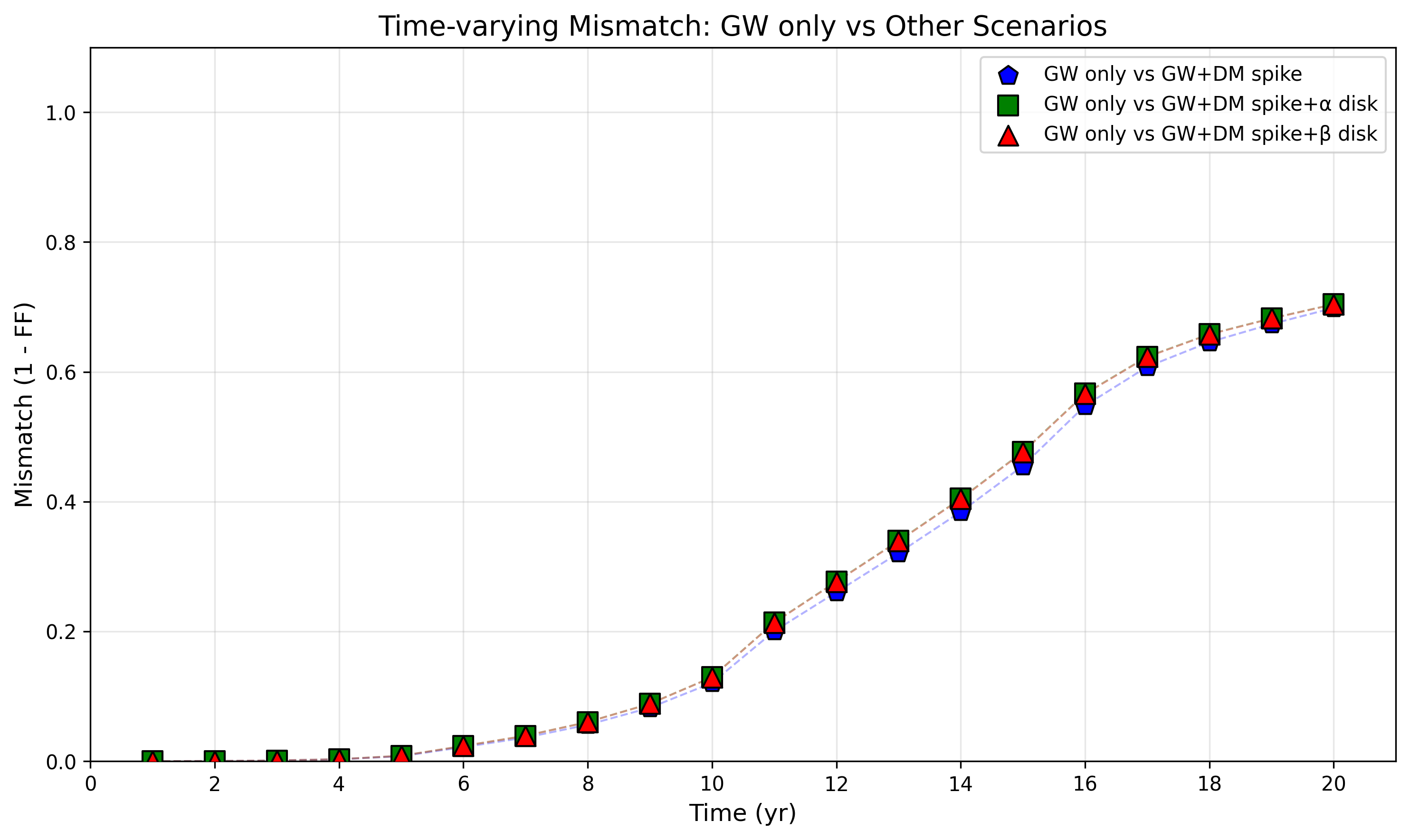}
    \\
    \includegraphics[height=0.2\textheight]{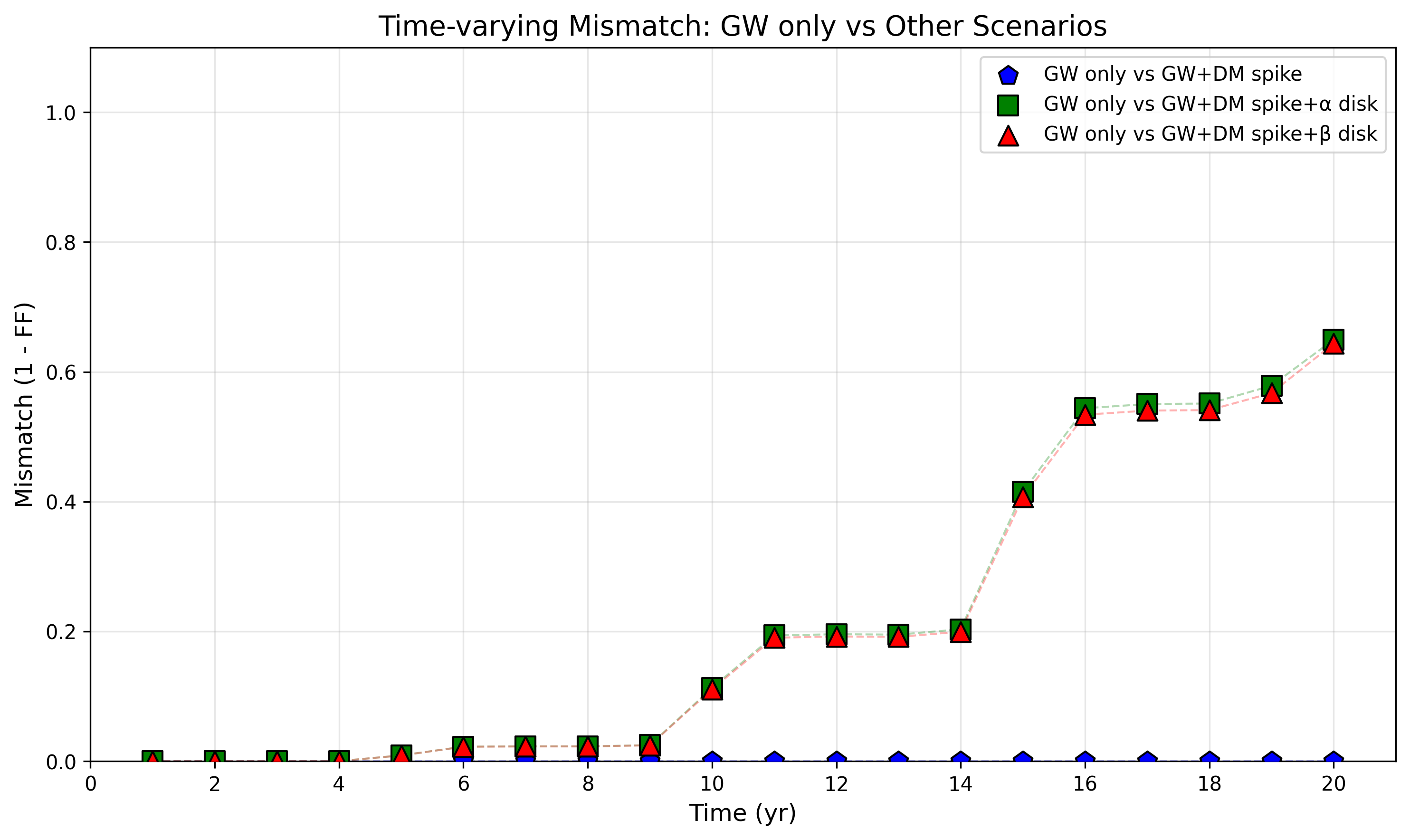}
    \caption{Mismatch as a function of observation time for the three environmental scenarios corresponding to Fig.~\ref{fig:GWLMC} and Fig.~\ref{fig:GWM31}. The top and bottom panels illustrate the results for the LMCC-PTA and M31C-PTA configurations, respectively.}
    \label{fig:mismatch}
\end{figure}

The relationship between the detectability and the environmental signatures is further elucidated by the Mismatch evolution shown in Fig.~\ref{fig:mismatch}. While a high SNR guarantees the robust detection of the GW source, the Mismatch characterizes the distinguishability of environmental effects from vacuum templates. Specifically, for Fig.~\ref{fig:GWLMC} with LMCC-PTA, a Mismatch of approximately 0.7 over a 20-year observation period indicates that the dynamical deviations induced by the DM spike and accretion disk are highly resolvable. In Fig.~\ref{fig:GWM31} with M31C-PTA case, although the DM spike alone results in a smaller Mismatch, the exceptionally high SNR, combined with the influence of the accretion disk, ensures that even subtle waveform departures could be identified in future high-precision data.

\section{Conclusions}
\label{sec:con}

In this work, we investigate the influence of environmental effects, including the DM spike and accretion disk, on the orbital evolution of inspiraling BBHs based on the post-Newtonian approximation theory. The orbital modifications induced by environmental effects lead to deviations in GW waveforms, and our analysis primarily focuses on the frequency band relevant to PTAs. 

Specifically, we employed the NFW profile to investigate the DM density distribution in two galaxies respectively. First, we determined the total DM mass (i.e., the virial mass $M_\mathrm{vir}$) within the virial radius $R_\mathrm{vir}$ of the LMC and M31. Under this constraint, we considered the DM spikes formed through adiabatic compression within the gravitational influence radii of the central SMBH and IMBH in the two galaxies, which are dense regions of DM distribution. Our analysis demonstrates that a larger spike index $\gamma_{\text{sp}}$ correlates with a higher DM density.

We then investigated the scenario where the secondary BH moves along an elliptical orbit using the post-Newtonian approximation. In the purely classical case without considering GW radiation, the orbit of the secondary BH around the central BH is strictly closed, conserving energy and angular momentum. However, when GW radiation is taken into account, the BBH system loses energy and angular momentum, causing the secondary BH to gradually approach the central BH and eventually merge. We found that for a fixed initial $p_0$, a larger initial $e_0$ results in a longer inspiral time. When considering the presence of DM spike, the secondary BH in the DM environment is also affected by DF and accretion. Our analysis shows that while GW radiation tends to reduce the orbital eccentricity $e$ and circularize the orbit, the presence of DM increases it. For the same initial eccentricity $e_0$ and semi-latus rectum $p_0$, this eccentricity enhancement effect becomes more pronounced with higher DM density (i.e., larger $\gamma_{sp}$). Given an initial semi-latus rectum value of $p_0 = 500R_s$ and an initial eccentricity of $e_0 = 0.3$, compared to the vacuum case, the secondary BH consumes significantly less time to reach the $r_{\rm{ISCO}}$ of the central BH within the DM spike.

Next, we considered the scenario in which a spherically symmetric DM spike and an axisymmetric accretion disk coexist around the central BH. It was assumed that the secondary BH moves entirely within the DM spike and the accretion disk. The accretion disk is divided into inner and outer regions, with different model distributions applied. In the inner region, the standard thin disk model ($\alpha$ disk and $\beta$ disk) was used, while in the outer region, the self-gravitating disk model was adopted. Comparing Fig.~\ref{fig:FINALDM} and Fig.~\ref{fig:acdfianl}, with $\gamma_{sp}=7/3$, $p_0=500R_s$, and $e_0=0.3$, it can be observed that introducing an accretion disk on basis of the DM spike significantly accelerates the merger of the BBH. For the BBH system at the center of each galaxy, the eccentricity $e$ increases markedly in the later stages of evolution, indicating that the orbit becomes increasingly flattened prior to the secondary BH reaching the $r_{\rm{ISCO}}$ of the central BH.

Based on the derived GW formulas \eqref{eq49} and \eqref{eq50}, we have plotted the waveforms for the first 20 years under environmental influences as shown in Figs.~\ref{fig:GWLMC} and~\ref{fig:GWM31}, and calculated their SNRs. Specifically, assuming a rigorous detection threshold of $\text{SNR} \ge 8$, we find that robust detection in the LMC by LMCC-PTA requires a secondary BH mass of $m_2 \gtrsim 10^3 M_\odot$ at specific intermediate orbits ($a_0 \approx 24$ AU). In contrast, the immense central mass in M31 allows M31C-PTA to detect significantly lighter companions ($m_2 \gtrsim 100 M_\odot$) across a broad orbital range ($a_0 \in [138, 2760]$ AU), while SKA-PTA requires much more massive companions ($m_2 \gtrsim 10^6 M_\odot$) to surpass the same threshold. Our results indicate that such BBH systems are highly detectable by future PTA observations, which firmly validates the astrophysical parameters adopted from~\cite{2025ApJ...978..104G}. More importantly, our Mismatch analysis reveals the distinguishability of these environmental signatures. After 20 years of observation, the Mismatches for the LMC scenarios reach approximately 0.7, meaning the environmental deviations are highly distinguishable from pure vacuum waveforms. For M31, although the Mismatch induced solely by the DM spike approaches 0 due to its lower density, the inclusion of an accretion disk introduces discernible Mismatch signatures. When combined with the extraordinarily high SNRs, we conclude that environmental effects in both galaxies leave robust and extractable imprints on the GW signals, providing a promising observational avenue for future multi-messenger astronomy.

\section{Discussions}
\label{sec:dis}

Our current dynamical framework provides a comprehensive baseline for evaluating the environmental effects on inspiraling BBHs. However, it employs several simplifying approximations. To contextualize our findings and guide future refinements, we focus our discussion on the limitations of the fundamental theoretical framework, the assumptions in our astrophysical modeling, the complexities of dynamic orbital interactions, and the ultimate observational prospects.

Fundamentally, our model relies on a classical and leading-order post-Newtonian baseline to maintain analytical tractability over astronomically long secular evolutions. A critical limitation of this approach is the decoupled treatment of GW radiation reaction and environmental effects. By assuming a Schwarzschild background for the central BH and deriving the secular evolution of orbital elements via leading-order post-Newtonian expansions, the GW fluxes are effectively evaluated using vacuum black-hole perturbation theory. However, a growing body of recent literature demonstrates that dark and baryonic environments couple naturally with gravitational radiation~\cite{Cardoso:2022whc,Datta:2025ruh,Destounis:2025tjn}. The presence of an astrophysical environment intrinsically perturbs the background spacetime and significantly modifies the resulting GW fluxes. Future iterations must transcend this approximation by self-consistently incorporating environmental density and flux perturbations directly into fully relativistic background fluxes.

Furthermore, as the secondary BH approaches the ISCO, strong-field effects inevitably take over. In our current framework, DF is approximated using purely classical Newtonian formulations. Recent theoretical advancements, however, have successfully pushed DF calculations to a fully relativistic level~\cite{Vicente:2022ivh,s4wh-x6c4}. In the strong-field regime, relativistic corrections to DF, extreme mass-ratio self-force, and the spin of the central BH (Kerr metric effects) will alter the magnitude and nature of drag forces. Implementing fully relativistic models such as relativistic Vlasov solvers for DM~\cite{Montalvo:2024iwq} or General Relativistic Magnetohydrodynamics for accretion disks~\cite{Ressler:2024mpx, Kim:2024zjb} remains a crucial next step, despite the insurmountable analytical challenges they currently pose for long-term orbital integrations.

Beyond the fundamental gravitational framework, the astrophysical modeling of both dark and baryonic environments in this study relies on idealized static assumptions that could be expanded to reflect greater physical realism. For the DM halo, we currently adopt the NFW profile as the matching condition for the DM spike. Future work could explore alternative density distributions, such as the Einasto profile $\rho_{\text{Ein}}(r) = \rho_0 \exp\left\{-\frac{2}{\alpha_{\text{Ein}}}\left[ \left(\frac{r}{r_0}\right)^{\alpha_{\text{Ein}}} - 1\right]\right\}$~\cite{1965TrAlm...5...87E}, the Cored Isothermal profile $\rho_{\text{Iso}}(r) = \frac{\rho_0}{1+(r/r_0)^2}$~\cite{10.1093/mnras/249.3.523}, or the Burkert profile $\rho_{\text{Bur}}(r) = \frac{\rho_0}{(1+r/r_0)(1+(r/r_0)^2)}$~\cite{Burkert_1995}. Additionally, we treat environmental influences as linear combinations, whereas these effects are dynamically coupled and require comprehensive numerical simulations to resolve non-linear interactions. Other complex phenomena, such as halo feedback mechanisms~\cite{PhysRevD.102.083006,PhysRevD.105.043009}, relativistic corrections to spike distributions~\cite{PhysRevD.106.044027}, and spikes around lower-mass primordial BHs~\cite{PhysRevD.107.083006}, also warrant future investigation.

Regarding the baryonic environment, modern astrophysics treats accretion disks as dynamic, multi-form systems characterized by state transitions. We utilized continuous thin disk profiles, but various models exist for different accretion states, including thick disks~\cite{1980AcA....30....1J,1981AcA....31..283P,1982ApJ...253..897P,10.1111/j.1365-2966.2006.10183.x}, slim disks~\cite{1988ApJ...332..646A}, and advection-dominated accretion flows (ADAFs)~\cite{narayan1998advectiondominatedaccretionblackholes}. Given that the central BHs in the LMC and M31 are currently under-luminous and not in highly active quasar states, ADAFs or truncated thin disks might better reflect their present-day physical realities. Combining these refined models to fit observational data will allow for more accurate inferences of the BH parameters and their surrounding environments.

The orbital evolution itself is fundamentally dictated by the complex interactions between the secondary BH and its surrounding medium. Our framework utilizes classical gas drag to provide a phenomenological baseline for the early inspiral phase. However, this omits fully relativistic disk-binary interactions, specifically relativistic Lindblad torques, which diverges from the state-of-the-art frameworks advanced by the LISA Science Team. In the strong-field regime, these relativistic effects can amplify torque magnitudes by several orders of magnitude~\cite{Duque:2025yfm}. More profoundly, they can undergo a sign reversal, potentially driving an outward orbital migration rather than a continuous inspiral.

Beyond planar drag forces, the geometry of the binary-disk interaction plays a critical role. An orbital misalignment between the BBH and the accretion disk implies that the secondary BH periodically passes through the disk, transforming continuous hydrodynamical drag into impulsive, discrete dynamical modifications. Furthermore, a sufficiently massive secondary BH can exert tidal torques that clear a distinct gap or cavity within the gas, a phenomenon with strong observational implications observed in systems like the quasar Mrk 231~\cite{Yan:2015mya}. The formation of such a cavity locally depletes the gas density, meaning that the continuous thin disk profiles we employed likely provide an upper bound for hydrodynamical drag and Bondi-Hoyle accretion rates. Future refinements should account for localized density depletion, as the interplay between gap-clearing processes and orbital evolution could introduce unique modulations in the resulting GW signals.

Translating these dynamical signatures into observational prospects requires analyzing how different environments dominate distinct orbital stages. In the spatially extended early inspiral phase, where the secondary BH moves at weakly relativistic speeds ($v \ll c$), gas drag from the accretion disk typically governs the orbital decay. As the binary shrinks into the late inspiral stages, the effects of the DM spike driven by DF and Bondi-Hoyle accretion become increasingly pronounced. Our comparative analysis demonstrates that these environments leave distinct imprints depending on galactic parameters: the high DM spike density in the LMC yields a significant phase Mismatch (approaching $0.7$ over a 20-year observation period), heavily driving the evolution. Conversely, in M31, the lower DM density allows the dense inner region of the accretion disk to introduce discernible waveform deviations. 

Given the high SNRs anticipated for such massive systems, combining future multi-band data from LISA, PTAs, and other GW facilities will enable matched-filtering analyses to break parameter degeneracies. Although tracking a complete BBH inspiral exceeds human timescales, the decades-long observational windows of PTAs provide crucial ``snapshots'' of the orbital phase. Precisely measuring the long-term phase and eccentricity evolution makes it feasible to isolate environmental signatures and reconstruct galactic DM distributions entirely independent of electromagnetic observations.

\acknowledgments 
This work is supported by the National Natural Science Foundation of China (Nos.12375059, 12503001), the National Key Research and Development Program of China (Nos. 2021YFC2203001, 2021YFC2201901), and the Project of National Astronomical Observatories, Chinese Academy of Sciences (No. E4TG6601).
We thank Hanjun Zou for helpful discussions.

\appendix

\section{Trend Chart}
\label{onlyDM}

\begin{figure}[htbp]
    \centering
    \includegraphics[width=0.4\textwidth]{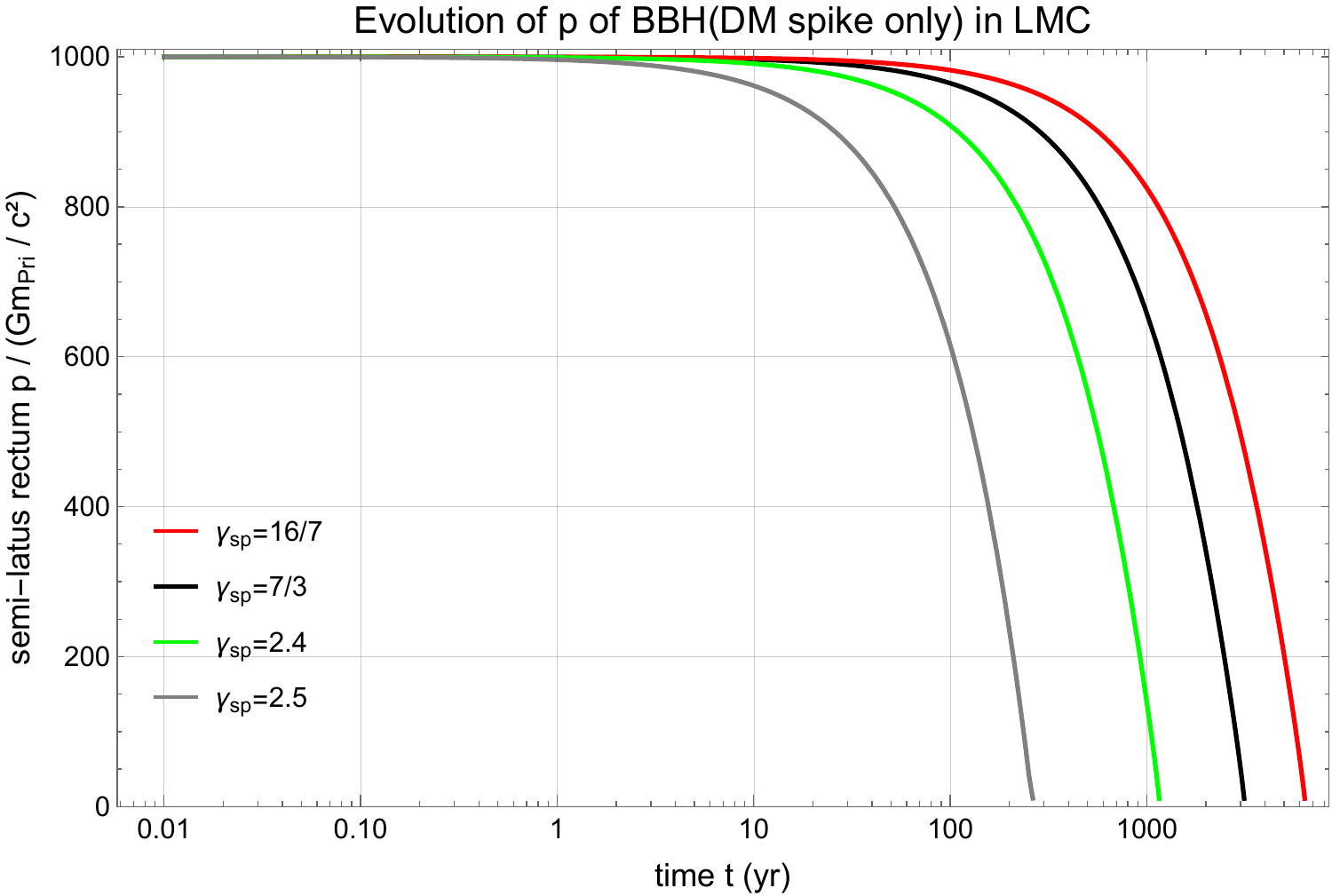}
    \\
    \includegraphics[width=0.4\textwidth]{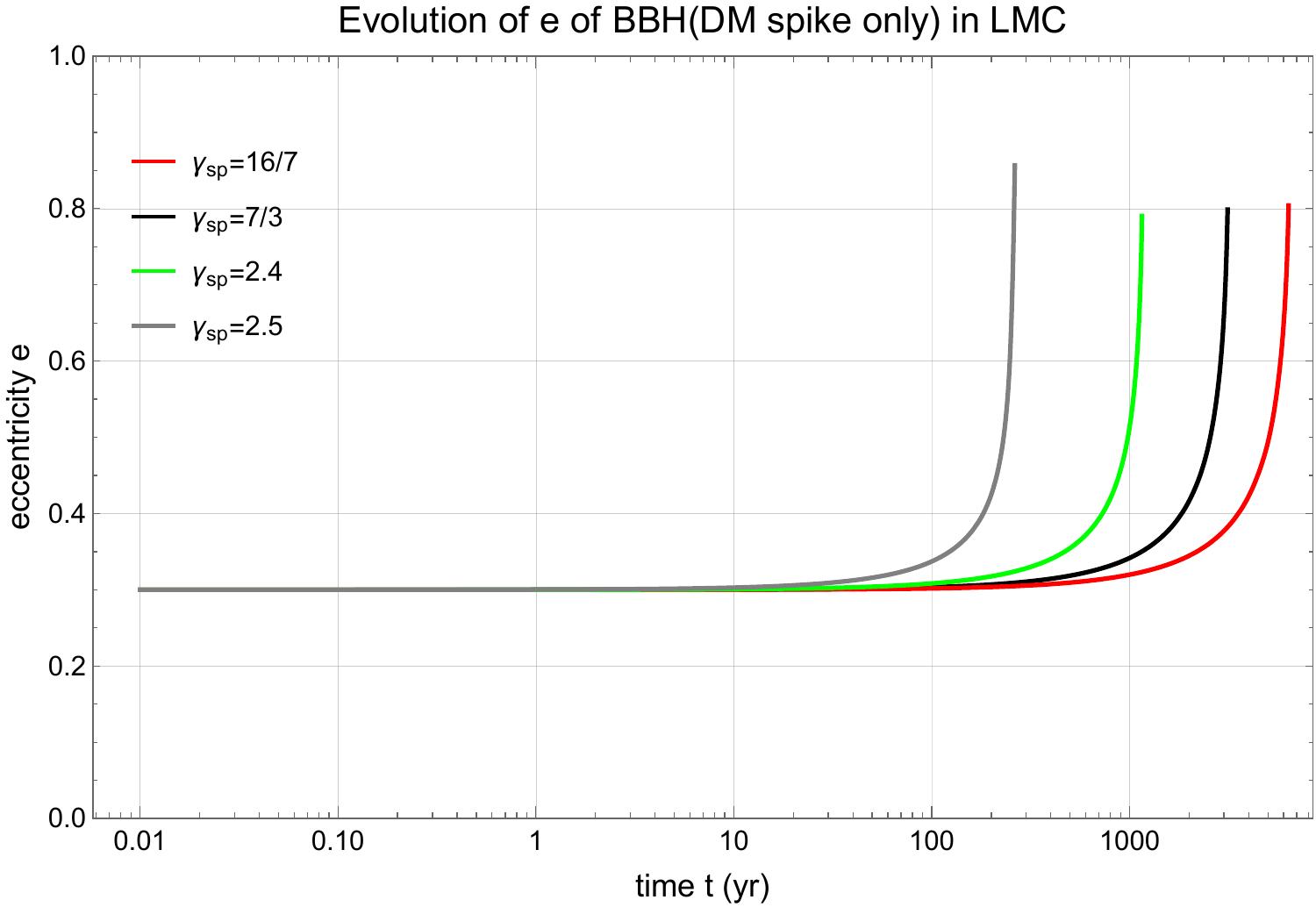}
    \caption{The plots are calculated using Eqs.~(\ref{eq23}) and~(\ref{eq27}).  The upper panel illustrates the influence of the DM spike on the temporal evolution of the semi-latus rectum $p$. Similarly, based on Eqs.~(\ref{eq24}) and~(\ref{eq28}), the lower panel depicts the impact of the DM spike on the evolution of the eccentricity $e$. In the absence of GW radiation, the combined effects of DF and accretion from the DM spike induce a secular decay in $p$ accompanied by a continuous growth in $e$.}
    \label{fig:onlyDMLMC}
\end{figure}

\begin{figure}[htbp]
    \centering
    \includegraphics[width=0.4\textwidth]{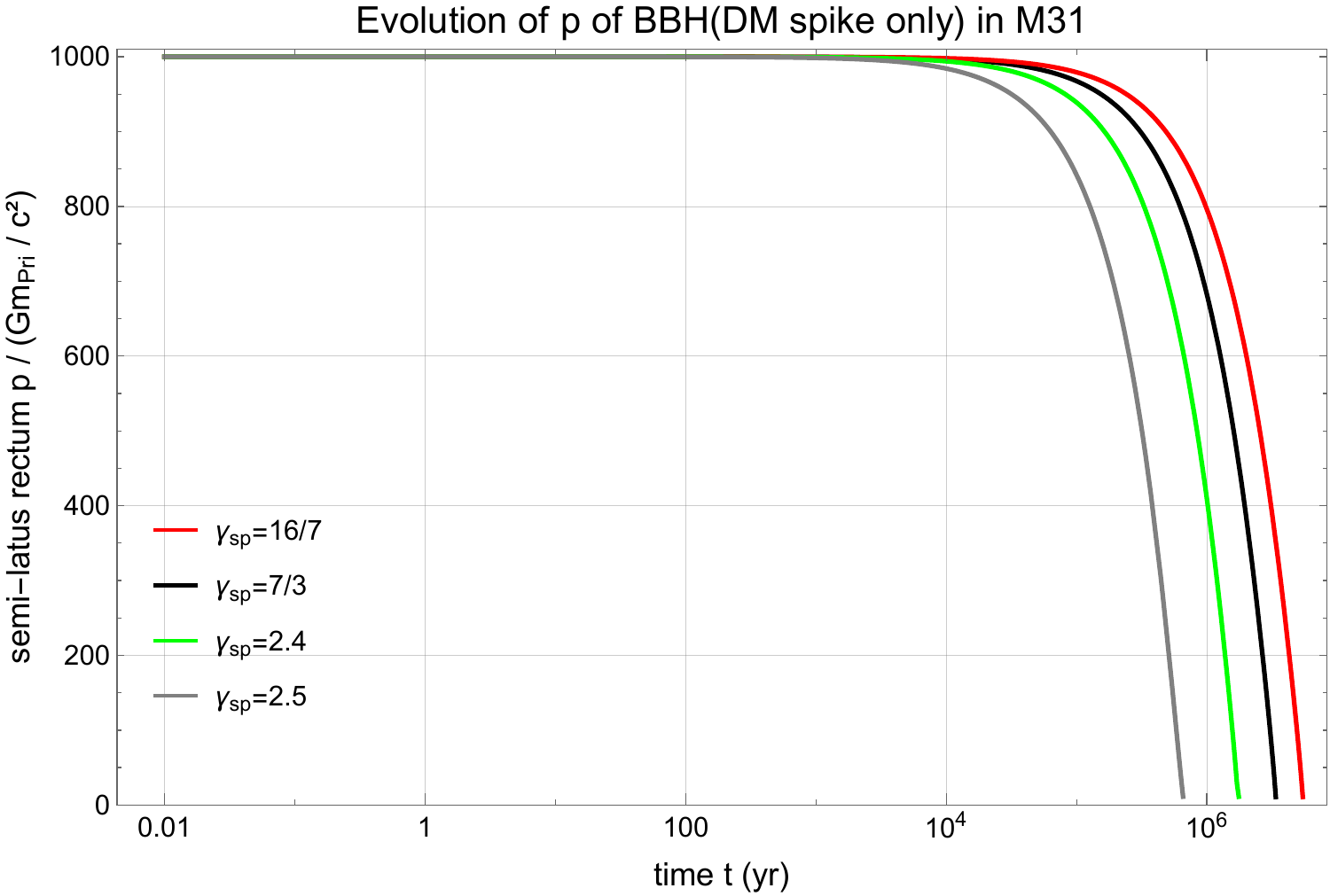}
    \\
    \includegraphics[width=0.4\textwidth]{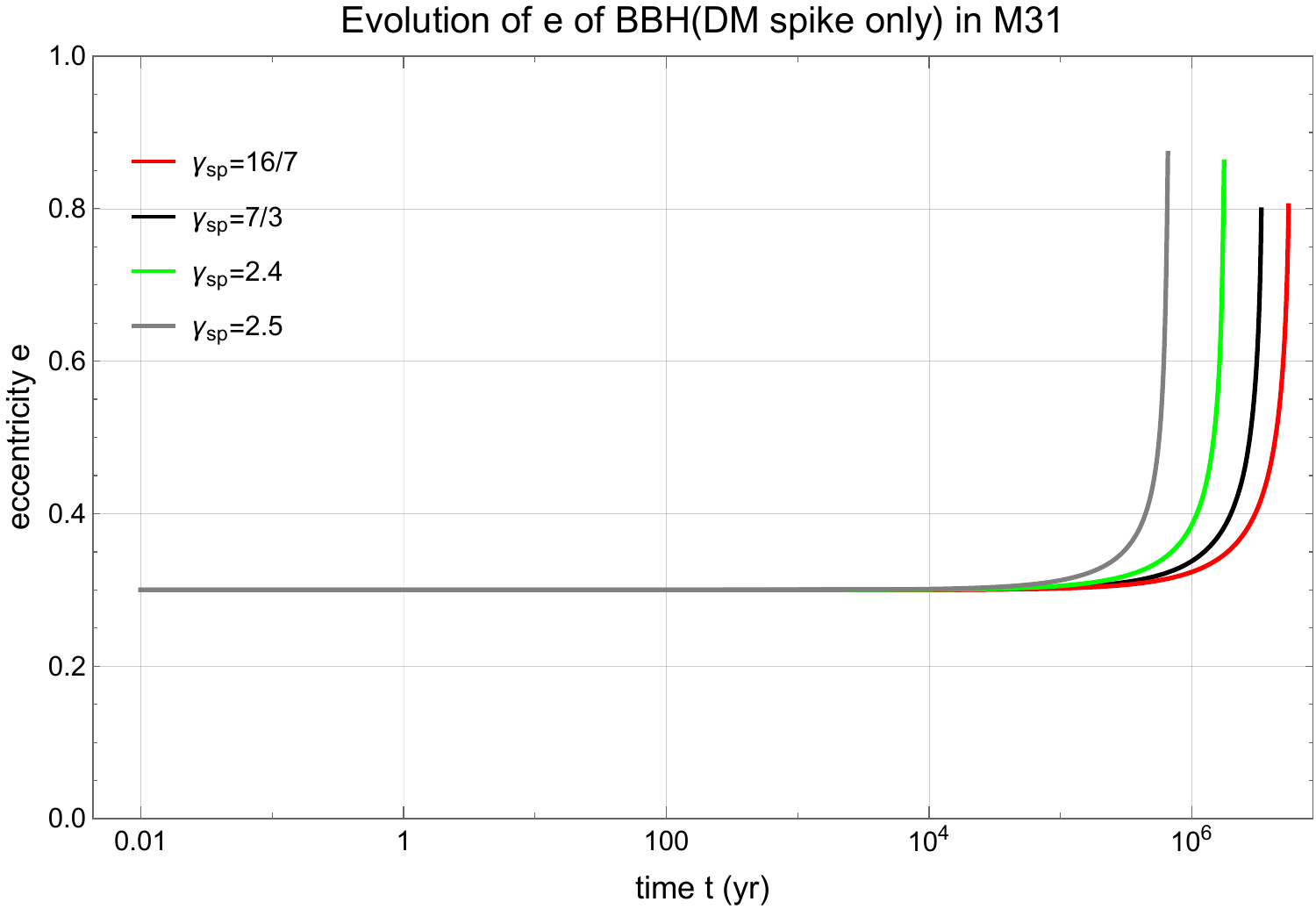}
    \caption{The top and bottom panels illustrate the isolated effect of the DM spike on the temporal evolution of the semi-latus rectum $p$ and orbital eccentricity $e$ for an inspiraling BBH system in M31. Compared to Fig.~\ref{fig:onlyDMLMC}, the orbital parameters in this scenario evolve over a significantly longer timescale.}
    \label{fig:onlyDMM31}
\end{figure}

Fig.~\ref{fig:onlyDMLMC} and Fig.~\ref{fig:onlyDMM31} illustrate the evolution of the semi-latus rectum $p$ and eccentricity $e$ as functions of time $t$ for BBHs in the LMC and M31, considering only the influence of the DM spike. GW reaction is neglected here to isolate and better observe the dynamic impact of the DM spike on the binary's orbital behavior. Due to DF and accretion, the orbital decay of BBH occurs more rapidly, and the presence of DM spikes effectively promotes the merger of systems. The initial conditions are set to $p_0=500R_s$ and $e_0=0.3$. It is evident that a steeper spike profile (i.e., a larger $\gamma_{sp}$) corresponds to a higher DM density, thereby significantly reducing the merger timescale.  The orbital evolution terminates once the secondary BH reaches the ISCO.

\section{Derivation details}
\label{A}

During the inspiral phase, two celestial bodies revolve around each other and gradually approach. The dynamic behavior at this stage is highly dependent on the post-Newtonian method. For analytical simplicity within this model, the accretion disk is assumed to be static. Consequently, the relative velocity $v_{rel}$ between the secondary BH and the surrounding gas is approximated as the orbital velocity $v$ of the secondary BH, maintaining a non-zero $v_{rel}$.

We can average the energy loss rate and the angular momentum loss rate due to gas's Ostriker DF with respect to orbital period:
\begin{align}
    \left\langle\frac{dE}{dt}\right\rangle_{Ostriker} &=\frac{1}{T}\int_{0}^{T}\frac{dE}{dt}|_{Ostriker}dt\nonumber\\&=\frac{1}{T}\int_{0}^{T}F_{Ostriker}vdt\nonumber\\&=\frac{1}{T}\int_{0}^{T}\frac{4 \pi G^2 m_{\text{2}}^2 \rho_{b}(r) I_b}{v}dt
    \nonumber\\&=\int_{0}^{2\pi}\frac{2G^{3/2}m_{\text{2}}^2I_b\rho_{b}(r)p^{1/2}}{(1-e^2)^{-3/2}m^{1/2}}\nonumber\\&\times
    \frac{1}{(1 + e \cos \varphi)^2 (e^2 + 2e \cos \varphi + 1)^{1/2}}d\varphi,
\end{align}

\begin{align}
    \left\langle\frac{dL}{dt}\right\rangle_{Ostriker} &=\frac{1}{T}\int_{0}^{T}\frac{dL}{dt}|_{Ostriker}dt\nonumber\\&=\frac{1}{T}\int_{0}^{T}rF_{Ostriker}\frac{r^2\dot{\varphi}}{v}dt\nonumber\\&=\frac{1}{T}\int_{0}^{T}\frac{4 \pi G^2 m_{\text{2}}^2 \rho_{b}(r) I_br^2\dot{\varphi}}{v^3}dt
    \nonumber\\&=\int_{0}^{2\pi}\frac{2Gm_{\text{2}}^2I_b\rho_{b}(r)p^{2}}{(1-e^2)^{-3/2}m}\nonumber\\&\times
    \frac{1}{(1 + e \cos \varphi)^2 (e^2 + 2e \cos \varphi + 1)^{3/2}}d\varphi.
\end{align}
The second step utilizes the relations $\frac{dE}{dt}=Fv$ and $\frac{dL}{dt}=r F\frac{r\dot{\varphi}}{v}$  respectively. The fourth step applies the relations  $\int_{0}^{T}\frac{dt}{T}(...)=(1-e^2)^{3/2}\int_{0}^{2\pi}\frac{d\varphi}{2\pi}(1 + e \cos \varphi)^{-2}(...)$, Eq.~(\ref{eq22}) and $\dot{\varphi}=\frac{d\varphi}{dt}=\sqrt{\frac{Gm}{p^3}}(1+e\cos\varphi)^{2} $.

For the accretion of gas in the accretion disk by the secondary BH mentioned in Eq.~(\ref{eq36}), we still consider Bondi-Hoyle accretion:
\begin{equation}
    \dot{\mu}=4\pi G^2\lambda_b\frac{\mu^2\rho_{\rm b}}{(v^2+c_b^2)^{3/2}},
\end{equation}
here $c_b$ stands for the sound speed of the gas in disk. We focus on the supersonic regime thus we assume $v \gg c_b$. Considering the influence of the gas accretion only, the orbital equation of motion is~\cite{2022PhRvD.106f4003D} 
\begin{equation}
    \mu \dot{\bm{v}} + \dot{\mu} \bm{v} = - \frac{G \mu m}{r^3} \bm{n},
\end{equation}
where $\bm{n}$ is the unit vector pointing from the central BH to the small BH. The accretion term $\dot{\mu} \bm{v}$ can be thought as a perturbation force:
\begin{equation}
    \bm{f}_{gas} \simeq - \frac{4 \pi G^2 \mu^2 \rho_{b} \lambda_b}{v^3} \bm{v}.
    \label{A5}
\end{equation}
Repeating the above calculation steps, we obtain the energy and angular momentum loss of the secondary BH due to gas accretion within the accretion disk. Since we are concerned with the long-term evolution behavior of the orbital parameters $p(t)$ and $e(t)$ under dissipative effects, we substitute the results obtained above into Eqs.~(\ref{eq13}) and~(\ref{eq14}) to obtain~(\ref{eq39}) and~(\ref{eq40}).

For the numerical results presented in Fig.~\ref{fig:acdfianl}, we adopt the parameter condition $I_b + \lambda_b = 1.5$ in Eqs.~(\ref{eq39}) and~(\ref{eq40}). This approximation is justified as follows. Analogous to the DM spike treatment, we restrict our analysis to the scenario where the secondary BH moves at supersonic velocities within the accretion disk. Given this supersonic motion through the thin disk ($v \gg c_b$), we obtain
\begin{equation}
    \mathcal{M} = \frac{v_{rel}}{c_b} \approx \frac{v}{c_b} \approx \frac{1}{(H/R)} \gg 1,
\end{equation}
where $\mathcal{M}$ is the Mach number. Since \( v_{rel}/c_b \) is very large, the term \( (v_{rel}/c_b)^{-2} \) is close to zero, therefore the term \(\log(1-(v_{rel}/c_b)^{-2})\) can be approximated as \(\log(1)=0\), and the formula can be simplified to
\begin{equation}
    I_b=\frac{1}{2}(\ln \left( 1 - (v_{\text{rel}}/c_b)^{-2} \right) + \ln \Lambda),
\end{equation}
\begin{equation}
    I_b \approx \frac{1}{2} \ln \Lambda.
\end{equation}
Formulas have the same form, therefore we stipulate $I_b$ + $\lambda_b$ = 1.5 for simplicity here. Given the structural similarity between Eqs.~(\ref{eq34}) and~(\ref{A5}), we set $I_{b}+\lambda_{b}=1.5$ to simplify the subsequent numerical analysis

In our derivation, particularly in Eqs.~(\ref{eq37})--(\ref{eq40}), we assumed the term $I_b + \lambda_b \approx 1.5$ based on the supersonic limit $v \gg c_b$. We acknowledge that as the BBH system evolves and the semi-latus rectum $p$ decreases, the orbital velocity $v$ increases significantly, which could potentially lead to a dynamic variation of the Coulomb logarithm $\ln \Lambda$. However, in the early stages of the inspiral, GW radiation has not yet become dominant, and treating $\ln \Lambda$ as a constant is a robust and reasonable approximation for the scientific objectives of this study.

\section{Inspiraling BBH in GC}
\label{GC}

\begin{figure}[htbp]
    \centering
    \includegraphics[width=0.4\textwidth]{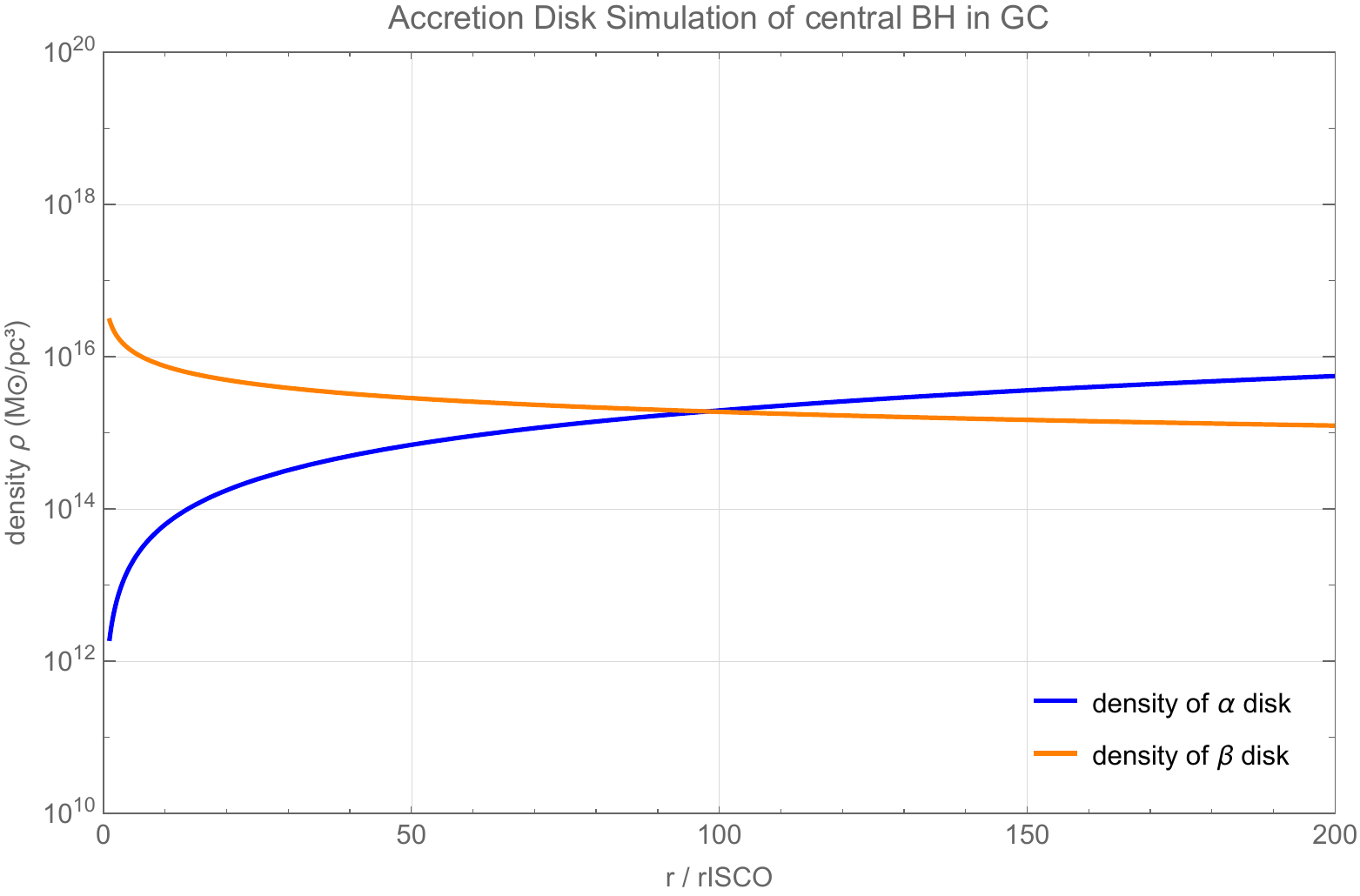}
    \\
    \includegraphics[width=0.4\textwidth]{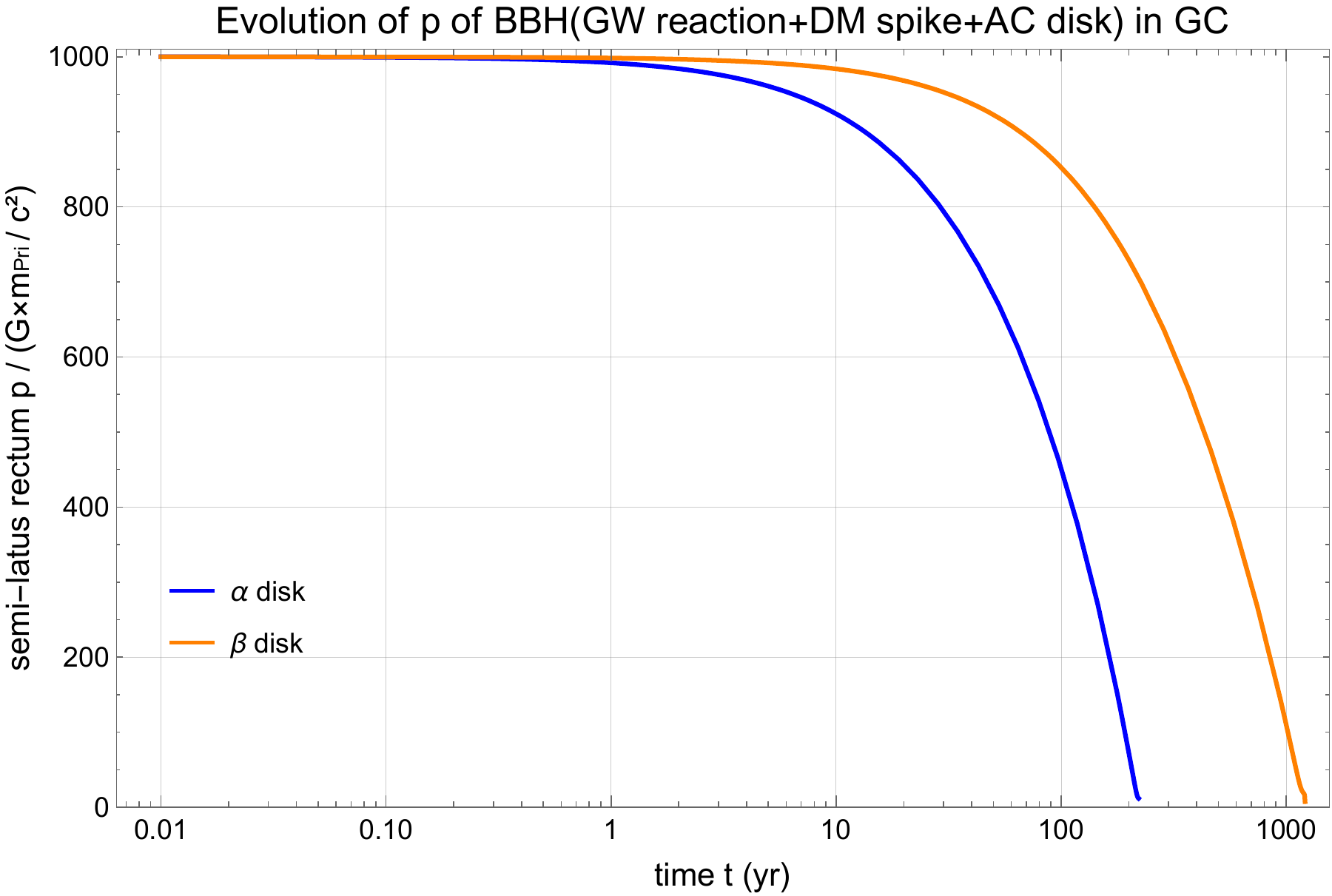}
    \\
    \includegraphics[width=0.4\textwidth]{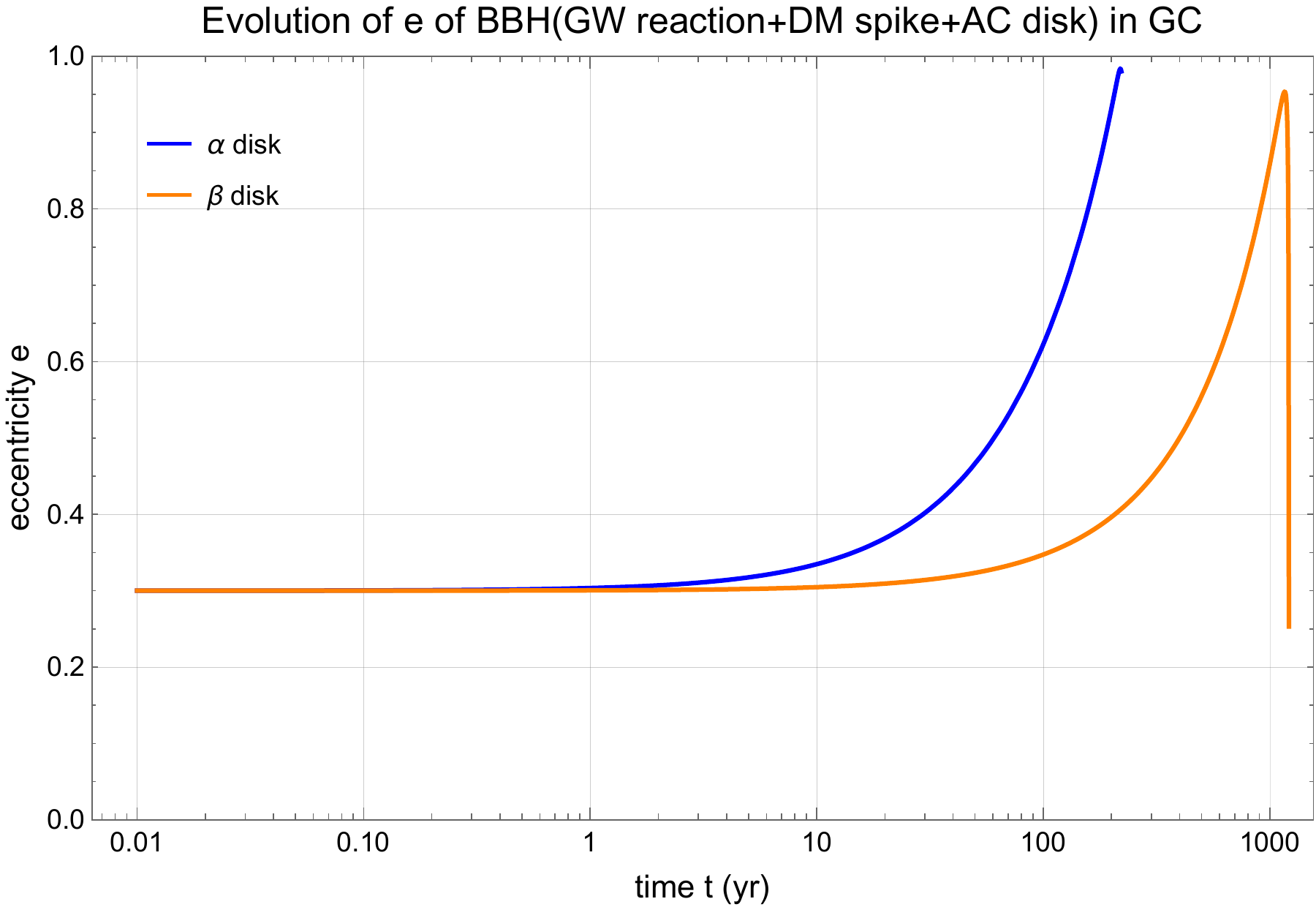}
    \caption{The top panel illustrates the density profiles of the $\alpha$ and $\beta$ accretion disks surrounding the central BH. The middle and lower panels depict the temporal evolution of the orbital parameters for the inspiraling BBH under combined environmental effects. We adopt a DM spike index of $\gamma_{sp}=7/3$. The initial orbital parameters are set to a semi-latus rectum $p_0 = 1000Gm_{\text{1}}/c^2=500R_s$ and an eccentricity $e_0=0.3$. The calculation of the orbital evolution is terminated once the secondary BH reaches the ISCO.}
    \label{fig:acdGC}
\end{figure}

The detectability of DM density distribution via GWs from BBHs in the Galactic Center was investigated in~\cite{Li:2025qtb}. This appendix examines the combined dynamical effects of environmental factors, specifically DM spikes and accretion disks. We have listed the relevant parameters of the inspiraling BBH in Table~\ref{tab:8}. Based on these parameters, we plot the results in Fig.~\ref{fig:acdGC}, calculate the numerical results in Table~\ref{tab:9}. 

\begin{table}
    \centering
    \begin{tabular}{cc}
    \hline\hline
        $m_{\text{1}}$  & $4.26\times10^6M_{\odot} $  \\
        \hline
        $m_{\text{2}}$ & $1000M_{\odot}$ \\
        \hline
        $R_s$ of central BH  & $4.07712\times10^{-7}\mathrm{pc}$ \\
        \hline
        $r_{\text{ISCO}}$ of central BH  & $12.23136\times10^{-7}\mathrm{pc}$ \\
        \hline
        $r_{sp}$  & $12.664\mathrm{pc}$ \\
        \hline
        $\rho_{sp}$ at $r_{sp}$  & $13.425M_{\odot}/\mathrm{pc}^3$ \\
        \hline
        $\gamma_{sp}$  & 7/3 \\
        \hline\hline
    \end{tabular}
    \caption{Adopted parameters for the inspiraling BBH system and its surrounding DM spike at GC.}
    \label{tab:8}
\end{table}

\begin{table}
    \centering
    \begin{tabular}{cccc}
    \hline\hline
        AC disk  & final time[$\mathrm{yr}$] & $p_{final}$ & $e_{final}$\\
        \hline
        $\alpha$ disk &$221$  & $5.95R_s$ & 0.9801\\
        \hline
        $\beta$ disk & $1214$ & $3.759R_s$ & 0.253\\
        \hline\hline
    \end{tabular}
    \caption{Final numerical results for the orbital evolution of the BBH system at the GC, accounting for GW radiation, a DM spike, and accretion disk models ($\alpha$ and $\beta$).}
    \label{tab:9}
\end{table}


\bibliography{refer.bib}

\end{document}